# Accuracy of hybrid functionals with non-self-consistent Kohn-Sham orbitals for predicting the properties of semiconductors

*Jonathan M. Skelton,[1,2,*] David S. D. Gunn,[3] Sebastian Metz[3,4] and Stephen C. Parker[2]*

[1] Department of Chemistry, University of Manchester, Oxford Road, Manchester M13 9PL, UK

[2] Department of Chemistry, University of Bath, Bath BA2 7AY, UK

[3] STFC Daresbury Laboratory, Keckwick Ln., Daresbury, Warrington WA4 4AD, UK

[4] Fraunhofer Institute for Solar Energy Systems ISE, Heidenhofstraße 2, 79110 Freiburg, Germany

Accurately modeling the electronic structure of materials is a persistent challenge to high-throughput screening. A promising means of balancing accuracy against computational cost are non-self-consistent calculations with hybrid density-functional theory, where the electronic band energies are evaluated using a hybrid functional from orbitals obtained with a less demanding (semi-)local functional. We have quantified the performance of this technique for predicting the physical properties of sixteen tetrahedral semiconductors with bandgaps from 0.2-5.5 eV. Provided the base functional predicts a non-metallic electronic structure, bandgaps within 5 % of the PBE0 and HSE06 gaps can be obtained with an order of magnitude reduction in computing time. The positions of the valence and conduction band extrema and the Fermi level are well reproduced, further enabling calculation of the band dispersion, density of states, and dielectric properties using Fermi's Golden Rule. While the error in the non-self-consistent total energies is ~50 meV atom$^{-1}$, the energy-volume curves are reproduced accurately enough to obtain the equilibrium volume and bulk modulus with minimal error. We also test the dielectric-dependent scPBE0 functional and obtain bandgaps and dielectric constants to within 2.5 % of the self-consistent results, which amount to a significant improvement over self-consistent PBE0 for a similar computational cost. We identify cases where the non-self-consistent approach is expected to perform poorly, and demonstrate that partial self-consistency provides a practical and efficient workaround. Finally, we perform proof-of-concept calculations on CoO and NiO to demonstrate the applicability of the technique to strongly-correlated open-shell transition-metal oxides.

## INTRODUCTION

First-principles modeling is an important part of contemporary materials science, and has rapidly evolved from an explanatory tool to support experimental characterisation to enabling theory-led materials design and optimisation to address some of the most pressing technological challenges of our time.[1] In particular, the balance of accuracy and computational cost provided by Kohn-Sham density-functional theory (KS DFT)[2,3] has allowed high-throughput modeling studies to access consistent datasets for hundreds to tens of thousands of materials.[4-6] This enables new structure-function relationships to be identified and explored,[7,8] and allows calculations to screen large pools of materials to select the most promising candidates for a particular application and thereby focus experimental synthesis and characterisation.[9-11] Theory-led materials discovery has shown a large amount of promise, including, for example, identifying novel candidate photovoltaics[12-14] and thermoelectrics[15-18] for renewable energy applications and photocatalysts for generating solar fuels.[19,20]

Contemporary materials-design problems invariably require accurate prediction of the electronic structure to estimate properties such as the bandgap, the electronic density of states and band dispersion/carrier effective masses, and the dielectric function.[11] However, (semi-)local DFT has a well-known tendency to underestimate bandgaps, which in extreme cases can lead to materials being incorrectly predicted to be metals.[1,21] At a fundamental level, approximate DFT functionals often fail to reproduce the variation in the exchange-correlation potential with the number of electrons, in particular the discontinuities at integer electron counts, and thus the gap between the highest-occupied and lowest-unoccupied KS electronic states typically differs from the true quasiparticle bandgap.[1,22-24] (Semi-)local DFT functionals can also display self-interaction error, whereby the Coulomb interaction of an electron with its own density is not fully cancelled by the approximate exchange and correlation.[25] This leads to over-delocalisation of electronic states, particularly spatially-localised d and f orbitals, and has been linked to a host of issues including a tendency towards metallicity[1,22] and an underestimation of reaction barriers.[26]

Within the materials-modeling community, these deficiencies are increasingly being addressed with so-called hybrid functionals, which include a fraction of the exact Hartree-Fock exchange to produce more accurate orbital energies and to counteract the self-interaction error in (semi-)local functionals. However, calculating the non-local exchange incurs a substantial overhead,[27] making hybrids prohibitively expensive for high-throughput screening studies. Various approaches are being taken to address this, including the development of new functionals[23,28] and alternative methods[29] optimised for predicting bandgaps, and more empirical models based on statistical trends and artificial intelligence.[30,31]

Recently, two-functional hybrid DFT calculations, where the hybrid is applied non-self-consistently to the density and Kohn-Sham orbitals from a less demanding (semi-)local functional, have emerged as a good balance of accuracy and cost. In general, the error in a DFT calculation arises from approximations in the exchange-correlation potential and errors in the self-consistent orbitals and density.[32,33] In cases where the leading error in a DFT functional is in the density, evaluating the orbitals with a different method and applying the functional non-self-consistently can yield improved results. For example, using orbitals from Hartree-Fock theory, which are formally free from self-interaction error, with semi-local DFT functionals has been shown to improve calculated reaction barriers while maintaining good predictions of the reaction energies.[26] If, on the other hand, the dominant error is in the potential, applying a hybrid functional to the orbitals from a less expensive (semi-)local functional should give improved orbital energies at a lower cost than a fully self-consistent hybrid calculation. This approach is reminiscent of the "single-shot" $G_0W_0$ variant of the $GW$ method,[34] and has been applied successfully to solid-state modeling problems including band alignment,[35] high-$k$ dielectrics,[36] and semiconductor alloys.[37]

However, there has been comparatively little systematic investigation into the general accuracy of this technique for predicting the properties of semiconductors. In this work, we present a systematic study of the performance of non-self-consistent hybrid functionals on a test set of sixteen tetrahedral semiconductors with bandgaps from 0.2-5.5 eV. We test several common base functionals spanning the local-density approximation (LDA), generalized-gradient approximation (GGA) and meta-GGA forms, and we consider bandgaps, electronic band dispersions and densities of states, dielectric responses, and structural and mechanical properties. We demonstrate that this approach can be used successfully with the widely-used PBE0[38] and HSE06 hybrids,[39] as well as with the dielectric-dependent scPBE0 functional.[40] We identify cases where this method can be expected to perform poorly, and show that partial self-consistency provides an effective and practical workaround. Finally, we also show that the technique can be used for strongly-correlated, open-shell transition-metal oxides, which have historically posed a challenge to (semi-)local DFT.

**METHODS**

**Density-Functional Theory Calculations.** Calculations were performed on a set of sixteen tetrahedral semiconductors using the plane-wave DFT formalism implemented in the Vienna *Ab initio* Simulation Package (VASP) code.[41] A list of compounds and the experimental room-temperature (300 K) lattice constants, bandgaps and high-frequency dielectric constants are collected in Table 1.

We tested a total of nine electronic structure methods: the LDA using the Ceperley-Alder parameterisation of the correlation energy;[42] the Perdew-Burke-Ernzerhof (PBE) GGA and the revised variant of PBE for solids (PBEsol);[43,44] the initial and revised Tao-Perdew-Staroverov-Scuseria (TPSS/revTPSS) and the newer Strongly-Constrained and Appropriately Normalized (SCAN) meta-GGA functionals;[45–48] pure Hartree-Fock (HF); and the PBE0 and 2006 Heyd-Scuseria-Ernzerhof (HSE06) hybrid functionals.[38,39] All calculations were performed at the 300 K lattice constants in Table 1, which, for these systems, completely specify the crystal structures.

The ion cores were modelled using the projector augmented-wave (PAW) method[49,50] with the outermost s and p electrons and the Ga, As and In semi-core d electrons included the valence region. The VASP code is distributed with PAW potentials generated from LDA and PBE reference configurations, of which only the former can be used for LDA calculations; we therefore used the LDA potentials for all our calculations. The plane-wave cutoffs were selected according to the largest values suggested by the PAW potentials used, and Γ-centered $k$-point meshes generated using the Monkhorst-Pack scheme[51] were chosen to give a minimum spacing between points of 0.25 Å⁻¹ along each of the reciprocal-lattice vectors.

These choices were made to mimic the manner in which these parameters might be set in a typical high-throughput screening study, but the convergence of various properties with PBE was checked by increasing the cutoff and $k$-point sampling to 1.5 and 2 × the base values shown in Table 1 (see Supporting Information, Section S1). These tests suggest that the chosen cutoffs are sufficient to converge the bandgaps, dielectric constants and the equilibrium total energy, volume and bulk moduli to within 1 % for the majority of the compounds, while the chosen $k$-point meshes are sufficient to converge the bandgaps to < 1 %, the dielectric constants to < 10 %, the equilibrium total energy and volume to < 1 %, and the bulk moduli to < 5 %.

**Table 1.** Tetrahedral semiconductors studied in this work with the measured room-temperature (300 K) lattice constants $a$, direct/indirect bandgaps $E_{g,dir}/E_{g,indir}$, high-frequency dielectric constants $\varepsilon_\infty$ and the plane-wave cutoffs and $k$-point sampling used in the calculations.[52–55]

| | $a$ [Å] | $E_{g,dir}$ [eV] | $E_{g,indir}$ [eV] [a] | $\varepsilon_\infty$ | Cutoff [eV] | $k$-points |
|---|---|---|---|---|---|---|
| C | 3.567 | - | 5.48 | 5.7 | 400 | 14×14×14 |
| Si | 5.431 | - | 1.12 | 11.85 | 245.7 | 10×10×10 |
| Ge | 5.658 | - | 0.66 | 16.6 | 174 | 8×8×8 |
| AlP | 5.451 | - | 2.45 | 7.54 | 255.2 | 8×8×8 |
| AlAs | 5.661 | - | 2.16 | 8.16 | 288.8 | 8×8×8 |
| AlSb | 6.135 | - | 1.58 | 10.24 | 241 | 8×8×8 |
| GaP | 5.451 | - | 2.26 | 9.11 | 282.8 | 8×8×8 |
| GaAs | 5.653 | 1.42 | - | 10.88 | 288.8 | 8×8×8 |
| GaSb | 6.096 | 0.73 | - | 14.44 | 282.8 | 8×8×8 |
| InP | 5.869 | 1.35 | - | 9.61 | 255.2 | 8×8×8 |
| InAs | 6.058 | 0.36 | - | 12.25 | 288.8 | 8×8×8 |
| InSb | 6.479 | 0.18 | - | 15.68 | 239.2 | 8×8×8 |
| ZnS | 5.41 | 3.68 | - | 5.13 | 276.8 | 10×10×10 |
| ZnSe | 5.668 | 2.71 | - | 6.2 | 276.8 | 8×8×8 |
| ZnTe | 6.101 | 2.39 | - | 7.28 | 276.8 | 8×8×8 |
| CdTe | 6.48 | 1.48 | - | 7.21 | 274.3 | 8×8×8 |

Self-consistent optimisation of the density and orbitals was carried out to a tolerance of 1 $\mu$eV ($10^{-6}$ eV) on the total energy. Integration of the Brillouin-zone was performed using the tetrahedron method with Blöchl corrections.[56] The precision of the charge-density grids was set automatically to avoid aliasing errors (`PREC = Accurate` in VASP). The PAW projection was performed in reciprocal space (`LREAL = .FALSE.`), and non-spherical contributions to the gradient corrections inside the PAW spheres were accounted for (`LASPH = .TRUE.`).

Band dispersions were obtained by calculating the eigenvalues along strings of 21 high-symmetry $k$-points between the $X = (1/2 , 0 , 1/2)$, $\Gamma = (0,0,0)$ and $L = (1/2 , 1/2 , 1/2)$ special points in the Brillouin zone, which were introduced as zero-weighted ("fake") $k$-points alongside the irreducible points on the sampling meshes in Table 1. For these calculations, the number of bands was set to 12, 24 and 36, respectively, for compounds with 8, 18 and 28 valence electrons, giving at least the same number of valence and conduction bands.

Energy-dependent dielectric functions were computed within the independent-particle random-phase approximation (IP-RPA) using the linear-optics routines in VASP.[57] We note that this implementation includes the additional commutator terms in the dipolar momentum matrix elements that arise from the non-local potential introduced by the PAW method. Since the IP-RPA relies on a sum over empty conduction states, the number of bands was set to 24, 48 and 60 for compounds with 8, 18 and 28 valence electrons, providing a 4-6 × excess of virtual states. The high-frequency dielectric constant $\varepsilon_\infty$ was obtained by averaging the diagonal elements of the real dielectric matrix at $E = 0$. As discussed in detail in Supporting Information, Section S2, we also compared the IP-RPA against two alternative methods of calculating $\varepsilon_\infty$, *viz.* using density-functional perturbation theory (DFPT)[57] and from the self-consistent response to finite fields.[58,59]

Finally, energy-volume curves were computed by performing a sequence of single-point energy calculations with the lattice constants in Table 1 adjusted to vary the cell volume from -5 to +10 % in steps of 1 %. As recommended in the VASP manual, these calculations were performed with larger plane-wave cutoffs of 1.5 × the base values in Table 1, although convergence testing showed this is not necessarily required (see Supporting Information, Section S1).

**Non-Self-Consistent Calculations.** Non-self-consistent calculations were performed in two steps: a self-consistent calculation was performed with the chosen base functional to obtain a set of orbitals (VASP `WAVECAR` file), which were then reloaded and the electron band energies recalculated using the hybrid functional without updating the orbitals. This was performed using the `ALGO = Eigenval` tag in VASP, in conjunction with `NELM = NELMIN = 1` to force a single evaluation of the band energies. This technique was tested for all the types of calculation outlined above, i.e. computation of bandgaps, electronic density of states and dispersion curves, IP-RPA dielectric functions, and energy-volume curves, with the technical parameters, *viz.* cutoff, *k*-points, and number of bands, set as described.

**Partially-Self-Consistent Calculations.** Partially-self-consistent calculations were performed as for the self-consistent calculations but with the number of orbital updates restricted by setting a reduced convergence threshold of 1 eV - 100 $\mu$eV ($10^0$-$10^{-4}$ eV) on the total energy, which is controlled using the `EDIFF` tag in VASP.

**Additional Test Calculations on CoO and NiO.** Calculations were performed on the low-temperature monoclinic phase of CoO reported in Ref. [60] (spacegroup $C2/m$, $a$ = 5.182 Å, $b$ = 3.018 Å, $c$ = 3.019 Å, $\beta$ = 125.6 °; ICSD # 191775) and cubic NiO with the experimental room-temperature lattice constant of 4.176 Å.[61] Both primitive cells contain four atoms and have antiferromagnetic ordering. A 400 eV plane-wave cutoff was used with 6×9×11 and 10×10×10 Γ-centered *k*-point meshes for CoO and NiO, respectively, along with PAW pseudopotentials including the O 2p and Co/Ni 4s, 4d and semi-core 3p electrons in the valence shell. We performed self-consistent calculations with PBE, HSE 06 and PBE + $U$ using the rotationally-invariant Dudarev model[62] and a correction $U_{\text{eff}}$ = 7 eV applied to the metal d orbitals, chosen based on the work in Ref. [63]. Non-self-consistent and partially-self-consistent HSE 06 calculations were also performed starting from the PBE and PBE + $U$ orbitals.

**RESULTS AND DISCUSSION**

**Calculated Bandgaps.** Fig. 1 compares the mean absolute error (MAE) in bandgaps calculated using the nine electronic-structure methods in our test to the experimental values in Table 1. The MAE is calculated as usual from:

$$\text{MAE} = \frac{1}{N} \sum_N \left| P_N^{\text{calc}} - P_N^{\text{ref}} \right| \tag{1}$$

where the sum runs over the $N$ compounds in the test set and $P_N^{\text{calc}}$ and $P_N^{\text{ref}}$ are the calculated and reference values of the property $P$ respectively.

The accuracy of the (semi-)local functionals falls in the order LDA < GGA < meta-GGA, with MAEs ranging from $0.92 \pm 0.44$ eV with LDA to $0.55 \pm 0.28$ eV with the SCAN meta-GGA. As expected, all six functionals consistently underestimate the bandgaps (Fig. S3.1). Interestingly, the newer SCAN functional shows a clear improvement over TPSS and revTPSS, with both a smaller MAE and a narrower spread of errors. Pure HF significantly overestimates the gaps and produces the largest MAE of $6.37 \pm 1.58$ eV. Of the two hybrids, HSE06 predicts the closest gaps to experiment, with a MAE of $0.17 \pm 0.1$ eV, while PBE0 tends to overestimate the gaps but nonetheless improves upon the most of the (semi-)local functionals.

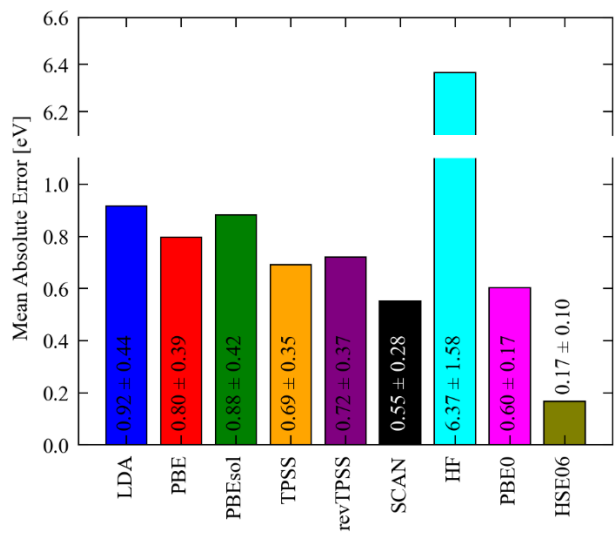

**Figure 1.** Mean absolute error (MAE; Eq. 1) in the calculated bandgaps of the 16 materials in Table 1 obtained using the LDA (blue), the PBE and PBEsol GGAs (red, green), the TPSS, revTPSS and SCAN meta-GGAs (orange, purple, black), Hartree-Fock (HF; cyan) and the PBE0 and HSE06 hybrids (magenta, olive). The text labels give the MAEs and standard deviations. All calculations were performed at the experimental lattice constants in Table 1. Note the break in the vertical axis to accommodate the large error with HF.

Of the sixteen semiconductors in the test set, four have bandgaps < 1 eV, *viz.* Ge (0.66 eV), GaSb (0.73 eV), InAs (0.36 eV) and InSb (0.18 eV). All six (semi-)local functionals incorrectly predict InAs and InSb to be metallic. LDA and PBEsol also predict Ge to be metallic, while PBE predicts a narrow gap of 28 meV. LDA also predicts GaSb to be a metal, while PBE and PBEsol predict gaps of 93 and 49 meV. Our test set therefore incorporates several compounds where some of the (semi-)local DFT methods erroneously predict a (near-)metallic electronic structure. InAs, InSb and GaSb are direct-gap semiconductors, and the error for all three is due to a singly-degenerate conduction band at the Brillouin-zone centre ($\Gamma$ point) being placed below a triply-degenerate occupied band and thus producing a negative bandgap. This also leads to LDA and PBEsol predicting Ge to be a metal, despite Ge being an indirect-gap semiconductor.

If the Kohn-Sham orbitals from the base functionals are reasonable and the leading error in the calculated bandgaps is due to approximations in the exchange-correlation potential, the error should be reduced by recalculating the band energies non-self-consistently using the hybrids. To test this, we performed a series of non-self-consistent calculations with HF, PBE0 and HSE06 using the density and orbitals obtained with the six (semi-)local functionals (Fig. 2, Fig. 3).

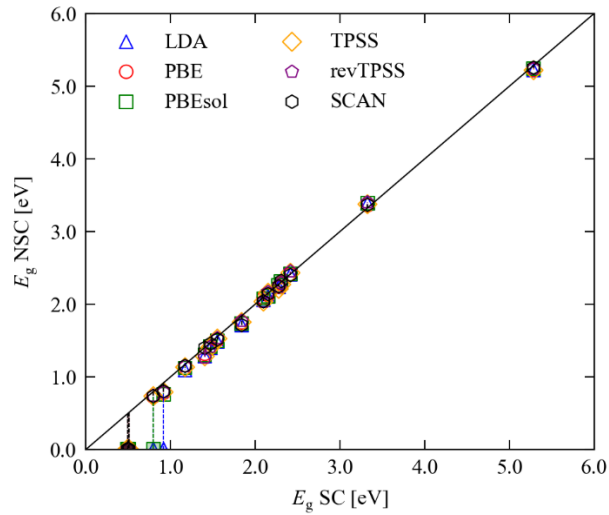

**Figure 2.** Comparison of self-consistent (SC) and non-self-consistent (NSC) HSE06 bandgaps calculated using orbitals from the LDA (blue triangles), PBE (red circles), PBEsol (green squares), TPSS (orange diamonds), revTPSS (purple pentagons) and SCAN (black hexagons) functionals. All calculations were performed at the experimental lattice constants in Table 1. Shaded markers indicate systems for which the base functional predicts a bandgap of < 10 meV.

The HSE06 bandgaps obtained from the self-consistent and non-self-consistent calculations show near-quantitative agreement save for a small number of clear outliers (Fig. 2) corresponding to calculations where the base functional predicts a metallic electronic structure with $E_g < 10$ meV, for which we would anticipate a larger error in the orbitals. The same trends are observed with PBE0 (Fig. S3.2). We also calculated the percentage mean absolute relative error (MARE) in the non-self-consistent and self-consistent PBE0 and HSE06 bandgaps according to (Fig. 3):

$$\text{MARE} = 100 \times \frac{1}{N} \sum_N \left| \frac{P_N^{\text{calc}} - P_N^{\text{ref}}}{P_N^{\text{ref}}} \right| \qquad (2)$$

Excluding the easily-identified calculations where the base functional predicts a metallic electronic structure, the MARE with respect to the self-consistent result is 3-4 % for both PBE0 and HSE06 with negligible variation among the base functionals.

The fundamental bandgap measured experimentally is the difference between the ionisation energy $I$ and electron affinity $A$:[64]

$$E_g = I - A \qquad (3)$$

$I$ and $A$ can be equated to the energy differences between the $N$-electron system, $E(N)$, and systems with a single electron added or removed:

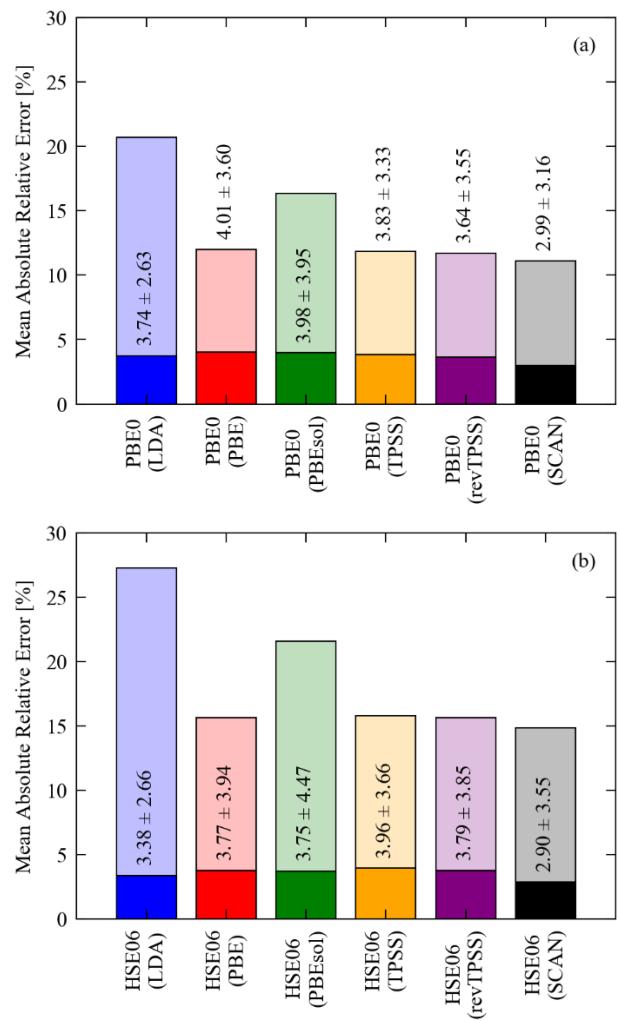

**Figure 3.** Mean absolute relative error (MARE; Eq. 2) in the PBE0 (a) and HSE06 (b) bandgaps calculated non-self-consistently with orbitals from the LDA (blue), PBE (red), PBEsol (green), TPSS (orange), revTPSS (purple) and SCAN (black) functionals. The solid bars show the MARE scores obtained excluding data points where the base functional predicts a bandgap of < 10 meV and the shaded bars show the scores calculated across all data points. The text labels indicate the averages and standard deviations depicted by the solid bars.

$$I = E(N-1) - E(N)$$
$$A = E(N) - E(N+1)$$

(4)

It can be shown that the electron density $\rho(N)$ and the total energy $E(N)$ should vary linearly with fractional changes in electron count but can change discontinuously at integer occupation numbers, which is known as the derivative discontinuity and is an intrinsic property of KS DFT.[65–67] In a semiconductor, this can be understood conceptually as each added (pair of) electrons occupying different bands and resulting in very different spatial electron densities.[68] The exact KS potential, and the resulting KS orbital energies, can therefore jump by constant values at each integer $N$.[68] The derivative discontinuity is not well reproduced in (semi-)local DFT functionals, and self-interaction error in such functionals can additionally lead to deviation from the expected linear change in orbital energies at fractional occupation.[67] Both issues ultimately result in the KS bandgaps obtained from (semi-)local functionals differing from the physical quasiparticle bandgaps by a system-dependent additive constant $\Delta$, i.e.:[69]

$$E_{\text{g}} = \epsilon_{\text{CBM}}^{\text{KS}} - \epsilon_{\text{VBM}}^{\text{KS}} + \Delta = E_{\text{g}}^{\text{KS}} + \Delta$$

(5)

where $\epsilon_{\text{CBM}}^{\text{KS}}$ and $\epsilon_{\text{VBM}}^{\text{KS}}$ correspond to the KS orbital energies of the unoccupied conduction-band minimum and occupied valence-band maximum, respectively, and the difference gives the KS bandgap $E_{\text{g}}^{\text{KS}}$. Hybrid functionals both correct for self-interaction error and include a better approximation of the derivative discontinuity.[67] The near-quantitative reproduction of the self-consistent hybrid bandgaps in the non-self-consistent calculations therefore suggests that the latter do incorporate the shift in orbital energies due to the derivative discontinuity in the hybrid exchange-correlation potential, and is thus consistent with the leading error in the majority of the (semi-)local calculations being in the potential rather than in the orbitals.

Less satisfactory results are obtained for non-self-consistent HF calculations. Where the base functional incorrectly predicts a metal, large errors are obtained as for PBE0 and HSE06. However, we obtain significant errors for AlSb and GaSb irrespective of the bandgap predicted by the base functional (Fig. S3.3). Compared to the other semiconductors, the self-consistent HF calculations give anomalous overestimates of the AlSb, GaSb and InSb bandgaps (Fig. S3.1). Excluding these data points, as well as those where the base functionals predict bandgaps below 10 meV, we obtain MAREs of 10-12 % between the self-consistent and non-self-consistent results. Given that HF differs substantially from PBE0 and HSE06, which replace fractions of the PBE exchange energy with the exact HF exchange, we ascribe the larger errors to larger differences between the (semi-)local and self-consistent HF orbitals.

Starting from the PBE orbitals, the hybrid calculations took an average 10-11 iterations of the conjugate-gradient algorithm to reach self-consistency. For the majority of the test systems, our calculations thus suggest that the PBE0 and HSE06 bandgaps can thus be obtained non-self-consistently to an accuracy of ~5 % with an order of magnitude reduction in the required computing time. It is also notable that the average error and spread in the non-self-consistent results fall as the gap predicted by the base functional increases, suggesting that the accuracy should generally improve for wider-gap semiconductors and insulators (Figs. S3.4-S3.6).

**Density of States, Band Dispersions and Dielectric Properties.** Fig. 4 and Figs. S4.1-S4.16 compare the band dispersions and electronic density of states predicted by self-consistent HSE06 and non-self-consistent HSE06 using PBE orbitals. Save for the two systems that PBE predicts to be metals, the non-self-consistent hybrid calculations reproduce the electronic structure in the vicinity of the band edges to a high degree of accuracy, with an absolute energy difference of $89 \pm 137$ meV between equivalent bands in the two sets of dispersions. On closer inspection, the errors are consistently larger for the occupied semi-core d bands present in some of the systems, with errors of $323 \pm 145$ meV for these bands and $40 \pm 62$ meV for the other s/p valence and conduction bands. This can most likely be explained by self-interaction error in PBE. In these test systems, the d bands are deep in energy relative to the band edges, so the larger error has a minimal effect on the bandgaps. However, for compounds with shallow d bands, for example transition metal oxides, we would expect self-interaction error to have a larger impact, and we return to this point later in the discussion.

We also compared the energies of the valence band maxima and conduction band minima ($E_{\text{VBM}}/E_{\text{CBM}}$) and the Fermi energies $E_{\text{F}}$ from the PBE and self-consistent/non-self-consistent HSE06 calculations (Tables S4.1 and S4.2). For all sixteen semiconductors, the calculated $E_{\text{F}}$ are close to the VBM (i.e. $E_{\text{F}} \approx E_{\text{VBM}}$). Relative to HSE06, the PBE $E_{\text{VBM}}$ are $520 \pm 106$ meV higher and the $E_{\text{CBM}}$ are $268 \pm 139$ meV lower, indicating that differences in the energies of the occupied bands have a larger impact on the bandgap underestimation. Again excluding InAs and InSb, the non-self-consistent $E_{\text{VBM}}$ f underestimate the self-consistent results by $3.4 \pm 61$ meV, while the $E_{\text{CBM}}$ are underestimated by $48 \pm 38$ meV. The non-self-consistent calculations therefore predict the VBM energy roughly correctly but tend to underestimate the position of the CBM.

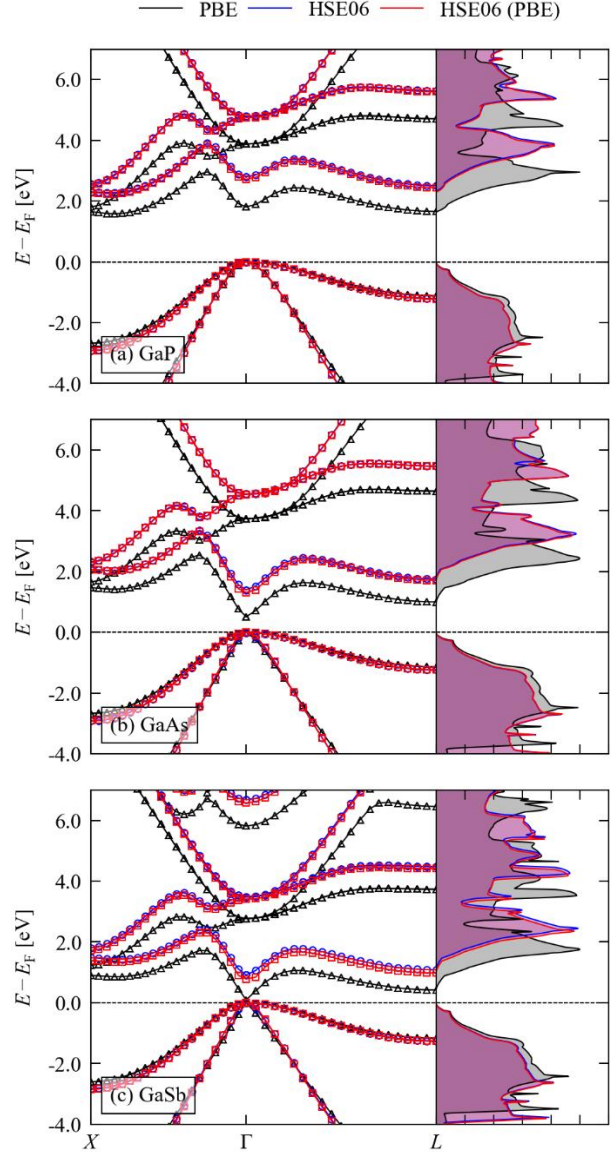

**Figure 4.** Calculated electronic band dispersion and density of states $g(E)$ for GaP (a), GaAs (b) and GaSb (c). Each plot compares the electronic structures obtained from self-consistent PBE and HSE06 calculations (black/blue) to non-self-consistent HSE06 calculations using the PBE orbitals (red). Note that the self-consistent and non-self-consistent HSE06 dispersions and density of states curves in the three subplots are almost superimposable.

Using the linear-optics routines in VASP, the imaginary part of the dielectric function $\varepsilon_{\text{im}}(E)$ can be evaluated in the independent-particle random-phase approximation (IP-RPA; more commonly known as Fermi's Golden Rule) from a summation over transitions between occupied and virtual orbitals, and the real part $\varepsilon_{\text{re}}(E)$ is then obtained from $\varepsilon_{\text{im}}(E)$ *via* the Kramers-Kronig relation. Although this method has some notable downsides, including a requirement to include and optimise a large number of virtual states in the calculations, and the neglect of local-field effects in the VASP implementation,[57] it is computationally undemanding and can be used to evaluate the energy-dependent dielectric function $\varepsilon(E) = \varepsilon_{\text{re}}(E) + i\varepsilon_{\text{im}}(E)$ from the non-self-consistent electronic structure. As shown in Fig. 5 and Figs. S4.17-S4.32, the non-self-consistent calculations give a quantitative reproduction of the $\varepsilon(E)$ and thus would allow dependent properties such as the wavelength-dependent refractive index $n(\lambda)$, extinction coefficient $\kappa(\lambda)$, absorption coefficient $\alpha(\lambda)$ and reflectance $R(\lambda)$, to be obtained at a reduced cost.

Of particular interest is the high-frequency dielectric constant $\varepsilon_{\infty}$, given by the real part of the dielectric function at $E = 0$, i.e. $\varepsilon_{\infty} = \varepsilon_{\text{re}}(E = 0)$. Fig. 6 compares the $\varepsilon_{\infty}$ obtained from the self-consistent and non-self-consistent calculations to the experimental measurements in Table 1. The mean absolute relative error in the HSE06 results is $9.13 \pm 2.89$ %. The error increases for larger $\varepsilon_{\infty}$, due to the inverse relationship between $\varepsilon_{\infty}$ and $E_{\text{g}}$ such that a small absolute error in a narrow bandgap translates to a large error in $\varepsilon_{\infty}$.

For all but one system the calculations underestimate $\varepsilon_{\infty}$. As discussed in Supporting Information, Section S2, this is likely a limitation of the IP-RPA, and alternative methods reduce the error to $5.56 \pm 4.74$ % but require self-consistent calculations.

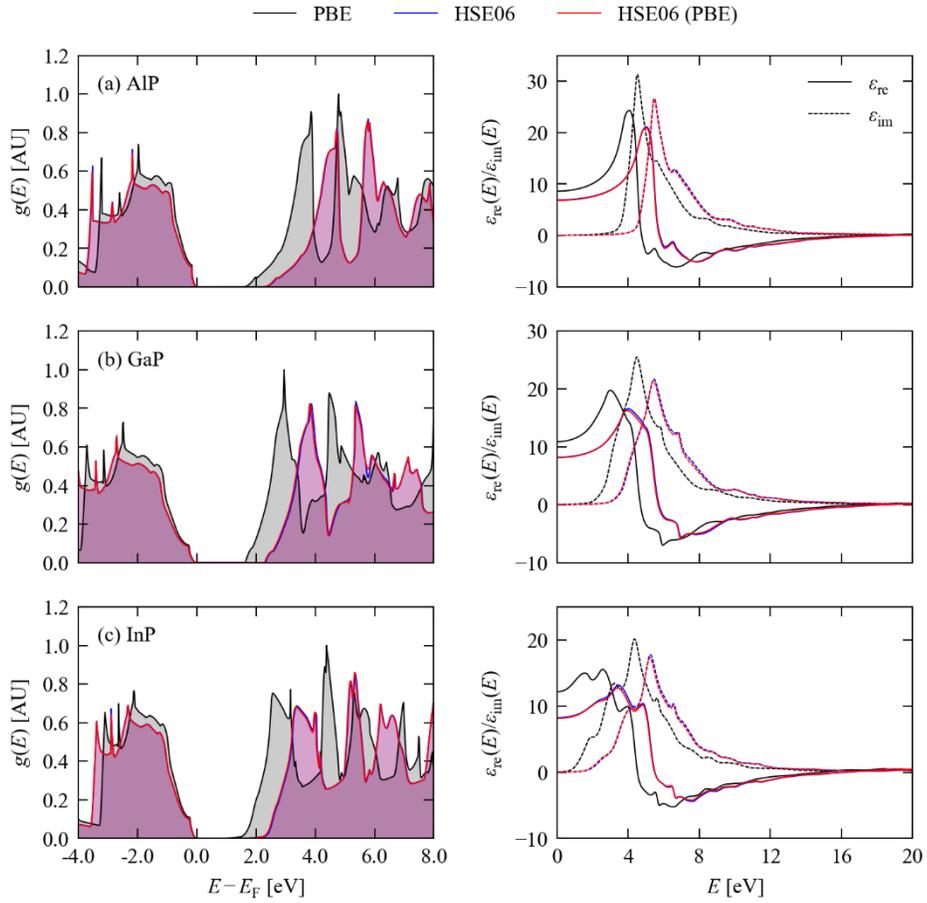

**Figure 5.** Density of states $g(E)$ (DoS; left) and real and imaginary energy-dependent dielectric functions $\varepsilon_{re}(E)/\varepsilon_{im}(E)$ of AlP (a), GaP (b) and InP (c). Each plot compares curves obtained from self-consistent PBE and HSE06 calculations (black/blue) to non-self-consistent HSE06 calculations using the PBE orbitals (red). Note that the self-consistent and non-self-consistent HSE06 density of states and dielectric functions in the six subplots are almost superimposable.

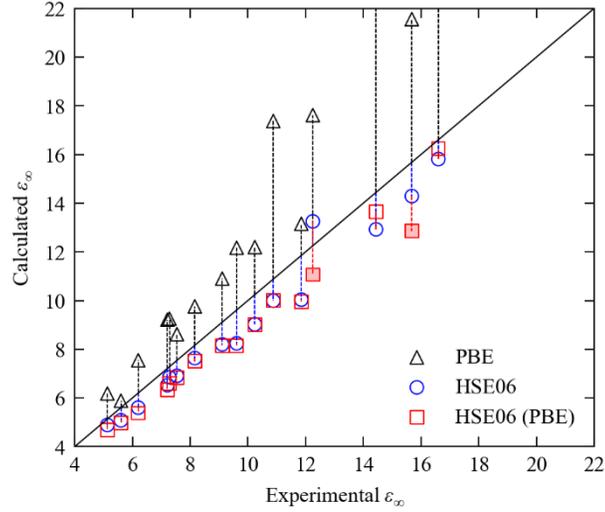

**Figure 6.** Comparison of the measured high-frequency dielectric constants $\varepsilon_\infty$ in Table 1 to calculations with self-consistent PBE and HSE06 (black triangles/blue circles) and non-self-consistent HSE06 calculations using the PBE orbitals (red circles). PBE predicts anomalously-large $\varepsilon_\infty$ of $2.0 \times 10^3$ and $1.5 \times 10^2$ for Ge and GaSb (see Supporting Information Section S2), and so these markers fall outside of the plot range. Shaded markers identify non-self-consistent calculations where PBE predicts a bandgap of < 10 meV.

Nevertheless, we consider the accuracy to be sufficient for high-throughput screening studies. The HSE06 IP-RPA $\varepsilon_\infty$ are in general a better match to experiment than PBE calculations with more accurate methods, although computing $\varepsilon_\infty$ from the self-consistent response to a finite field[58,59] with PBE gave good results (see Supporting Information, Section S2).

Excluding the two systems for which PBE predicts a metallic electronic structure, the $\varepsilon_\infty$ obtained from the non-self-consistent calculations reproduce the self-consistent values, with a MARE of 2.21 $\pm$ 1.64 %. The good reproduction of $\varepsilon_\infty$, particularly for materials with wider bandgaps, suggests the possibility of non-self-consistent dielectric-dependent hybrid functionals, which we investigate further below.

Evaluating the dispersion/density of states and IP-RPA dielectric functions typically require additional $k$-points and virtual states, respectively, to be included in the calculations, and therefore the absolute saving in computation time using the non-self-consistent approach will typically be larger than for calculating the bandgap. As well as allowing for more compounds to be included in a high-throughput screening study, for example, these time savings could also be used for a denser Brillouin-zone sampling ($k$-point mesh) for improved accuracy in "production" calculations.

**Total Energies and Equations of State.** Since the non-self-consistent hybrid calculations are not variational, we would not expect the Hellmann-Feynman forces to be reasonable, and we might also expect significant errors in the total energies. To test this, we compared energy-volume curves ($E/V$ curves) calculated with PBE0 to non-self-consistent calculations using the PBE orbitals (Fig. 7, Figs. S5.1-S5.16). The energy-volume curves were calculated over expansions of -5 to +10 % about the volumes in Table 1 in steps of 1 %, and the data was fit to the Birch-Murnaghan equation of state (EoS; Eq. 6)[70] to estimate for each compound an equilibrium total energy $E_0$, cell volume $V_0$, bulk modulus $B_0$, and pressure derivative $B_0' = (\partial B_0 / \partial p)_{p=0}$.

$$E(V) = \frac{9}{16} V_0 B_0 \left\{ \left[ \left( \frac{V_0}{V} \right)^{\frac{2}{3}} - 1 \right]^3 B_0' + \left[ \left( \frac{V_0}{V} \right)^{\frac{2}{3}} - 1 \right]^2 \left[ 6 - 4 \left( \frac{V_0}{V} \right)^{\frac{2}{3}} \right] \right\} \qquad (6)$$

Unsurprisingly, given their small experimental bandgaps, we find that PBE predicts a metallic electronic structure at one or more cell volumes for Ge, GaSb, InAs and InSb. Excluding these compounds, the non-self-consistent calculations show a mean absolute error of 111 meV in the fitted $E_0$ compared to the self-consistent results (Table 2); although this corresponds to a small relative error of 1.19 %, this is not accurate enough for thermochemistry. However, as can be seen in the $E/V$ curves in Fig. 7, the error is remarkably consistent with respect to volume change, resulting in a small shift in the calculated $V_0$ and almost identical curvature. This leads to very small relative errors of 0.15, 0.61 and 0.3 % in the fitted $V_0$, $B_0$ and $B_0'$, respectively, indicating that the non-self-consistent technique can be used to evaluate these properties. These small errors are consistent over a wide range of $V_0$ and $B_0$. Diamond (C) has the smallest predicted $V_0$ of 11.1 $\text{Å}^3$ and the hardest $B_0$ of 494 GPa, and we obtain negligible errors of $0.7 \times 10^{-2}$ $\text{Å}^3$ and 1.2 GPa from the non-self-consistent calculations (0.63/0.24 %). CdTe has the largest $V_0$ of 70.3 $\text{Å}^3$ and the smallest $B_0$ of 40.6 GPa, and we obtain errors of $8.3 \times 10^{-2}$ $\text{Å}^3$ and 0.31 GPa (0.12/0.75 %).

**Table 2.** Mean absolute error (MAE; Eq. 1) and mean absolute relative error (MARE; Eq. 2) in the equilibrium total energy $E_0$, cell volume $V_0$ and bulk modulus and derivative $B_0/B_0'$ obtained for a subset of the test compounds by fitting Eq. 6 to energy-volume curves obtained from PBE0 and non-self-consistent PBE0 using the PBE orbitals.

| Parameter | MAE | MARE [%] |
|---|---|---|
| $E_0$ [meV] | 111 | 1.19 |
| $V_0$ [$\text{Å}^3$] | $7.2 \times 10^{-2}$ | 0.15 |
| $B_0$ [GPa] | 0.51 | 0.61 |
| $B_0'$ | $1.3 \times 10^{-2}$ | 0.30 |

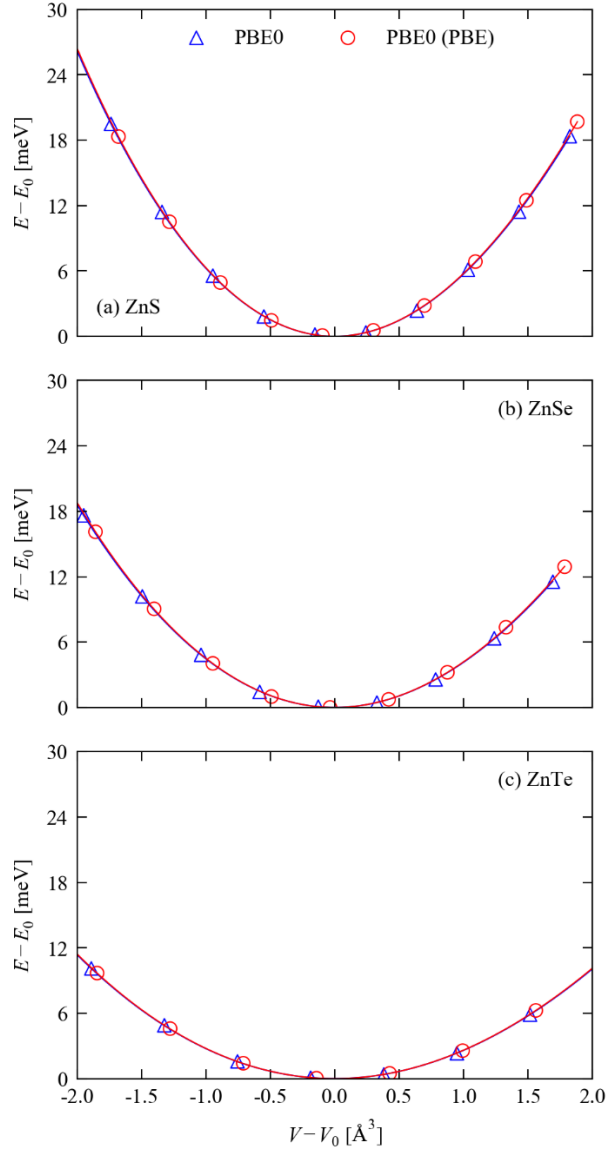

**Figure 7.** Calculated energy-volume curves for ZnS (a), ZnSe (b) and ZnTe (c) obtained using the PBE0 hybrid functional. The blue and red lines show the curves obtained from self-consistent total energies (blue) and non-self-consistent energies calculated using the PBE orbitals (red). On each plot, the markers show the calculated energies and the solid lines are fits to the Birch-Murnaghan equation of state (Eq. 6; Ref. [70]). Note that the curves are shifted relative to the equilibrium volume and energy ($V_0/E_0$) predicted from each set of calculations.

As the total energy is calculated from the band energies, we checked for correlation between the error in $E_0$ and the presence of semi-core d bands. Although relatively small errors are indeed obtained for C (29 meV), Si (50 meV) and AlP (58 meV), for which our chosen PAW potentials do not include d electrons, the largest error of 274 meV is obtained for AlSb which also does not include d states in the valence shell.

Inability to compute accurate forces would formally make the non-self-consistent method unsuitable for systems with more degrees of freedom than the cell volume, but it could for example be combined with the approach in Ref. [71] where a (semi-)local functional is used to optimise the atomic positions and cell shape at each volume and single-point energies are then calculated with a more accurate method to obtain the energy-volume curve. It has been shown that a good estimate of the equilibrium volume is the most important factor in predicting accurate phonon frequencies,[72] so a similar approach may also be of use for obtaining phonon frequencies.

**Dielectric-Dependent Hybrid Functionals: scPBE0.** The PBE0 and HSE06 hybrid functionals are both derived from the PBE GGA by replacing a fraction $\alpha$ of the PBE exchange energy with the exact Hartree-Fock exchange:[38,39]

$$E_{xc} = \alpha E_x^{HF,SR}(\omega) + (1-\alpha)E_x^{PBE,SR}(\omega) + E_x^{PBE,LR}(\omega) + E_c^{PBE} \qquad (7)$$

Here $E_{xc}$ is the exchange-correlation energy, $E_x^{HF}$ and $E_x^{PBE}$ are the Hartree-Fock and PBE exchange energies, respectively, and $E_c^{PBE}$ is the PBE correlation energy. The exchange interaction is screened such that a fraction $\alpha$ of $E_x^{PBE}$ is replaced by $E_x^{HF}$ at short range, denoted by the subscript SR, changing smoothly to the pure $E_x^{PBE}$ at long range (LR). This range separation is controlled by the adjustable parameter $\omega$. For PBE0, $\alpha = 0.25$ and there is no range separation, corresponding to 25 % HF and 75 % PBE exchange at all distances. For HSE 06, $\alpha$ is again set to 0.25, and a range separation $\omega = 0.2$ is set so that the 25 % HF/75 % PBE exchange at short range decays to 100 % PBE exchange at long range.

Although the $\alpha = 0.25$ in PBE0 has a physical basis,[38] various studies have shown that improved property predictions can be obtained by optimising it,[73,74] and it has been suggested that an optimum *ab initio* choice is the inverse of the high-frequency dielectric constant $\varepsilon_\infty$.[40] Despite the theoretical drawback of effectively making the Hamiltonian of the (quantum) system dependent on a macroscopic quantity, proof-of-concept studies using so-called self-consistent PBE0 (scPBE0), where $\alpha$ and $\varepsilon_\infty$ are iterated to self-consistency, have shown good results for metal oxides.[75–77]

The downside to these dielectric-dependent functionals is the overhead of an additional self-consistency loop on top of determining the ground-state density and orbitals. If, however, non-self-consistent calculations can reproduce the dependence of $\varepsilon_\infty$ on $\alpha$, it should be possible to perform non-self-consistent scPBE0 calculations at a much lower cost, and indeed non-self-consistent calculations with dielectric-dependent functionals have been highlighted as a possible means to obtain accurate bandgaps and band levels in high-throughput studies.[35]

Fig. 8 shows the typical behaviour of the calculated $\varepsilon_\infty$ and bandgap as the $\alpha$ in PBE0 is varied from 0 to 1 (see also Figs. S6.1-S6.16). In general, the bandgap increases linearly with $\alpha$, producing a corresponding non-linear decrease in $\varepsilon_\infty$. AlSb, GaSb and InSb are notable exceptions, with steep non-linear increases in $E_g$ with $\alpha$ leading to a more pronounced widening of the gap at $\alpha = 1$. The non-self-consistent calculations give good results at small $\alpha$, but the error in the bandgap grows systematically as $\alpha$ is increased. The error in the calculated high-frequency dielectric constants also increases, but the absolute deviation is smaller due to the inverse relationship between $E_g$ and $\varepsilon_\infty$. This can be explained by larger deviation between the PBE and self-consistent hybrid orbitals at larger $\alpha$. Interestingly, the non-self-consistent calculations predict a linear increase in $E_g$ with $\alpha$ for the three antimonides, giving a large deviation for $\alpha > 0.5$ (Figs. S6.6, S6.9 and S6.12), and this dichotomy mirrors the anomalously-large Hartree-Fock bandgaps calculated for these systems and consequent larger deviation between the self-consistent and non-self-consistent calculations.

With reference to the data in Table 1, the experimental dielectric constants and bandgaps of the sixteen test compounds are recovered for $\alpha < 0.4$, which falls into the region where the error in the non-self-consistent calculations is small. To test the accuracy of scPBE0, we developed an *ad hoc* implementation using a wrapper code written in the Python programming language.[78] Starting with the PBE0 value of $\alpha = 0.25$, $\varepsilon_\infty$ is calculated and used to iteratively update $\alpha$ according to:

$$\alpha(1) = 0.25; \; \alpha(n+1) = \frac{1}{\varepsilon_\infty(n)} \tag{8}$$

While this method gave excellent results for the majority of the test set, we encountered issues with the four narrow-gap compounds, *viz.* Ge, GaSb, InAs and InSb. For Ge, the calculations converged to $\alpha = 6.7 \times 10^{-4}$, giving a bandgap of 33 meV and an $\varepsilon_\infty$ of $1.5 \times 10^3$, compared to the experimental values of 0.66 eV and 16.6 respectively.

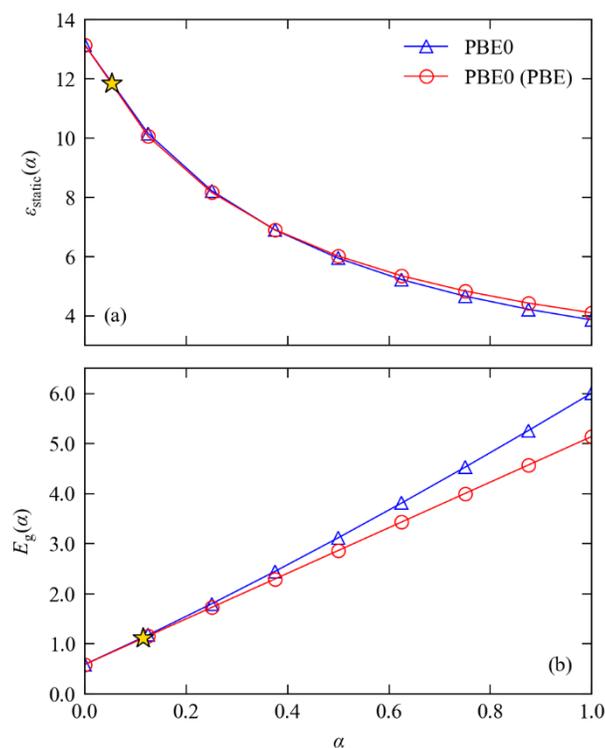

**Figure 8.** Dependence of the calculated high-frequency dielectric constant $\varepsilon_\infty$ and bandgap $E_g$ of Si on the fraction of exact exchange $\alpha$ in the PBE0 hybrid functional (c.f. Eq. 7). The self-consistent values and the non-self-consistent values obtained using the PBE orbitals are shown as blue triangles and red circles, respectively, and the experimental values from Table 1 are overlaid as gold stars.

The errors for the other three systems are less severe, but they were clear outliers, with the bandgaps underestimated by 0.1-0.5 eV (70-80 %) and the dielectric constants overestimated by 10-30 (60-200 %). We therefore deemed the scPBE0 functional, at least in our implementation, to be unsuitable for narrow-gap semiconductors[76] and opted to exclude these four compounds from our analysis.

Fig. 9 compares the bandgaps and $\varepsilon_\infty$ calculated with PBE0, scPBE0 and HSE06 to the experimental data in Table 1. scPBE0 improves significantly on PBE0, reducing the mean absolute error in the bandgaps and $\varepsilon_\infty$ from $0.56 \pm 0.17$ eV and $2.05 \pm 0.88$ to $0.13 \pm 0.11$ eV and $0.76 \pm 0.17$ respectively. scPBE0 also improves slightly on HSE06 which, as noted above, performs well for this test set.

Performing non-self-consistent scPBE0 calculations, where $\varepsilon_\infty$ is calculated non-self-consistently from the PBE orbitals as $\alpha$ is updated, results in only a marginal increase in error (Fig. 9). Comparing the self-consistent and non-self-consistent values of $\alpha$, including the four systems for which scPBE0 does not work well, also shows near-quantitative agreement (Fig. 10). This suggests that another useful application of the non-self-consistent approach might be to accelerate the convergence of scPBE0 by determining $\alpha$ during the early iterations of a fully self-consistent procedure.

In our testing, a typical scPBE0 calculation required seven updates of $\alpha$ (i.e. seven self-consistent PBE0 calculations) to converge $\varepsilon_\infty$ to $10^{-2}$, and despite re-using the orbitals between cycles was 3-4 $\times$ more expensive than PBE0. The non-self-consistent calculations also required an average of seven calculations, meaning a non-self-consistent scPBE0 result can be obtained for less than the cost of a single PBE0 calculation requiring a typical 10 electronic steps to converge.

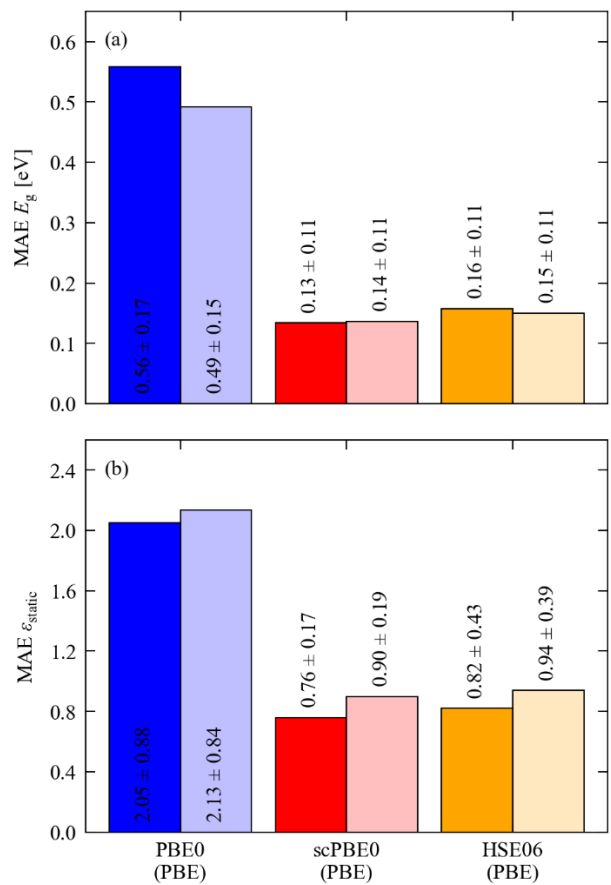

**Figure 9.** Mean absolute error (MAE; Eq. 1) in the calculated bandgaps $E_g$ and high-frequency dielectric constants $\varepsilon_\infty$ of a subset of the compounds in Table 1 obtained with PBE0 (blue), scPBE0 (red) and HSE06 (orange). For each method, the solid (left) and shaded (right) bars respectively show the results from self-consistent calculations and non-self-consistent calculations using the PBE orbitals. The text labels indicate the averages and standard deviations.

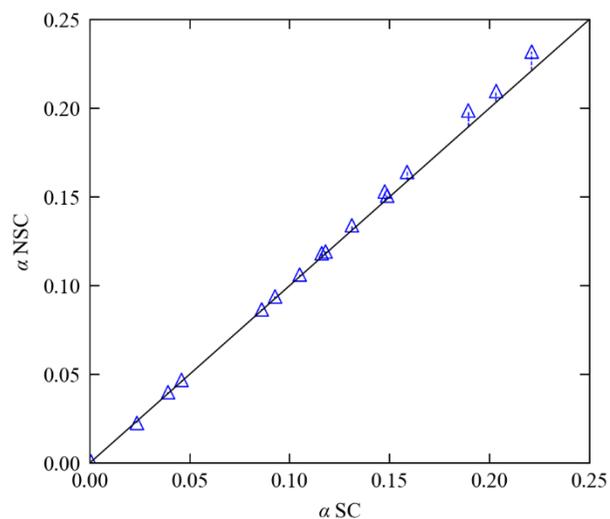

**Figure 10.** Comparison of the optimised scPBE0 exact exchange fraction $\alpha$ for the sixteen semiconductors in Table 1 obtained from self-consistent calculations and non-self-consistent calculations using the PBE orbitals.

As discussed in Supporting Information, Section S2, more accurate predictions of $\varepsilon_\infty$ can be obtained using alternative methods to the IP-RPA. With bare PBE0, dielectric constant obtained from the self-consistent response to finite fields reduces the MAE in $\varepsilon_\infty$ in Fig. 9b from 2.05 $\pm$ 0.88 to 0.41 $\pm$ 0.33. scPBE0 with the finite-field method did not improve the calculated bandgaps, but reduced the error in the converged $\varepsilon_\infty$ to 0.29 $\pm$ 0.29 and yielded values of $\alpha$ on average 11.4 % smaller than those obtained using the IP-RPA. This method also gave significantly better results for Ge and GaSb, although the results were outliers compared to other systems, but again failed for InAs and InSb. However, the improved performance needs to be set against the additional computational overhead of computing the response to finite fields.

**Partial Self-Consistency.** The results in the previous sections have identified two notable cases where the non-self-consistent approach can be expected to perform poorly, namely:

(1) Narrow-gap semiconductors where the base (semi-)local functional erroneously predicts a metallic electronic structure; and

(2) Systems with localised d (or f) electrons for which self-interaction error in the base functional may result in significant errors in the density.

In both cases the leading error is likely to be in the orbitals from the base functional rather than approximations in the exchange-correlation potential. A natural solution, in the spirit of the present study, is to perform partially-self-consistent calculations, i.e. to update the initial orbitals with the hybrid functional a fixed number of times, or to converge the total energy to a loose threshold. Since differences between codes will likely mean fixed numbers of orbital updates produce different levels of convergence, we chose the total energy as our control parameter.

Fig. 11 shows the change in the mean absolute error in the HSE06 bandgap after updating the initial PBE orbitals using a conjugate-gradient algorithm to converge the energy to thresholds of 1 eV to 100 $\mu$eV ($10^0$ - $10^{-4}$ eV). The initial non-self-consistent $E_g$ are in error by $107 \pm 149$ meV, falling to $52 \pm 35$ meV if InAs and InSb are excluded. VASP performs a minimum of two iterations when optimising the orbitals, after which the total energies are converged to $10^0$-$10^{-1}$ eV and the two sets of errors are substantially reduced to $12 \pm 7.3$ and $10 \pm 6.4$ meV. This loose convergence threshold is obviously reached much faster than our fully self-consistent threshold of $10^{-6}$ eV (7 - 13 iterations, average 10.4), demonstrating that partial self-consistency is a viable solution to Case (1). The reduced spread in errors may also be desirable for avoiding false positives/negatives in screening studies.

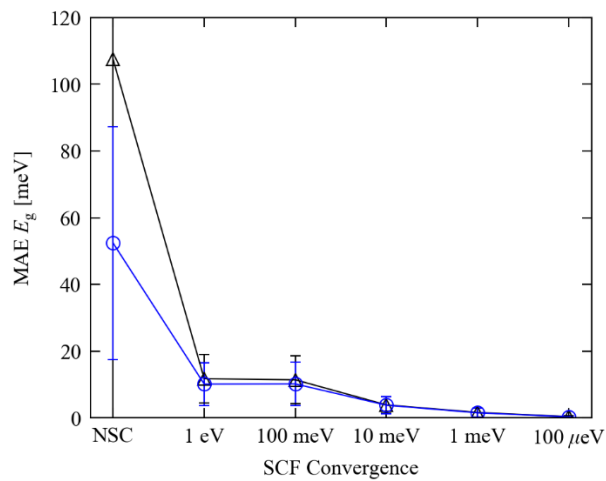

**Figure 11.** Mean absolute error (MAE; Eq. 1) in the HSE06 bandgaps calculated using PBE orbitals and orbitals updated to converge the total energy to increasingly tighter thresholds of 1 eV to 100 $\mu$eV. The black curve shows the MAE scores obtained across all sixteen semiconductors, whereas the blue curve excludes systems for which PBE predicts a bandgap of < 10 meV. Error bars show ± one standard deviation.

To examine Case (2), we performed partially self-consistent calculations of the band dispersions of our sixteen test compounds and examined the convergence of the semi-core d bands (Fig. 12). For InSb (Fig. 12a), the d bands in the non-self-consistent calculation are ~0.3 eV below the self-consistent positions. Partial convergence to 1 eV places them ~20 meV below, and tolerances lower than 100 meV yield results visually indistinguishable to the self-consistent dispersion on the 0.8 eV energy scale in the plot. Updating the orbitals appears to lead to rigid shifts, suggesting the non-self-consistent calculations correctly predict the splitting and dispersion of the d bands, and the orbital updates mainly refine the positions relative to the Fermi energy.

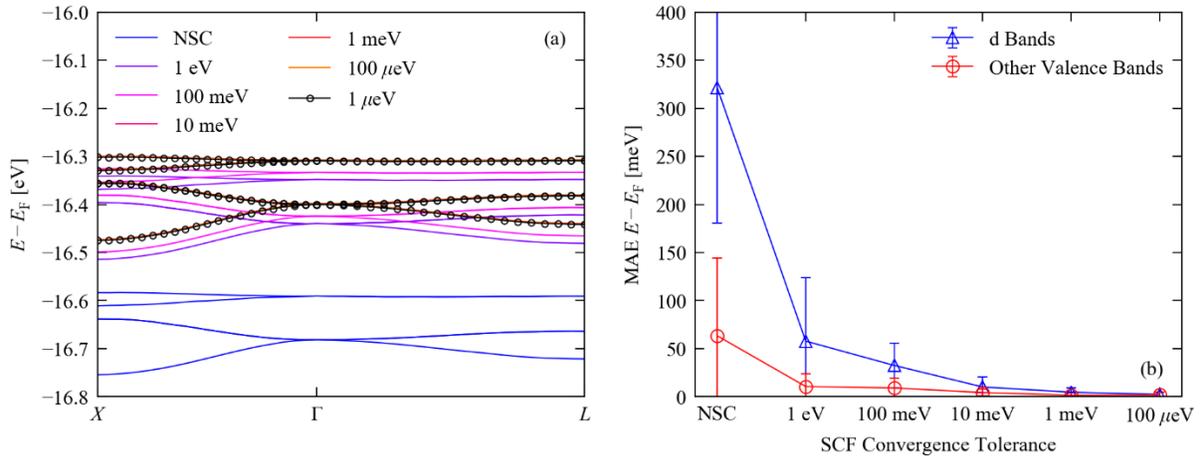

**Figure 12.** (a) Change in the dispersion of the semi-core d bands in InSb calculated using non-self-consistent and partially-self-consistent HSE06 starting from the PBE orbitals with SCF tolerances of 1 eV - 100 $\mu$eV ($10^0$-$10^{-4}$ eV) on the total energy. The lines are colour coded from blue to orange to indicate successively tighter convergence thresholds, and the black lines with markers show the reference dispersion obtained from a fully self-consistent calculation performed to a tolerance of 1 $\mu$eV ($10^{-6}$ eV). (b) Variation in the mean absolute error (MAE; Eq. 1) in the energies of the semi-core d bands (blue) and other valence and conduction bands (red) as the initial orbitals are updated to successively tighter tolerances.

Similar rigid shifts were also observed in the other systems with d bands in our test set (Figs. S7.1-S7.13). Fig. 12b shows the error in the energies of the d bands and other s/p bands as the orbitals are updated to successively tighter tolerances. In the initial non-self-consistent calculations the d band energies have an initial error of 321 $\pm$ 141 meV compared to 63 $\pm$ 81 meV for the other bands. Converging the HSE06 total energy to 1 eV, with two orbital updates, reduces the error and the spread substantially, but whereas the s/p band energies are converged to < 10 meV at a tolerance of 100 meV, the same level of accuracy is only reached for the semi-core d bands when the convergence is tightened by 1-2 orders of magnitude. This provides strong evidence for larger errors in the initial PBE d bands.

These results thus indicate that partial self-consistency can be used to treat both Cases (1) and (2) while maintaining a significant reduction in computing compared to fully self-consistent calculations. We also note that both issues could easily be identified as part of a high-throughput workflow - Case (1) by checking the bandgap obtained with the base functional, and Case (2) from the chemical composition and choice of pseudopotential if used.

**Transition-Metal Oxides: CoO and NiO.** Finally, we assessed the performance of the non-self-consistent hybrid functionals for open-shell systems by performing test calculations on the transition-metal (TM) oxides CoO and NiO. In TM oxides, the strong electron correlation in the spatially-localised d orbitals typically leads to large self-interaction error in (semi-)local functionals.[79] To counter this, DFT + $U$ method, where a modified DFT Hamiltonian is used to disfavour fractional occupation of specific atomic orbitals without adding significantly to the computational cost , has emerged as a practical alternative to more demanding hybrid DFT.

CoO and NiO are widely used as examples to illustrate self-interaction error and its mitigation using DFT + $U$.[63,79] As shown in Table 3, PBE predicts CoO to be metallic, whereas the experimental gap is 2.4 eV, [80] and underestimates the 4.3 eV bandgap of NiO[81] by 4.5 ×. DFT + $U$ using the method proposed by Dudarev *et al.*[62] and a $U_{\text{eff}}$ = 7 eV applied to the Co/Ni d orbitals[63] yields significantly-improved results. Comparison of the magnetic moments to experiment[82] also suggest better localisation of the transition-metal d electrons with PBE + $U$.

**Table 3.** Comparison of the bandgaps $E_g$ and transition metal magnetic moment $|M|$ of CoO and NiO obtained with PBE, PBE + $U$, HSE06, and non-self-consistent HSE06 using the PBE and PBE + $U$ orbitals.[a]

| | CoO | | NiO | |
|---|---|---|---|---|
| XC | $E_g$ [eV] | $|M|$ [BM] | $E_g$ [eV] | $|M|$ [BM] |
| Expt. | 2.4[80] | 3.80[82] | 4.3[81] | 1.9[82] |
| PBE | Metallic | 2.30 | 0.95 | 1.37 |
| PBE + $U$ | 2.23 | 2.85 | 3.32 | 1.76 |
| HSE06 | 1.49 | 2.76 | 4.51 | 1.69 |
| HSE06 (PBE) | Metallic | - | 3.97 (-12 %) | - |
| HSE06 (PBE + $U$) | 1.70 (+14 %) | - | 4.62 (+2.4 %) | - |

[a] For the two non-self-consistent calculations the percentage difference to the self-consistent HSE06 bandgaps are shown in parentheses for comparison.

HSE06 underestimates the experimental bandgap of CoO by 0.91 eV (37 %) but reproduces the NiO gap to within 5 %. As expected from the metallic electronic structure predicted by PBE, non-self-consistent HSE06 calculations using the PBE orbitals fail to predict a gap for CoO, while using the DFT + $U$ orbitals reproduces the self-consistent HSE06 bandgap to within 14 %. For NiO, the non-self-consistent HSE06 calculation using the PBE orbitals widens by gap by ~3 eV which, although a 12 % underestimate of the self-consistent bandgap, is a very significant improvement, and using the PBE + $U$ orbitals yields a bandgap within 2.5 % of the self-consistent value.

Fig. 13 compares the electronic density of states (DoS) curves obtained for the two TM oxides with the five functionals. For CoO, PBE predicts a markedly different DoS to HSE06, with a high density of states extending to ~0.25 eV above the Fermi energy (Fig. 13a). The non-self-consistent calculations starting from the PBE orbitals show some improvement, but the calculations still predict a non-zero density of states above $E_F$ and metallic behaviour.

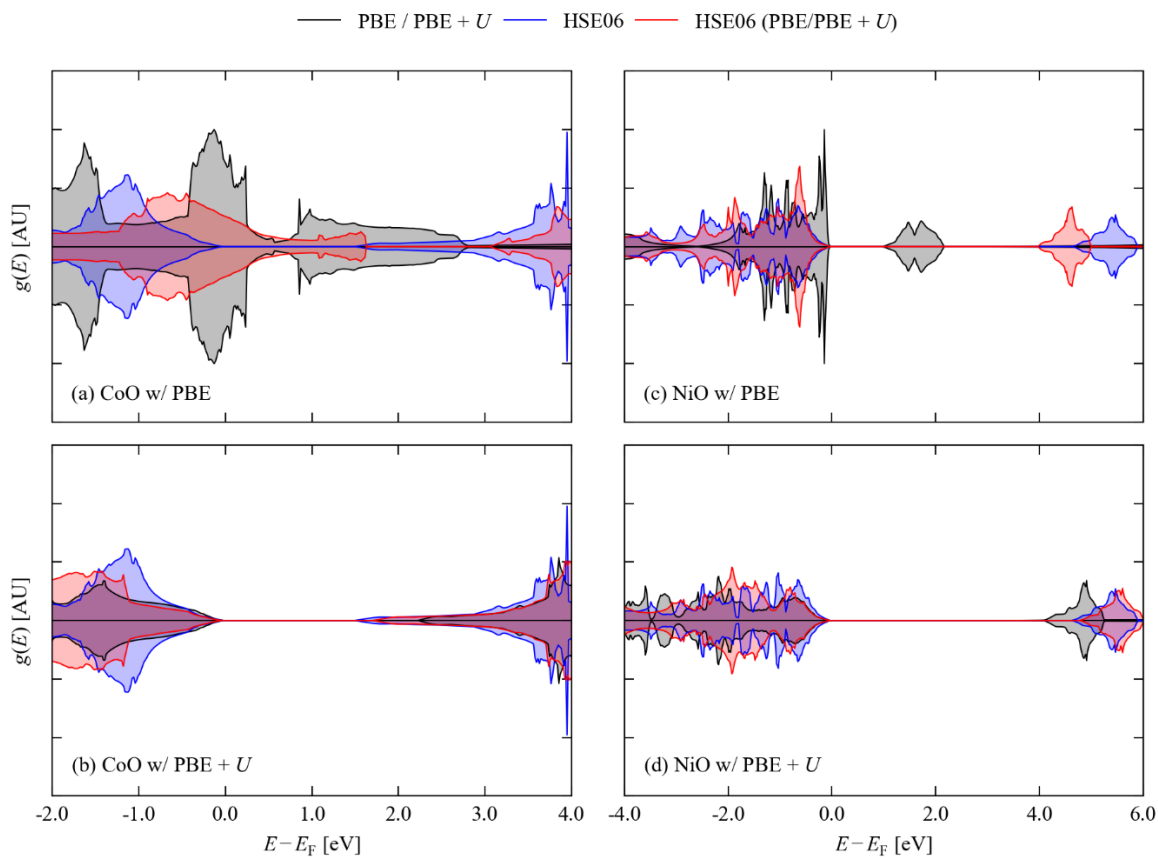

**Figure 13.** Comparison of the electronic density of states $g(E)$ (DoS) of (a)/(b) CoO and (c)/(d) NiO calculated using (a)/(c) PBE or (b)/(d) PBE + $U$, HSE06 and non-self-consistent HSE 06 starting from the PBE or PBE + $U$ orbitals. In each subplot the $g(E)$ for two separate spin channels are shown as positive and negative values.

The PBE + $U$ and corresponding non-self-consistent HSE06 calculations ( Fig. 13b) both correctly predict a bandgap, but the shape of the valence and conduction band DoS in the three calculations nonetheless differ more than for the sixteen s/p semiconductors in our test set (c.f. Fig. 4 and Figs. S4.1-S4.16).

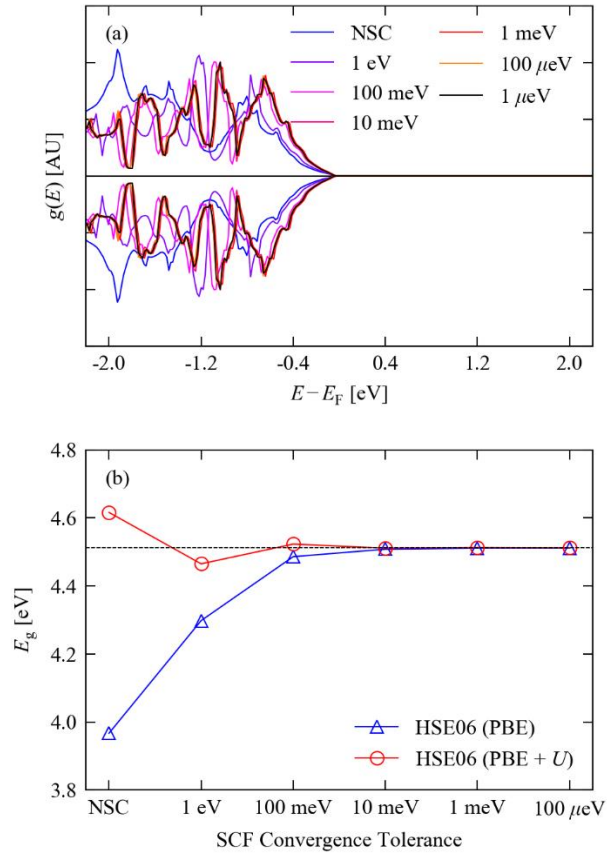

**Figure 14.** Partially-self-consistent HSE06 calculations on NiO starting from PBE + $U$ orbitals. (a) Change in the valence band density of states $g(E)$ (DoS) with SCF tolerances of 1 eV - 100 $\mu$eV ($10^0$-$10^{-4}$ eV) on the total energy. The lines are colour coded from blue to orange to indicate successively tighter convergence thresholds, and the black line shows the reference DoS obtained from the fully self-consistent calculation performed to a tolerance of 1 $\mu$eV ($10^{-6}$ eV). (b) Convergence of the bandgap $E_g$ with SCF tolerance. The self-consistent HSE06 bandgap is shown as a dashed black line for reference.

For NiO, all the functionals predict a bandgap, although PBE again predicts a markedly different DoS to HSE06, and whereas PBE + $U$ and the non-self-consistent calculations show a better correspondence to the self-consistent DoS, there are some clear differences. Projection of the DoS onto atomic states (Figs. S8.1-S8.4) shows that the valence and conduction band edges in both oxides are formed predominantly of metal states,[83] and we would therefore expect the description of the strongly-correlated d states to have a significant effect on the shape of the DoS, in contrast to the semiconductors in our main test set where the d states are core-like and the band edges are formed of s and p states.

Following on from the previous section, to confirm whether partial self-consistency can mitigate this issue we performed calculations on both oxides to a series of different tolerances on the total energy, starting from the PBE and PBE + $U$ orbitals (Fig. 14, Figs. S8.5/S8.6). With the exception of CoO, for which starting from the PBE orbitals led to convergence issues at tighter tolerances, the DoS converges quickly as the orbitals are updated, such that the partially-self-consistent result is almost indistinguishable with the self-consistent result at tolerances < 100 meV. Similarly, starting from the PBE and PBE + $U$ orbitals, respectively, loose tolerances of 100 meV and 1 eV are sufficient to converge the calculated bandgap of NiO to within ~1 % of the self-consistent result. Perhaps unsurprisingly, and as shown in Fig. 14b, starting from the PBE + $U$ orbitals leads to faster convergence than starting from the PBE orbitals. Reaching a convergence threshold of 100 meV requires 3-5 iterations of the conjugate gradient algorithm compared to 15 iterations to reach the self-consistent threshold of 1 $\mu$eV, representing a substantial saving in computing time.

**CONCLUSIONS**

We have performed systematic tests of the accuracy of non-self-consistent hybrid functionals for modeling the electronic structure, dielectric and structural properties of sixteen tetrahedral semiconductors.

Our results show that the non-self-consistent calculations can reproduce the PBE0 and HSE06 bandgaps to within an average 5 % at an order of magnitude lower computational cost, with the accuracy improving for wider-gap systems. The dispersion, density of states, and dielectric properties are also reproduced to a high degree of accuracy, allowing for the calculation of other electronic and optical properties. While the typical error in the absolute total energies is significant, the non-self-consistent calculations reproduce the energy/volume curves sufficiently accurately to allow the equilibrium volume and bulk modulus to be obtained with minimal error. Our testing further shows that the non-self-consistent approach can also be used in conjunction with dielectric-dependent functionals such as scPBE0, and non-self-consistent calculations with scPBE0 were found to give better predictions of the experimental bandgaps and high-frequency dielectric constants than self-consistent PBE0 at a comparable or lower cost. This highlights the broad potential applicability of this technique to contemporary high-throughput screening studies.

Our studies identify two cases where non-self-consistent calculations are likely to perform poorly, *viz.* when the base functional incorrectly predicts a metallic electronic structure, and when the system contains strongly-correlated electronic states that may be affected by self-interaction error. Both are easily identified and can be practically addressed by using a method such as DFT + $U$ to correct the density, or with partial self-consistency, in both cases retaining significant cost savings versus fully self-consistent calculations. This approach appears to work particularly well for transition-metal oxides, which are well known as challenging systems in solid-state physics. Partial self-consistency to a loose total-energy tolerance also reduces average and spread of errors with respect to the self-consistent results, providing a means to improve accuracy where required.

Finally, we note that, notwithstanding cases where the base functional incorrectly predicts a metallic electronic structure, or produces large errors in the orbitals due to self-interaction error, our testing suggests that the base functional used to obtain the initial orbitals does not have a large effect on the accuracy of the non-self-consistent calculations, and so this technique could be used alongside (meta-)GGAs optimised for structural properties, such as PBEsol, in studies requiring e.g. accurate lattice dynamics in addition to electronic and optical properties.

## ASSOCIATED CONTENT

**Supporting Information**. Electronic supporting information is available with the following: (S1) tests of the plane-wave cutoff and $k$-point convergence with PBE; (S2) a comparison of methods for calculating the high-frequency dielectric constant and performance tests with the dielectric-dependent scPBE0 functional; (S3) comparison of the bandgaps predicted by all nine functionals to experimental measurements and additional comparison of the self-consistent and non-self-consistent HF, PBE0 and HSE06 bandgaps obtained using orbitals from the six (semi-)local base functionals; (S4) a full set of band dispersions, density of states curves, and dielectric functions obtained using PBE, self-consistent HSE06 and non-self-consistent HSE06 using the PBE orbitals; (S5) a full set of energy/volume curves calculated using self-consistent PBE0 and non-self-consistent PBE0 using the PBE orbitals; (S6) dependence of the bandgap and high-frequency dielectric constants of the sixteen compounds on the fraction of exact exchange in PBE0, obtained using self-consistent and non-self-consistent calculations using the PBE orbitals; (S7) change in the dispersion of the valence d bands as the orbitals are updated to increasingly tighter total-energy tolerances in partially-self-consistent HSE06 calculations on AlAs, GaAs, GaSb, InP, InAs, InSb, ZnS, ZnSe and ZnTe; (S8) atom-projected electronic density of states curves for NiO and CoO obtained with the five functionals tested and convergence of the DoS as a function of the SCF tolerance in partially-self-consistent calculations; and (S9) supplementary references.

**Data-Access Statement.** A full set of raw data from the calculations, including structures, calculated electronic density of states, band dispersions and dielectric functions and sample input files for the Vienna *Ab initio* Simulation Package (VASP) code are available in an online repository at [URL to be added on acceptance]. Data may also be obtained from the authors on request.

## AUTHOR INFORMATION

**Corresponding Author.** * Correspondence should be addressed to JMS. E-mail: jonathan.skelton@manchester.ac.uk

**Author Contributions.** The manuscript was written through contributions of all authors. All authors have given approval to the final version of the manuscript.

**Funding Sources.** UK Engineering and Physical Sciences Research Council (EP/P007821/1, EP/L000202, EP/R029431).

## ACKNOWLEDGEMENTS

The authors are grateful to L. A. Burton and J. M. Frost for helpful suggestions on parts of this work. JMS is grateful for the support of a Presidential Fellowship from the University of Manchester. This work was also supported by the UK Engineering and Physical Sciences Research Council (EP/P007821/1). Calculations were performed on the UK Archer facility *via* the UK Materials Chemistry Consortium, which is funded by the EPSRC (EP/L000202, EP/R029431), and on the Balena HPC cluster at the University of Bath, which is maintained by Bath University Computing Services.

## ABBREVIATIONS

DFT - density-functional theory; MAE - mean absolute error; MARE - mean absolute relative error.

**TOC ENTRY**

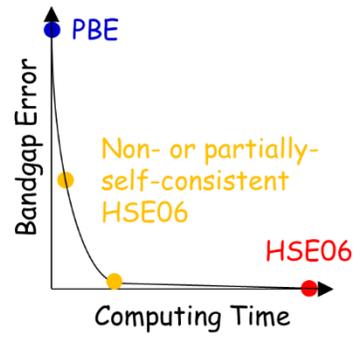

# Accuracy of hybrid functionals with non-self-consistent Kohn-Sham orbitals for predicting the properties of semiconductors

## Electronic supporting information

J. M. Skelton, D. S. D. Gunn, S. Metz and S. C. Parker

**S1. Convergence Testing**

As discussed in the text, the plane-wave cutoffs and *k*-point sampling for each of the compounds tested were chosen in a manner representative of how these parameters would be chosen in a typical high-throughput modeling study. To check the parameters, we performed a series of explicit convergence tests with the PBE exchange-correlation functional[1] in which the base cutoffs in Table 1 in the text were increased by 1.5 and 2 × and the *k*-point sampling meshes were similarly increased to 1.5 and 2 × the base number of subdivisions along each reciprocal lattice vector.

Tables S1.1 and S1.2 show the convergence of the PBE bandgap $E_g$ with respect to the cutoff and *k*-point sampling. Barring Ge, for which PBE with the base parameters predicts a near-metallic bandgap of 28 meV, these tests suggest the bandgaps are converged to within 1 % with the base cutoff. Similarly, with the exception of GaP, where a 1.4 % change in the gap is observed on increasing the *k*-point density, the tests suggest the base *k*-point meshes are sufficient to converge the calculated gaps to within 1 %.

Tables S1.3 and S1.4 show the convergence of the high-frequency dielectric constant $\varepsilon_\infty$ with respect to the cutoff and *k*-point sampling. As shown in Table S1.1, PBE predicts InAs and InSb to be metallic and Ge and GaSb to have very narrow gaps of < 100 meV respectively. We would therefore anticipate large variability in the calculated $\varepsilon_\infty$ for these compounds. With the exception of Ge, the tests indicate the calculated dielectric constants to be converged to < 1 % with respect to the cutoff. Excluding the four compounds identified above, the $\varepsilon_\infty$ of eight compounds are converged to < 1 % with respect to *k*-point sampling, the dielectric constants of a further three are converged to < 5 %, and the remaining compound, GaAs, is converged to 8.2 %.

Finally, Tables S1.5 and S1.6 show the convergence of the equilibrium total energies $E_0$, volumes $V_0$ and bulk moduli $B_0$ with respect to cutoff and *k*-point sampling. As described in the text, these parameters are obtained by fitting energy/volume curves to the Birch-Murnaghan equation of state (Eq. 1 in the text).[2] Our tests suggest that $E_0$ and $V_0$ are converged to well within 1 % with the base parameters. $B_0$ is obtained from the curvature of the energy/volume curve and is thus more sensitive to small changes in the calculated energies. Nonetheless, our tests indicate that the $B_0$ for nine of the sixteen compounds are converged to < 1 % with respect to the cutoff, a further five are converged to < 2.5 %, and the remaining two are converged to 2.72 and 5.75 %. The base cutoff was increased by 1.5 × for the production calculations presented in the text, and the $B_0$ obtained with this cutoff are all converged to less than or around 1 % with respect to the higher 2 × plane-wave cutoff. The $B_0$ appear to be more sensitive to the *k*-point sampling, but nonetheless the data in Table S1.6 suggests that the for seven of the compounds the calculated $B_0$ are converged to < 1 % with the base *k*-point meshes, a further four are converged to < 2.5 %, and the remainder are converged to < 5 %.

| System | $E_g$ [eV] | | | Range [meV] | Std. [%] |
|--------|------|--------|------|-------|---------|
|        | Base | 1.5 × | 2 × |       |         |
| C      | 4.109 | 4.123 | 4.125 | 16.7 | 0.18 |
| Si     | 0.591 | 0.585 | 0.584 | 7.0  | 0.53 |
| Ge     | 0.028 | 0.031 | 0.031 | 3.1  | 5.24 |
| AlP    | 1.569 | 1.568 | 1.568 | 1.3  | 0.04 |
| AlAs   | 1.447 | 1.447 | 1.447 | 0.7  | 0.02 |
| AlSb   | 1.216 | 1.215 | 1.215 | 0.9  | 0.03 |
| GaP    | 1.637 | 1.636 | 1.636 | 1.5  | 0.04 |
| GaAs   | 0.513 | 0.513 | 0.513 | 0.2  | 0.02 |
| GaSb   | 0.093 | 0.094 | 0.094 | 1.1  | 0.48 |
| InP    | 0.679 | 0.685 | 0.685 | 5.9  | 0.40 |
| InAs   | 0.000 | 0.000 | 0.000 | 0.0  | -    |
| InSb   | 0.000 | 0.000 | 0.000 | 0.0  | -    |
| ZnS    | 2.098 | 2.102 | 2.103 | 5.1  | 0.10 |
| ZnSe   | 1.285 | 1.286 | 1.287 | 1.6  | 0.05 |
| ZnTe   | 1.245 | 1.246 | 1.247 | 2.2  | 0.07 |
| CdTe   | 0.764 | 0.768 | 0.770 | 5.5  | 0.30 |

**Table S1.1** Convergence of the bandgap $E_g$ with respect to plane-wave cutoff with the PBE exchange-correlation functional. The bandgaps obtained with the base cutoffs listed in Table 1 in the text and those calculated with the cutoffs increased by 1.5 and 2 × this value are shown together with the ranges of the three gaps in meV and the standard deviations as percentages of the gaps obtained with the base cutoffs.

| System | $E_g$ [eV] | | | Range [meV] | Std. [%] |
|--------|------|-------|-----|-------------|----------|
|        | Base | 1.5 × | 2 × |             |          |
| C      | 4.109 | 4.116 | 4.109 | 7.5  | 0.09 |
| Si     | 0.591 | 0.591 | 0.591 | 0.2  | 0.01 |
| Ge     | 0.028 | 0.028 | 0.028 | 0.1  | 0.09 |
| AlP    | 1.569 | 1.569 | 1.569 | 0.5  | 0.02 |
| AlAs   | 1.447 | 1.448 | 1.448 | 0.7  | 0.02 |
| AlSb   | 1.216 | 1.207 | 1.200 | 16.2 | 0.55 |
| GaP    | 1.637 | 1.589 | 1.590 | 48.2 | 1.37 |
| GaAs   | 0.513 | 0.514 | 0.514 | 0.8  | 0.07 |
| GaSb   | 0.093 | 0.094 | 0.094 | 1.0  | 0.46 |
| InP    | 0.679 | 0.680 | 0.680 | 0.6  | 0.04 |
| InAs   | 0.000 | 0.000 | 0.000 | 0.0  | -    |
| InSb   | 0.000 | 0.000 | 0.000 | 0.0  | -    |
| ZnS    | 2.098 | 2.098 | 2.098 | 0.0  | 0.00 |
| ZnSe   | 1.285 | 1.286 | 1.286 | 0.4  | 0.01 |
| ZnTe   | 1.245 | 1.245 | 1.245 | 0.4  | 0.01 |
| CdTe   | 0.764 | 0.765 | 0.765 | 0.5  | 0.03 |

**Table S1.2** Convergence of the bandgap $E_g$ with respect to $k$-point sampling with the PBE exchange-correlation functional. The bandgaps obtained with the base $k$-point meshes listed in Table 1 in the text and values obtained using meshes increased by 1.5 and 2 × along each reciprocal lattice vector are shown together with the ranges of the three gaps in meV and the standard deviations as percentages of the gaps obtained with the base $k$-point meshes.

| System | $\varepsilon_\infty$ Base | 1.5 × | 2 × | Range | Std. [%] |
|---|---|---|---|---|---|
| C | 5.88 | 5.88 | 5.87 | $9.20 \times 10^{-3}$ | 0.07 |
| Si | 13.15 | 13.20 | 13.21 | $5.47 \times 10^{-2}$ | 0.19 |
| Ge | $2.03 \times 10^3$ | $1.64 \times 10^3$ | $1.64 \times 10^3$ | - | - |
| AlP | 8.62 | 8.64 | 8.64 | $2.45 \times 10^{-2}$ | 0.13 |
| AlAs | 9.76 | 9.76 | 9.76 | $3.60 \times 10^{-3}$ | 0.02 |
| AlSb | 12.19 | 12.19 | 12.19 | $2.70 \times 10^{-3}$ | 0.01 |
| GaP | 10.90 | 10.91 | 10.91 | $1.61 \times 10^{-2}$ | 0.07 |
| GaAs | 17.37 | 17.37 | 17.38 | $3.90 \times 10^{-3}$ | 0.01 |
| GaSb | $1.53 \times 10^2$ | $1.51 \times 10^2$ | $1.50 \times 10^2$ | 3.13 | 0.85 |
| InP | 12.17 | 12.16 | 12.15 | $1.75 \times 10^{-2}$ | 0.06 |
| InAs | 17.62 | 17.70 | 17.69 | $7.94 \times 10^{-2}$ | 0.20 |
| InSb | 21.56 | 21.66 | 21.70 | 0.14 | 0.28 |
| ZnS | 6.18 | 6.19 | 6.19 | $1.19 \times 10^{-2}$ | 0.09 |
| ZnSe | 7.54 | 7.54 | 7.54 | $3.60 \times 10^{-3}$ | 0.02 |
| ZnTe | 9.26 | 9.25 | 9.25 | $5.40 \times 10^{-3}$ | 0.02 |
| CdTe | 9.21 | 9.19 | 9.18 | $3.14 \times 10^{-2}$ | 0.14 |

**Table S1.3** Convergence of the high-frequency dielectric constant $\varepsilon_\infty$ with respect to plane-wave cutoff with the PBE exchange-correlation functional. The dielectric constants obtained with the base cutoffs listed in Table 1 in the text and values obtained with the cutoffs increased by 1.5 and 2 × are shown together with the ranges of the $\varepsilon_\infty$ and the standard deviations as percentages of the values obtained with the base cutoffs.

| System | $\varepsilon_\infty$ | | | Range | Std. [%] |
|--------|------|------|------|-------|----------|
| | Base | 1.5 × | 2 × | | |
| C | 5.88 | 5.92 | 5.93 | $4.91 \times 10^{-2}$ | 0.35 |
| Si | 13.15 | 13.18 | 13.29 | 0.13 | 0.44 |
| Ge | $2.03 \times 10^3$ | $7.80 \times 10^2$ | $4.24 \times 10^2$ | - | - |
| AlP | 8.62 | 8.59 | 8.65 | $5.36 \times 10^{-2}$ | 0.26 |
| AlAs | 9.76 | 9.65 | 9.73 | 0.10 | 0.45 |
| AlSb | 12.19 | 11.85 | 11.91 | 0.34 | 1.22 |
| GaP | 10.90 | 10.67 | 10.73 | 0.23 | 0.89 |
| GaAs | 17.37 | 14.66 | 14.10 | 3.27 | 8.22 |
| GaSb | $1.53 \times 10^2$ | 68.28 | 44.32 | - | - |
| InP | 12.17 | 11.35 | 11.25 | 0.92 | 3.39 |
| InAs | 17.62 | 15.65 | 15.48 | - | - |
| InSb | 21.56 | 18.59 | 18.15 | - | - |
| ZnS | 6.18 | 6.22 | 6.24 | $6.36 \times 10^{-2}$ | 0.42 |
| ZnSe | 7.54 | 7.46 | 7.49 | $8.56 \times 10^{-2}$ | 0.47 |
| ZnTe | 9.26 | 9.10 | 9.14 | 0.16 | 0.71 |
| CdTe | 9.21 | 8.87 | 8.85 | 0.36 | 1.80 |

**Table S1.4** Convergence of the high-frequency dielectric constant $\varepsilon_\infty$ with respect to $k$-point sampling with the PBE exchange-correlation functional. The dielectric constants obtained with the base $k$-point meshes listed in Table 1 in the text and $\varepsilon_\infty$ calculated using meshes increased by 1.5 and 2 × along each reciprocal lattice vector are shown together with the ranges of the $\varepsilon_\infty$ and the standard deviations as percentages of the values obtained with the base $k$-point meshes.

| System | $E_0$ [eV] | | | Range [meV] | Std. [%] | $V_0$ [Å³] | | | Range [$10^{-3}$ Å³] | Std. [%] | $B_0$ [GPa] | | | Range [GPa] | Std. [%] |
|---|---|---|---|---|---|---|---|---|---|---|---|---|---|---|---|
| | Base | 1.5 × | 2 × | | | Base | 1.5 × | 2 × | | | Base | 1.5 × | 2 × | | |
| C | -18.21 | -18.20 | -18.21 | 13.0 | -0.03 | 11.39 | 11.39 | 11.38 | 11.0 | 0.04 | 434.9 | 435.6 | 433.9 | 1.64 | 0.16 |
| Si | -8.979 | -8.994 | -8.994 | 15.5 | -0.08 | 48.31 | 48.16 | 48.15 | 154 | 0.15 | 55.14 | 56.80 | 56.72 | 1.65 | 1.38 |
| Ge | -10.90 | -10.92 | -10.93 | 26.4 | -0.11 | 40.79 | 40.68 | 40.67 | 116 | 0.13 | 88.39 | 89.51 | 89.55 | 1.16 | 0.61 |
| AlP | -10.39 | -10.40 | -10.40 | 11.1 | -0.05 | 41.65 | 41.55 | 41.55 | 94.4 | 0.11 | 83.58 | 83.27 | 83.18 | 0.39 | 0.20 |
| AlAs | -9.395 | -9.402 | -9.403 | 7.8 | -0.04 | 46.89 | 46.85 | 46.86 | 35.4 | 0.03 | 67.57 | 67.33 | 67.56 | 0.24 | 0.17 |
| AlSb | -8.222 | -8.226 | -8.227 | 5.0 | -0.03 | 60.30 | 60.26 | 60.25 | 44.7 | 0.03 | 49.59 | 49.74 | 49.84 | 0.25 | 0.21 |
| GaP | -9.107 | -9.114 | -9.116 | 9.5 | -0.04 | 41.72 | 41.68 | 41.67 | 43.0 | 0.05 | 76.18 | 75.96 | 76.42 | 0.46 | 0.25 |
| GaAs | -8.237 | -8.243 | -8.246 | 9.4 | -0.05 | 47.50 | 47.52 | 47.50 | 18.3 | 0.02 | 62.85 | 60.26 | 61.07 | 2.59 | 1.72 |
| GaSb | -7.343 | -7.347 | -7.350 | 6.4 | -0.04 | 60.09 | 60.09 | 60.09 | 5.74 | 0.00 | 41.75 | 41.78 | 41.55 | 0.24 | 0.25 |
| InP | -8.401 | -8.405 | -8.406 | 4.8 | -0.02 | 52.77 | 52.77 | 52.77 | 3.76 | 0.00 | 62.31 | 60.01 | 59.76 | 2.55 | 1.84 |
| InAs | -7.690 | -7.698 | -7.697 | 7.5 | -0.04 | 59.16 | 59.22 | 59.22 | 65.2 | 0.05 | 51.91 | 49.10 | 49.36 | 2.81 | 2.44 |
| InSb | -6.932 | -6.936 | -6.939 | 6.9 | -0.04 | 72.77 | 72.79 | 72.82 | 46.3 | 0.03 | 38.95 | 38.66 | 38.27 | 0.68 | 0.72 |
| ZnS | -6.849 | -6.856 | -6.857 | 7.9 | -0.05 | 40.38 | 40.33 | 40.32 | 58.7 | 0.07 | 68.26 | 69.14 | 69.53 | 1.28 | 0.78 |
| ZnSe | -6.023 | -6.025 | -6.026 | 3.2 | -0.02 | 47.20 | 47.18 | 47.19 | 16.0 | 0.01 | 55.02 | 56.77 | 56.78 | 1.76 | 1.50 |
| ZnTe | -5.178 | -5.181 | -5.182 | 3.8 | -0.03 | 59.01 | 58.94 | 58.95 | 73.8 | 0.06 | 41.23 | 43.75 | 43.44 | 2.52 | 2.72 |
| CdTe | -4.786 | -4.797 | -4.802 | 16.3 | -0.14 | 72.32 | 72.53 | 72.53 | 218 | 0.14 | 40.10 | 35.19 | 35.22 | 4.90 | 5.75 |

**Table S1.5** Convergence of the equilibrium total energies $E_0$, equilibrium volumes $V_0$ and bulk moduli $B_0$, obtained by fitting energy/volume curves calculated with the PBE exchange-correlation functional to the Birch-Murnaghan equation of state (Eq. 6 in the text),[2] with respect to plane-wave cutoff. The parameters obtained with the base cutoffs listed in Table 1 in the text and those calculated with the cutoffs increased by 1.5 and 2 × this value are shown together with the ranges and the standard deviations as percentages of the values obtained with the base cutoffs. Note that the production calculations described in the text were performed with 1.5 × the base cutoffs.

| System | $E_0$ [eV] | | | Range [meV] | Std. [%] | $V_0$ [Å³] | | | Range [$10^{-3}$ Å³] | Std. [%] | $B_0$ [GPa] | | | Range [GPa] | Std. [%] |
|---|---|---|---|---|---|---|---|---|---|---|---|---|---|---|---|
| | Default | 1.5 × | 2 × | | | Default | 1.5 × | 2 × | | | Default | 1.5 × | 2 × | | |
| C | -18.21 | -18.21 | -18.21 | 0.0 | 0.00 | 11.39 | 11.39 | 11.38 | 0.69 | 0.00 | 434.9 | 432.9 | 432.9 | 2.08 | 0.22 |
| Si | -10.90 | -10.90 | -10.90 | 0.8 | 0.00 | 40.79 | 40.79 | 40.79 | 0.88 | 0.00 | 88.39 | 88.53 | 88.49 | 0.14 | 0.07 |
| Ge | -8.979 | -8.989 | -8.989 | 10.6 | -0.05 | 48.31 | 48.27 | 48.18 | 129 | 0.11 | 55.14 | 55.93 | 60.27 | 5.12 | 4.08 |
| AlP | -10.39 | -10.39 | -10.39 | 0.9 | 0.00 | 41.65 | 41.64 | 41.64 | 8.06 | 0.01 | 83.58 | 83.48 | 83.32 | 0.26 | 0.13 |
| AlAs | -9.395 | -9.397 | -9.397 | 1.7 | -0.01 | 46.89 | 46.89 | 46.89 | 5.26 | 0.00 | 67.57 | 66.85 | 66.89 | 0.72 | 0.49 |
| AlSb | -8.222 | -8.225 | -8.225 | 2.6 | -0.01 | 60.30 | 60.30 | 60.30 | 3.44 | 0.00 | 49.59 | 49.93 | 49.84 | 0.34 | 0.29 |
| GaP | -9.107 | -9.109 | -9.109 | 2.6 | -0.01 | 41.72 | 41.71 | 41.72 | 9.79 | 0.01 | 76.18 | 76.83 | 76.29 | 0.66 | 0.37 |
| GaAs | -8.237 | -8.242 | -8.242 | 5.7 | -0.03 | 47.50 | 47.57 | 47.58 | 76.9 | 0.07 | 62.85 | 60.12 | 59.75 | 3.10 | 2.20 |
| GaSb | -7.343 | -7.349 | -7.350 | 7.0 | -0.04 | 60.09 | 59.96 | 60.00 | 133 | 0.09 | 41.75 | 45.48 | 43.91 | 3.74 | 3.67 |
| InP | -8.401 | -8.404 | -8.404 | 2.4 | -0.01 | 52.77 | 52.81 | 52.81 | 45.3 | 0.04 | 62.31 | 59.70 | 59.77 | 2.61 | 1.95 |
| InAs | -7.690 | -7.694 | -7.694 | 4.1 | -0.02 | 59.16 | 59.22 | 59.23 | 73.4 | 0.06 | 51.91 | 49.24 | 48.87 | 3.04 | 2.61 |
| InSb | -6.932 | -6.937 | -6.937 | 5.6 | -0.04 | 72.77 | 72.81 | 72.90 | 132 | 0.08 | 38.95 | 37.65 | 36.26 | 2.69 | 2.82 |
| ZnS | -6.849 | -6.849 | -6.849 | 0.2 | 0.00 | 40.38 | 40.37 | 40.38 | 9.21 | 0.01 | 68.26 | 69.18 | 68.53 | 0.92 | 0.57 |
| ZnSe | -6.023 | -6.024 | -6.023 | 0.9 | -0.01 | 47.20 | 47.20 | 47.20 | 0.74 | 0.00 | 55.02 | 56.87 | 56.23 | 1.85 | 1.39 |
| ZnTe | -5.178 | -5.180 | -5.180 | 1.2 | -0.01 | 59.01 | 58.96 | 58.96 | 50.0 | 0.04 | 41.23 | 43.09 | 43.18 | 1.95 | 2.18 |
| CdTe | -4.786 | -4.788 | -4.787 | 1.9 | -0.02 | 72.32 | 72.57 | 72.49 | 253 | 0.15 | 40.10 | 35.32 | 36.93 | 4.78 | 4.95 |

**Table S1.6** Convergence of the equilibrium total energies $E_0$, equilibrium volumes $V_0$ and bulk moduli $B_0$, obtained by fitting energy/volume curves calculated with the PBE exchange-correlation functional to the Birch-Murnaghan equation of state (Eq. 6 in the text),[2] with respect to $k$-point sampling. The parameters obtained with the base $k$-point sampling listed in Table 1 in the text and those calculated with the meshes increased by 1.5 and 2 × along each of the reciprocal lattice vectors this value are shown together with the ranges and the standard deviations as percentages of the values obtained with the base $k$-point meshes.

**S2. Calculation of High-Frequency Dielectric Constants**

In this section we discuss the calculation of the high-frequency dielectric constant $\varepsilon_\infty$ and compare the values obtained using the independent-particle random-phase approximation (IP-RPA) method employed for the calculations in the text to those obtained using density-functional perturbation theory (DFPT) and from the self-consistent response to finite fields.

VASP implements several methods of calculating $\varepsilon_\infty$ *viz.* using the IP-RPA (the `LOPTICS` tag), using DFPT (`LEPSILON`),[3] and from the response to a finite electric field (`LCALCEPS`).[4] All three methods are available for use with the LDA and GGA functionals, whereas DFPT is not implemented for meta-GGA and non-local hybrid functionals. More sophisticated methods of evaluating the dielectric properties are also available, including time-dependent density-functional theory (TD-DFT), *GW* theory and the Bethe-Salpeter equation (BSE),[5–7] but we consider the present implementations to be too computationally demanding for high-throughput modeling and therefore beyond the scope of this work.

$\varepsilon_\infty$ is obtained in the IP-RPA formalism from the frequency-dependent dielectric function $\varepsilon(E) = \varepsilon_{re}(E) + i\varepsilon_{im}(E)$ as $\varepsilon_\infty = \varepsilon_{re}(E = 0)$. The imaginary part of the dielectric function $\varepsilon_{im}(E)$ is evaluated as a weighted sum of transitions between occupied and virtual states using Fermi's Golden Rule, and the real part $\varepsilon_{re}(E)$ is then obtained from the Kramers-Kronig relation. This method requires a significant number of empty conduction states to be included in the calculation in order to converge, and the VASP implementation also neglects local-field effects. The DFPT and finite-field approaches do not require additional virtual states to be included in calculations. The DFPT implementation calculates $\varepsilon_\infty$ both with and without local-field effects, while the finite-field method includes local-field effects implicitly. However, DFPT and the finite-field method both involve iterative processes that are incompatible with non-self-consistent calculations.

To compare the accuracy of the different approaches, we computed the dielectric constants of the sixteen semiconductors in Table 1 in the text using PBE[1] with the DFPT, finite-field and IP-RPA methods, and using PBE0[8] and HSE06[9] with the finite-field and IP-RPA methods (Table S2.1).

Comparing the PBE DFPT calculations with and without local-field effects shows that these effects reduce $\varepsilon_\infty$ by an average 4.05 %. DFPT predicts anomalously-large dielectric constants for Ge and GaSb, which we attribute to the small but finite PBE bandgaps of 28 and 93 meV (see text). The metallic electronic structures of InAs and InSb result in $\varepsilon_\infty$ being overestimated by 1.2-3.7 × the experimental values, although these results are far more realistic than those for Ge and GaSb.

The finite-field method is prone to failure in narrow-gap systems due to interband (Zener) tunnelling.[10] The $\varepsilon_\infty$ obtained for Ge, GaSb, InAs and InSb using this method are all significantly overestimated at 1.4-9.6 × the experimental values, although the method does not divergence for Ge and GaSb as DFPT does. Excluding these four systems, the $\varepsilon_\infty$ obtained from the finite-field and IP-RPA methods are an average 13 and 10.7 % smaller than the equivalent DFPT values. Particularly large differences are observed for GaAs, InP and CdTe, and further excluding these data points reduces the average differences to 5.68 and 5.43 %. In principle, the IP-RPA and DFPT results excluding local-field effects should match, but given the clearly anomalous results obtained for some compounds with DFPT the origin of the discrepancy is not clear.

To verify the convergence of the IP-RPA results with respect to the number of virtual states, we compared the $\varepsilon_\infty$ obtained with our chosen number of virtual states (see text) to those calculated using orbital energies from exact diagonalization of the Hamiltonian in the full basis spanned by the plane-wave cutoff, ranging from 216 bands for diamond (a 54 × excess of virtual states) to 725 for CdTe (~80 × excess; Table S2.2). The two sets of results are well within 1 % of each other, indicating that the IP-RPA calculations are converged and an insufficient number of unoccupied states is unlikely to explain the discrepancy between the DFPT and IP-RPA results.

In the finite-field calculations, we found that the $\varepsilon_\infty$ converge much more slowly than the total energy. By default, the convergence criteria are tightened by two orders of magnitude during the field-polarised calculations, so that our chosen tolerance of $10^{-6}$ eV on the total energy implies a tolerance of $10^{-8}$ eV under the applied field. To check convergence with respect to the tolerance, we performed a further set of calculations with a tighter $10^{-8}$ eV tolerance (i.e. $10^{-10}$ eV under the field). As shown in Table S2.2, this tighter tolerance is necessary to converge the calculated $\varepsilon_\infty$ to < 0.1, and is particularly important for the small-gap systems. We therefore used the tighter $10^{-8}$ eV tolerance during all our finite-field calculations.

Returning to the data in Table S2.1, using PBE0 with the IP-RPA method underestimates the $\varepsilon_\infty$ obtained with the finite-field method by an average 21.3 %. For HSE06 the average difference is a much smaller 3.2 %, although a general tendency to underestimate is masked by large overestimates for InAs and InSb, without which the average underestimation rises to 5.6 %. The finite-field values are consistently closer to the experimental measurements (Table S2.3; c.f. Table 1 in the text and Table S2.1), with mean average relative errors (MAREs; Eq. 2 in the text) of $7.16 \pm 5.76$ and $5.56 \pm 4.74$ % for PBE0 and HSE06, respectively, compared to errors of $26.8 \pm 7.04$ and $9.13 \pm 2.89$ % using the IP-RPA method. With the exception of the PBE0 IP-RPA results, the dielectric constants predicted by the hybrid functionals are closer to the experimental measurements than the three sets of values predicted with PBE, although excluding the four zero/small-gap compounds the PBE finite-field calculations show good performance, with a MARE OF $12.8 \pm 4.82$ %.

The accuracy with which $\varepsilon_\infty$ can be calculated is important for the dielectric-dependent scPBE0 calculations. Table S2.4 compares the optimised exact-exchange fractions $\alpha$, bandgaps and dielectric constants obtained using the IPA-RPA and finite-field approaches to calculate $\varepsilon_\infty$. The narrow bandgaps of InSb and InAs made the finite-field calculations unstable after the initial self-consistent updates of $\alpha$, so these systems were excluded from testing. These two compounds, together with Ge and GaSb, were similarly excluded from the scPBE0 calculations using the IP-RPA method due to the divergence of $\varepsilon_\infty$ as $\alpha \to 0$.

Across the 12 systems included in both calculations, the IP-RPA overestimates $\alpha$ by an average 13 % compared to the finite-field values, leading to differences in the converged bandgaps and high-frequency dielectric constants of 4.9 and -11.4 % respectively (Table S2.4). The $\varepsilon_\infty$ obtained from the finite-field scPBE0 calculations are generally a better match to experiment, with a smaller MARE of $4.73 \pm 4.38$ % compared to $9.65 \pm 2.71$ % (Table S2.3). Both sets of results are however a significant improvement over bare PBE0 and are comparable to the HSE06 calculations, with a slightly smaller MARE for the finite-field scPBE0 results and a slightly larger one for the corresponding IP-RPA values.

| System | PBE | | | | HSE06 | | PBE0 | |
|---|---|---|---|---|---|---|---|---|
| | DFPT[a] | DFPT[b] | Finite Field | IP-RPA | Finite Field | IP-RPA | Finite Field | IP-RPA |
| C | 5.96 | 5.85 | 5.84 | 5.88 | 5.61 | 5.08 | 5.59 | 4.72 |
| Si | 13.89 | 13.27 | 12.80 | 13.15 | 11.11 | 10.04 | 10.86 | 8.23 |
| Ge | $1.00 \times 10^5$ | $9.30 \times 10^4$ | 24.35 | $2.02 \times 10^3$ | 14.75 | 15.81 | 14.28 | 10.95 |
| AlP | 8.96 | 8.32 | 8.06 | 8.62 | 7.22 | 6.91 | 7.14 | 5.86 |
| AlAs | 10.36 | 9.78 | 9.18 | 9.76 | 8.11 | 7.63 | 8.00 | 6.37 |
| AlSb | 13.22 | 12.66 | 11.36 | 12.19 | 9.65 | 9.02 | 9.46 | 7.25 |
| GaP | 11.75 | 11.21 | 10.20 | 10.90 | 9.65 | 8.20 | 8.70 | 6.80 |
| GaAs | 29.99 | 29.57 | 13.14 | 17.37 | 10.50 | 9.99 | 10.31 | 7.89 |
| GaSb | $2.01 \times 10^3$ | $1.98 \times 10^3$ | 20.69 | $1.53 \times 10^2$ | 12.89 | 12.92 | 12.47 | 9.39 |
| InP | 15.63 | 15.22 | 10.67 | 12.17 | 8.85 | 8.24 | 8.68 | 6.67 |
| InAs | 15.71 | 15.18 | 24.09 | 17.61 | 11.47 | 13.25 | 10.93 | 8.32 |
| InSb | 57.95 | 56.66 | 150.71 | 21.56 | 12.75 | 14.29 | 12.15 | 9.10 |
| ZnS | 6.31 | 5.93 | 5.90 | 6.18 | 5.12 | 4.89 | 5.09 | 4.38 |
| ZnSe | 8.15 | 7.78 | 7.11 | 7.54 | 5.95 | 5.60 | 5.91 | 4.85 |
| ZnTe | 10.05 | 9.62 | 8.71 | 9.26 | 7.31 | 6.85 | 7.22 | 5.75 |
| CdTe | 10.84 | 10.47 | 8.47 | 9.21 | 6.99 | 6.52 | 6.88 | 5.42 |

**Table S2.1** High-frequency dielectric constants $\varepsilon_\infty$ for the sixteen tetrahedral semiconductors in Table 1 in the text calculated with PBE, PBE0 and HSE06 using the density-functional theory (DFPT), finite-field and independent-particle random-phase approximation (IP-RPA) summation methods. [a] DFPT values excluding local-field effects. [b] DFPT values including local-field effects.

| System | IP-RPA | IP-RPA (Exact) | Finite Field $(10^{-6})$ | Finite Field $(10^{-8})$ |
|---|---|---|---|---|
| C | 5.88 | 5.87 | 5.46 | 5.84 |
| Si | 13.15 | 13.16 | 12.66 | 12.80 |
| Ge | $2.02 \times 10^3$ | $2.02 \times 10^3$ | 17.74 | 24.35 |
| AlP | 8.62 | 8.62 | 8.16 | 8.06 |
| AlAs | 9.76 | 9.75 | 9.32 | 9.18 |
| AlSb | 12.19 | 12.20 | 11.44 | 11.36 |
| GaP | 10.90 | 10.91 | 9.99 | 10.20 |
| GaAs | 17.37 | 17.39 | 12.31 | 13.14 |
| GaSb | 152.6 | 152.7 | 16.66 | 20.69 |
| InP | 12.17 | 12.20 | 10.16 | 10.67 |
| InAs | 17.61 | 17.64 | 23.10 | 24.09 |
| InSb | 21.56 | 21.64 | 131.9 | 150.7 |
| ZnS | 6.18 | 6.19 | 5.82 | 5.90 |
| ZnSe | 7.54 | 7.55 | 6.98 | 7.11 |
| ZnTe | 9.26 | 9.28 | 8.63 | 8.71 |
| CdTe | 9.21 | 9.27 | 8.22 | 8.47 |

**Table S2.2** Comparison of high-frequency dielectric constants $\varepsilon_\infty$ calculated using PBE with the independent-particle random-phase approximation (IP-RPA) and finite-field methods. The second and third columns compare values obtained from IP-RPA calculations using the numbers of virtual states listed in the text and with the virtual states obtained by exact diagonalization of the Hamiltonian in the full orbital basis spanned by the plane-wave cutoff. The fourth and fifth columns compare finite-field results obtained with tolerances of $10^{-6}$ and $10^{-8}$ eV on the total energy during the initial electronic-structure optimisation, which correspond to tighter tolerances of $10^{-8}$ and $10^{-10}$ eV in the subsequent field-polarised calculations.

| XC Functional | Method | MAE | MARE [%] |
|---|---|---|---|
| PBE[a] | DFPT[b] | $3.86 \pm 4.79$ | $42.4 \pm 42.6$ |
| | DFPT[c] | $3.41 \pm 4.80$ | $36.8 \pm 43.2$ |
| | Finite Field | $0.89 \pm 0.34$ | $12.8 \pm 4.82$ |
| | IP-RPA | $1.95 \pm 1.48$ | $22.7 \pm 12.9$ |
| HSE06 | Finite Field | $0.69 \pm 0.91$ | $5.56 \pm 4.74$ |
| | IP-RPA | $0.91 \pm 0.43$ | $9.13 \pm 2.89$ |
| PBE0 | Finite Field | $0.88 \pm 0.95$ | $7.16 \pm 5.76$ |
| | IP-RPA | $2.86 \pm 1.67$ | $26.8 \pm 7.04$ |
| scPBE0[d] | Finite Field | $0.54 \pm 0.76$ | $4.73 \pm 4.38$ |
| | IP-RPA | $0.76 \pm 0.17$ | $9.65 \pm 2.71$ |
| scPBE0 (PBE)[e] | IP-RPA | $0.90 \pm 0.19$ | $11.6 \pm 3.71$ |

**Table S2.3** Comparison of the mean absolute error (MAE; Eq. 1 in the text) and mean absolute relative error (MARE; Eq. 2 in the text) in the high-frequency dielectric constants $\varepsilon_\infty$ of the sixteen tetrahedral semiconductors in Table 1 in the text calculated using the different approaches described above. [a] The PBE results exclude InP, InAs, Ge and GaSb due to the zero or very narrow PBE bandgaps and resultant large error in the calculated $\varepsilon_\infty$ (see above for explanation). [b] DFPT values excluding local-field effects. [c] DFPT values including local-field effects. [d] scPBE0 finite-field results exclude InSb and InAs as the narrow bandgaps caused the finite-field calculations to become unstable after the initial self-consistent updates of $\alpha$ (see text). [e] These results exclude InP, InAs, Ge and GaSb due to the divergence of $\varepsilon_\infty$ as $\alpha \rightarrow 0$ (see text).

| System | Finite Field | | | Summation (IPA) | | |
|---|---|---|---|---|---|---|
| | $\alpha$ | $E_g$ [eV] | $\varepsilon_\infty$ | $\alpha$ | $E_g$ [eV] | $\varepsilon_\infty$ |
| C | 0.177 | 5.45 | 5.66 | 0.203 | 5.65 | 4.92 |
| Si | 0.083 | 0.98 | 12.06 | 0.093 | 1.03 | 10.79 |
| Ge | 0.053 | 0.29 | 18.97 | - | - | - |
| AlP | 0.133 | 2.29 | 7.52 | 0.149 | 2.38 | 6.72 |
| AlAs | 0.117 | 2.04 | 8.56 | 0.131 | 2.11 | 7.63 |
| AlSb | 0.095 | 1.66 | 10.56 | 0.105 | 1.71 | 9.52 |
| GaP | 0.106 | 2.19 | 9.47 | 0.118 | 2.26 | 8.48 |
| GaAs[a] | 0.084 | 0.98 | 11.92 | 0.086 | 0.99 | 11.63 |
| GaSb | 0.062 | 0.40 | 16.24 | - | - | - |
| InP | 0.104 | 1.23 | 9.64 | 0.116 | 1.29 | 8.62 |
| InAs | - | - | - | - | - | - |
| InSb | - | - | - | - | - | - |
| ZnS | 0.191 | 3.51 | 5.24 | 0.221 | 3.74 | 4.52 |
| ZnSe | 0.160 | 2.38 | 6.26 | 0.189 | 2.58 | 5.28 |
| ZnTe | 0.128 | 1.99 | 7.84 | 0.148 | 2.11 | 6.78 |
| CdTe | 0.134 | 1.47 | 7.48 | 0.159 | 1.61 | 6.30 |

**Table S2.4** Comparison of the optimised fraction of exact exchange $\alpha$, bandgap $E_g$ and high-frequency dielectric constants $\varepsilon_\infty$ for the sixteen semiconductors in Table 1 in the text obtained from scPBE0 calculations using the finite-field and independent-particle random-phase approximation (IP-RPA) methods. [a] Using the finite-field method, it was not possible to converge $\varepsilon_\infty$ to < $10^{-2}$, even with the tighter optimisation tolerance of $10^{-8}$ eV (see text); this value is therefore the average of the last six optimisation cycles during which $\alpha$ stabilised to a value of 0.084.

**S3. Calculated Bandgaps**

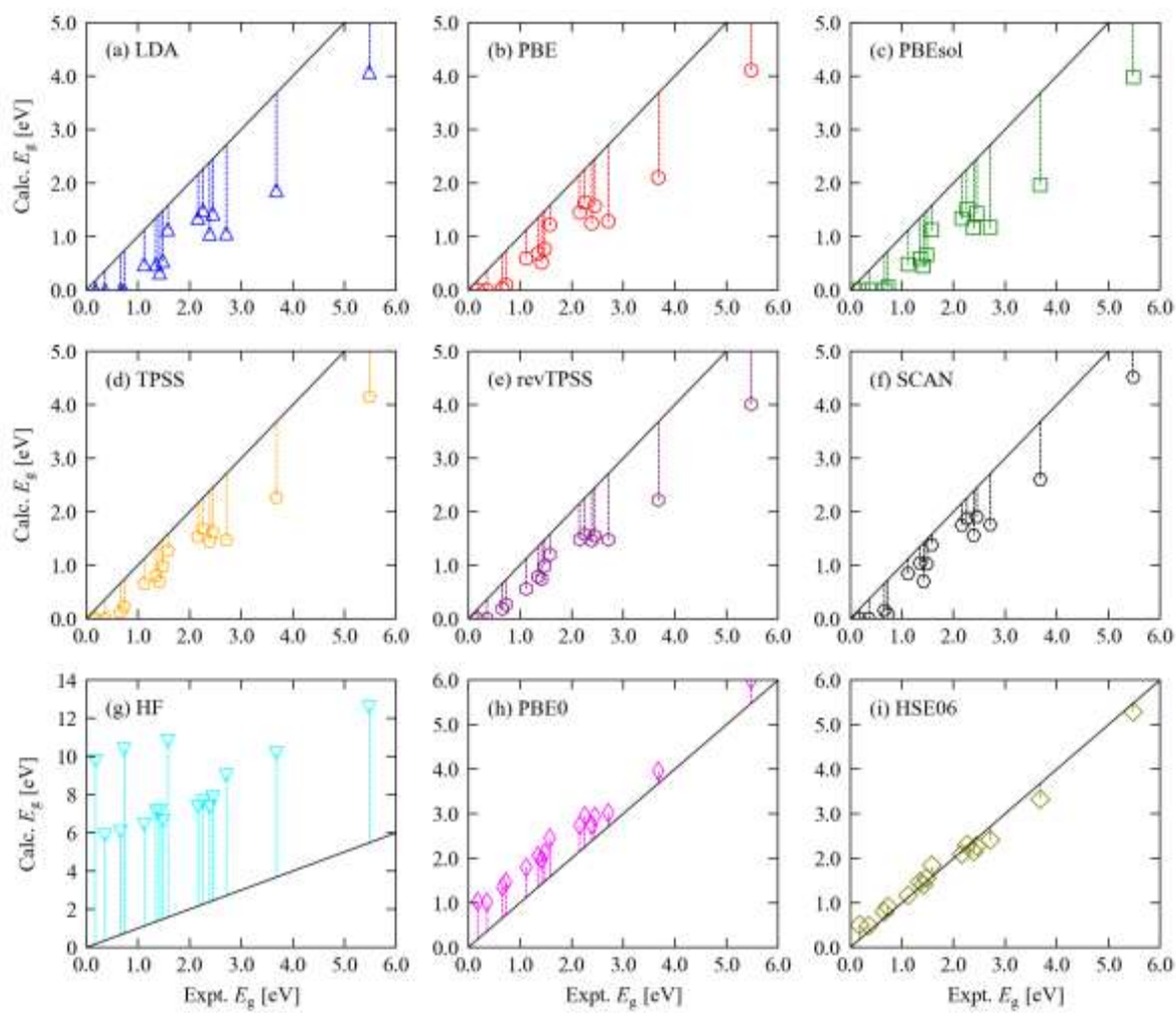

**Figure S3.1** Comparison of the experimental bandgaps $E_\text{g}$ of the sixteen tetrahedral semiconductors examined in this work to bandgaps calculated with the LDA (a),[11] the PBE (b)[1] and PBEsol (c)[12] GGA functionals, the TPSS (d),[13] revTPSS (e)[14] and SCAN (f)[15] meta-GGA functionals, Hartree-Fock (g) and the PBE0 (h)[8] and HSE06 (i)[9] hybrid functionals.

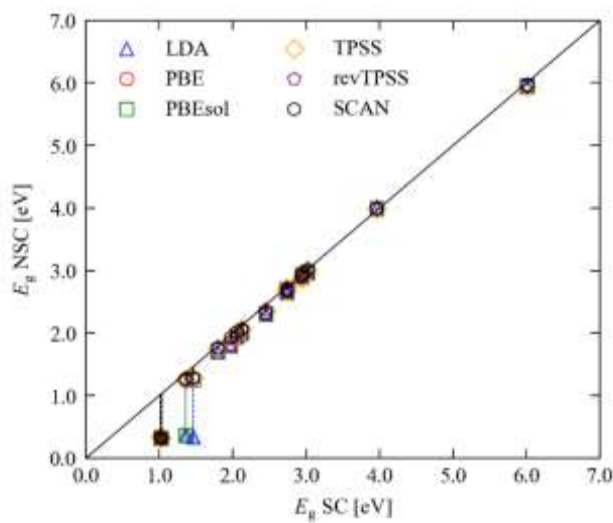

**Figure S3.2** Comparison of the self-consistent (SC) PBE0 bandgaps to non-self-consistent (NSC) PBE0 gaps calculated with orbitals from LDA (blue triangles), PBE (red circles), PBEsol (green squares), TPSS (orange diamonds), revTPSS (purple pentagons) and SCAN (black hexagons). Shaded markers denote systems for which the base functional predicts a bandgap of < 10 meV.

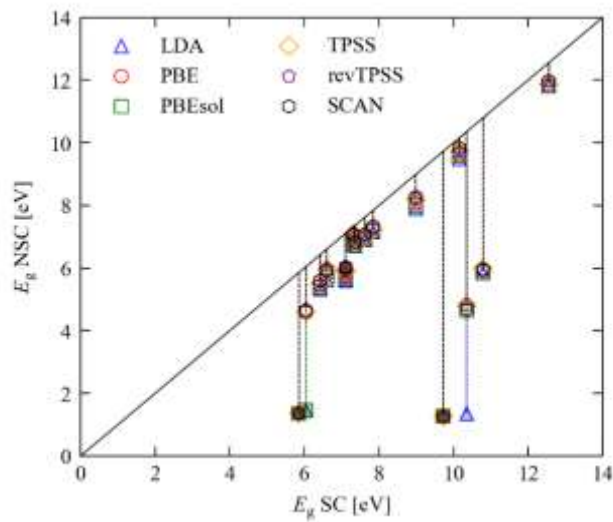

**Figure S3.3** Comparison of the self-consistent (SC) Hartree-Fock bandgaps to non-self-consistent (NSC) HF gaps calculated with orbitals from LDA (blue triangles), PBE (red circles), PBEsol (green squares), TPSS (orange diamonds), revTPSS (purple pentagons) and SCAN (black hexagons). Shaded markers denote systems for which the base functional predicts a bandgap of < 10 meV.

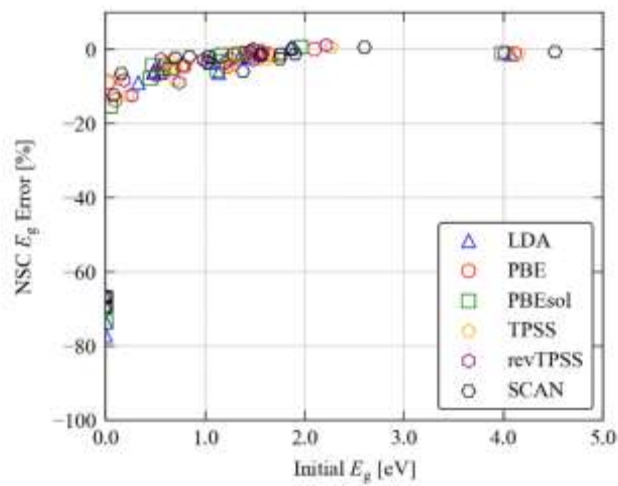

**Figure S3.4** Percentage error on non-self-consistent PBE0 bandgaps calculated using wavefunctions from the LDA, PBE and PBEsol GGA and TPSS, revTPSS and SCAN meta-GGA functionals as a function of the bandgap predicted by the base functional.

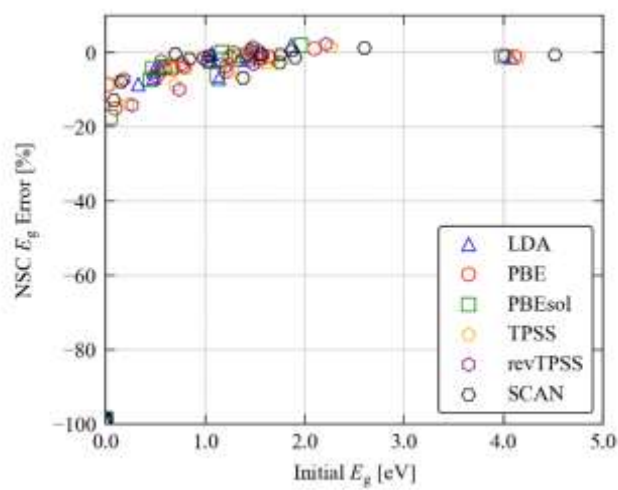

**Figure S3.5** Percentage error on non-self-consistent HSE06 bandgaps calculated using orbitals from the LDA, PBE and PBEsol GGA and TPSS, revTPSS and SCAN meta-GGA functionals as a function of the bandgap predicted by the base functional.

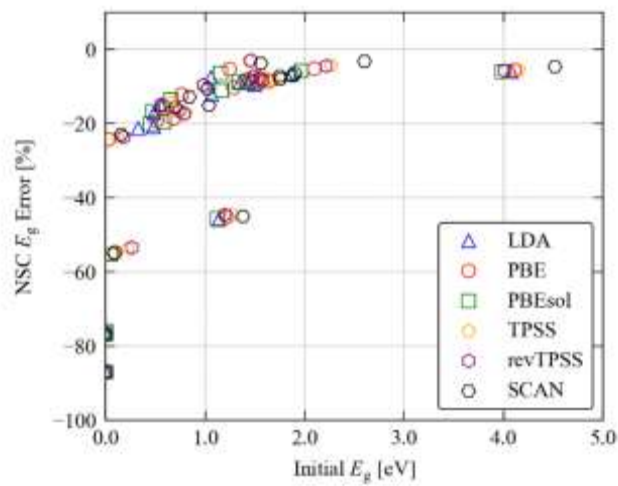

**Figure S3.6** Percentage error on non-self-consistent Hartree-Fock bandgaps calculated using orbitals from the LDA, PBE and PBEsol GGA and TPSS, revTPSS and SCAN meta-GGA functionals as a function of the bandgap predicted by the base functional.

## S4. Density of States, Band Dispersions and Dielectric Properties

| System | PBE $E_F$ [eV] | PBE $E_{VBM}$ [eV] | PBE $E_{CBM}$ [eV] | HSE06 (PBE) $E_F$ [eV] | HSE06 (PBE) $E_{VBM}$ [eV] | HSE06 (PBE) $E_{CBM}$ [eV] | HSE06 $E_F$ [eV] | HSE06 $E_{VBM}$ [eV] | HSE06 $E_{CBM}$ [eV] |
|--------|-------|-------|-------|-------|-------|-------|-------|-------|-------|
| C | 9.973 | 9.969 | 14.08 | 9.423 | 9.418 | 14.64 | 9.347 | 9.339 | 14.62 |
| Si | 5.928 | 5.929 | 6.511 | 5.607 | 5.603 | 6.715 | 5.571 | 5.566 | 6.727 |
| Ge | 4.067 | 4.068 | 4.096 | 3.761 | 3.756 | 4.484 | 3.696 | 3.691 | 4.487 |
| AlP | 4.664 | 4.663 | 6.232 | 4.194 | 4.190 | 6.432 | 4.173 | 4.173 | 6.449 |
| AlAs | 3.207 | 3.198 | 4.646 | 2.750 | 2.743 | 4.826 | 2.761 | 2.754 | 4.850 |
| AlSb | 4.892 | 4.889 | 6.089 | 4.527 | 4.525 | 6.255 | 4.448 | 4.447 | 6.264 |
| GaP | 4.733 | 4.727 | 6.314 | 4.287 | 4.280 | 6.523 | 4.282 | 4.279 | 6.552 |
| GaAs[a] | 3.290 | 3.287 | 3.800 | 2.862 | 2.860 | 4.176 | 2.889 | 2.882 | 4.286 |
| GaSb | 5.089 | 5.087 | 5.181 | 4.751 | 4.750 | 5.529 | 4.681 | 4.676 | 5.594 |
| InP | 5.205 | 5.199 | 5.879 | 4.787 | 4.782 | 6.189 | 4.769 | 4.768 | 6.238 |
| InAs | 3.920 | 3.919 | 3.919 | 3.699 | 3.516 | 3.516 | 3.538 | 3.525 | 4.016 |
| InSb | 5.187 | 5.188 | 5.188 | 5.017 | 4.853 | 4.853 | 4.783 | 4.779 | 5.286 |
| ZnS | 2.162 | 2.156 | 4.254 | 1.301 | 1.296 | 4.656 | 1.424 | 1.423 | 4.747 |
| ZnSe | 1.806 | 1.805 | 3.090 | 1.049 | 1.041 | 3.454 | 1.150 | 1.142 | 3.556 |
| ZnTe | 3.318 | 3.315 | 4.560 | 2.737 | 2.732 | 4.858 | 2.779 | 2.773 | 4.924 |
| CdTe | 2.456 | 2.450 | 3.214 | 1.933 | 1.926 | 3.433 | 1.932 | 1.930 | 3.482 |

**Table S4.1** Fermi energies $E_F$ and energies of the valence-band minima and conduction-band maxima ($E_{VBM}/E_{CBM}$) obtained with PBE,[1] non-self-consistent HSE06 using the PBE orbitals, and self-consistent HSE06.[9] The $E_F$ are obtained from the calculation used to generate the DoS in Figs. S4.1 - S4.16, and the $E_{VBM}$ and $E_{CBM}$ are obtained along the $k$-point path in the dispersion curves.

| System | PBE | | | HSE06 (PBE) | | |
|---|---|---|---|---|---|---|
| | $\Delta E_F$ [meV] | $\Delta E_{VBM}$ [meV] | $\Delta E_{CBM}$ [meV] | $\Delta E_F$ [meV] | $\Delta E_{VBM}$ [meV] | $\Delta E_{CBM}$ [meV] |
| C | 626 | 631 | -539 | 75.9 | 79.1 | 20.5 |
| Si | 357 | 363 | -217 | 35.5 | 37.2 | -12.2 |
| Ge | 371 | 377 | -392 | 65.0 | 65.1 | -3.6 |
| AlP | 492 | 491 | -217 | 21.6 | 16.8 | -17.0 |
| AlAs | 446 | 445 | -204 | -11.4 | -10.2 | -23.0 |
| AlSb | 444 | 442 | -175 | 79.3 | 78.0 | -9.0 |
| GaP | 452 | 448 | -238 | 5.1 | 1.0 | -29.4 |
| GaAs[a] | 401 | 405 | -486 | -26.4 | -22.3 | -110 |
| GaSb | 407 | 411 | -413 | 69.7 | 73.4 | -64.6 |
| InP | 436 | 432 | -359 | 17.8 | 14.9 | -49.1 |
| InAs | 381 | 394 | -97.0 | - | - | - |
| InSb | 405 | 409 | -98.4 | - | - | - |
| ZnS | 738 | 733 | -493 | -123 | -127 | -91.1 |
| ZnSe | 656 | 663 | -466 | -102 | -101 | -102 |
| ZnTe | 539 | 542 | -364 | -41.7 | -41.1 | -66.4 |
| CdTe | 524 | 520 | -268 | 0.9 | -3.4 | -48.4 |
| Average [meV] | 524 | 520 | -268 | 0.9 | -3.4 | -48.4 |
| Std. Dev [meV] | 107 | 106 | 139 | 60.3 | 61.1 | 38.1 |

**Table S4.2** Differences in the Fermi energies $E_F$ and energies of the valence-band minima and conduction-band maxima ($E_{VBM}/E_{CBM}$) obtained with PBE and non-self-consistent HSE06 to fully-self-consistent HSE06 calculations. The average and standard deviation (std. dev.) of the values in each column are given in the bottom two rows of the table.

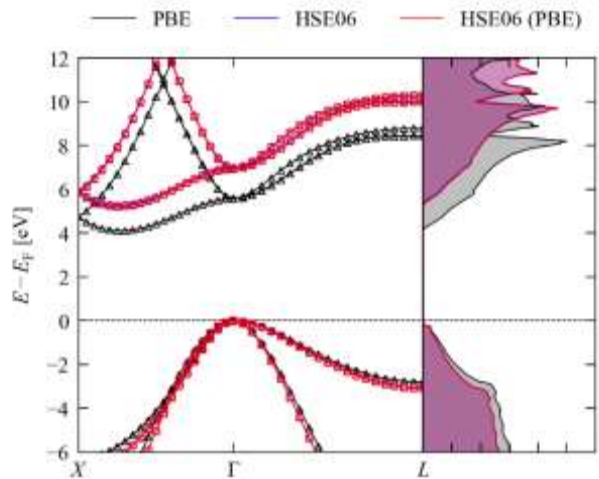

**Figure S4.1** Comparison of the electronic band dispersion and density of states $g(E)$ for C calculated using PBE (black), HSE06 (blue) and non-self-consistent HSE06 calculations using the PBE orbitals (red).

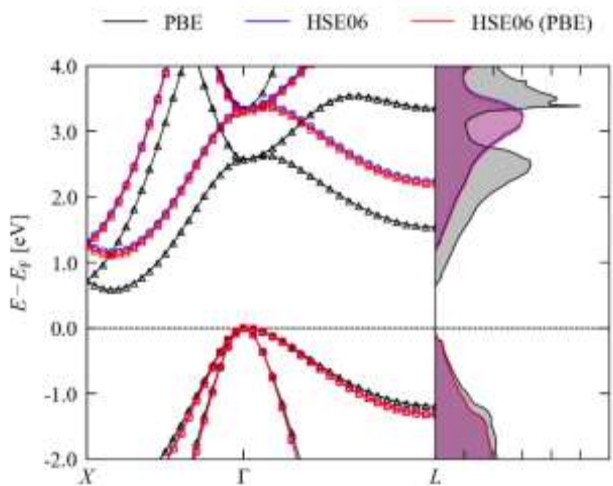

**Figure S4.2** Comparison of the electronic band dispersion and density of states $g(E)$ for Si calculated using PBE (black), HSE06 (blue) and non-self-consistent HSE06 calculations using the PBE orbitals (red).

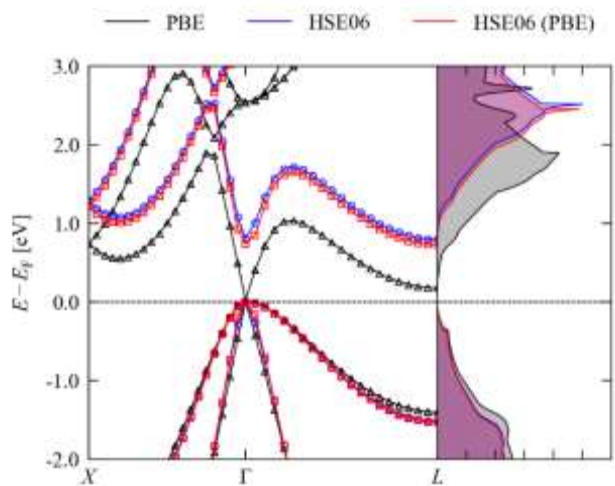

**Figure S4.3** Comparison of the electronic band dispersion and density of states $g(E)$ for Ge calculated using PBE (black), HSE06 (blue) and non-self-consistent HSE06 calculations using the PBE orbitals (red).

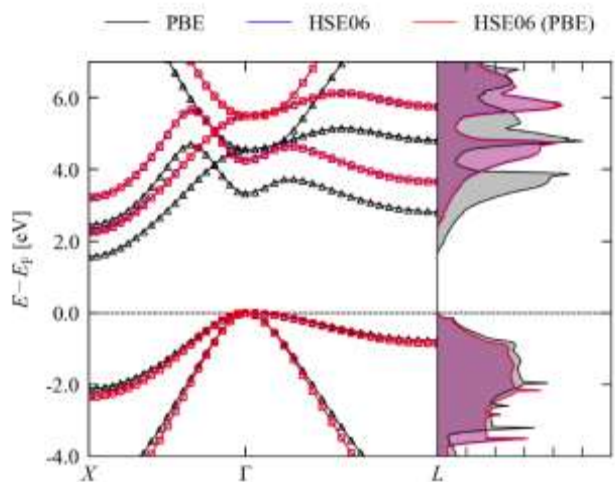

**Figure S4.4** Comparison of the electronic band dispersion and density of states $g(E)$ for AlP calculated using PBE (black), HSE06 (blue) and non-self-consistent HSE06 calculations using the PBE orbitals (red).

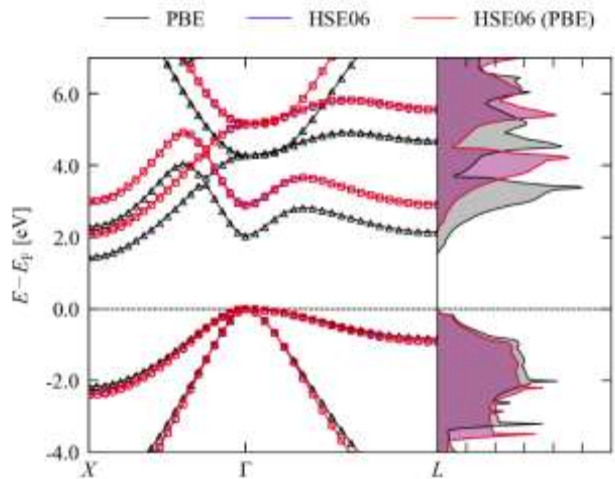

**Figure S4.5** Comparison of the electronic band dispersion and density of states $g(E)$ for AlAs calculated using PBE (black), HSE06 (blue) and non-self-consistent HSE06 calculations using the PBE orbitals (red).

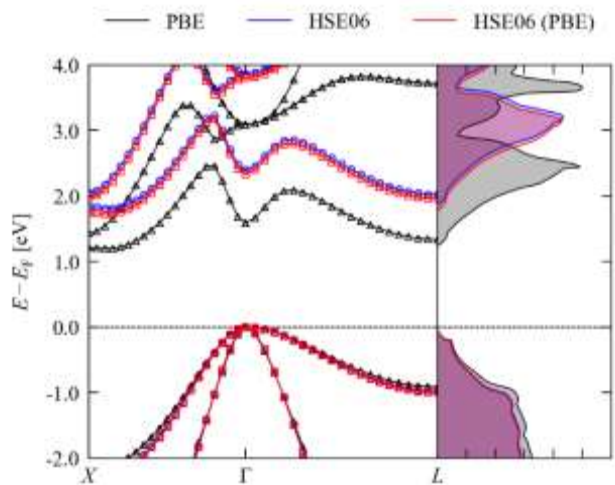

**Figure S4.6** Comparison of the electronic band dispersion and density of states $g(E)$ for AlSb calculated using PBE (black), HSE06 (blue) and non-self-consistent HSE06 calculations using the PBE orbitals (red).

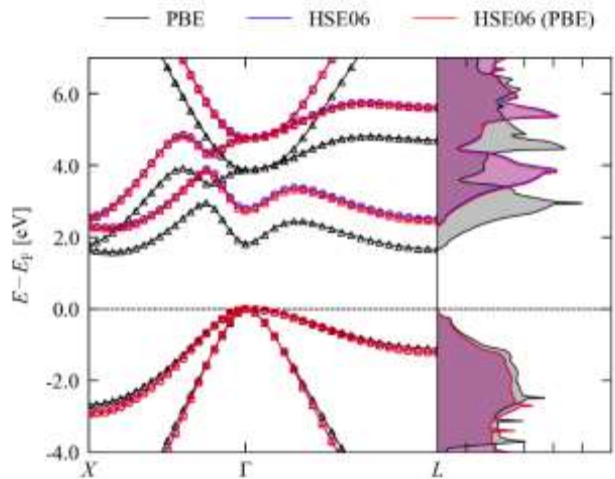

**Figure S4.7** Comparison of the electronic band dispersion and density of states $g(E)$ for GaP calculated using PBE (black), HSE06 (blue) and non-self-consistent HSE06 calculations using the PBE orbitals (red).

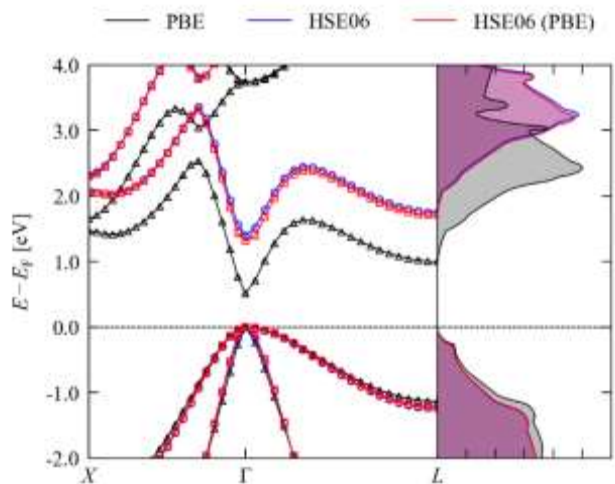

**Figure S4.8** Comparison of the electronic band dispersion and density of states $g(E)$ for GaAs calculated using PBE (black), HSE06 (blue) and non-self-consistent HSE06 calculations using the PBE orbitals (red).

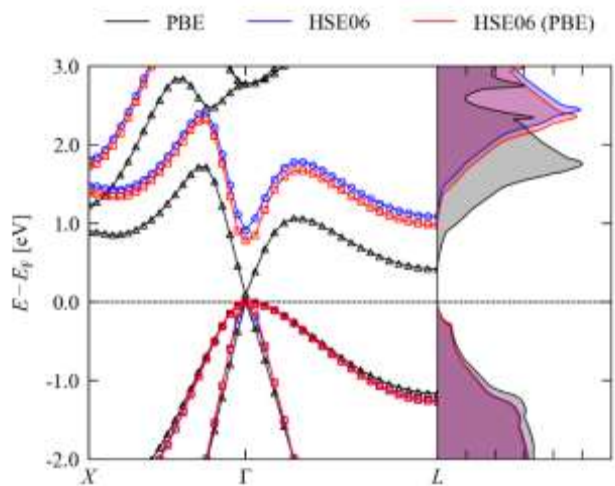

**Figure S4.9** Comparison of the electronic band dispersion and density of states $g(E)$ for GaSb calculated using PBE (black), HSE06 (blue) and non-self-consistent HSE06 calculations using the PBE orbitals (red).

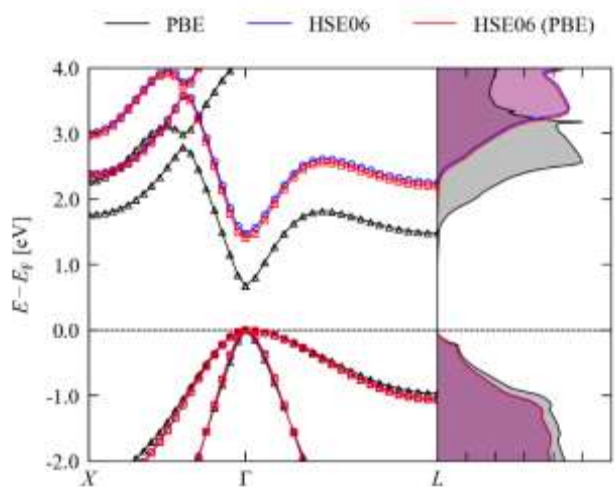

**Figure S4.10** Comparison of the electronic band dispersion and density of states $g(E)$ for InP calculated using PBE (black), HSE06 (blue) and non-self-consistent HSE06 calculations using the PBE orbitals (red).

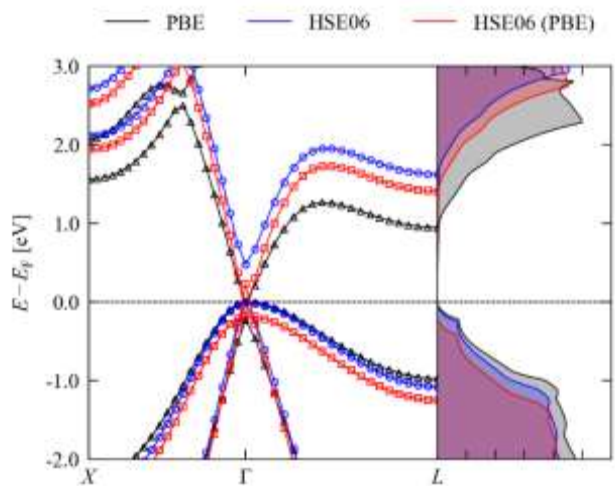

**Figure S4.11** Electronic band dispersion and density of states $g(E)$ for InAs calculated using PBE (black), HSE06 (blue) and non-self-consistent HSE06 calculations using the PBE orbitals (red). The PBE functional predicts a metallic electronic structure for this compound, resulting in notably larger errors in the non-self-consistent calculations (see text).

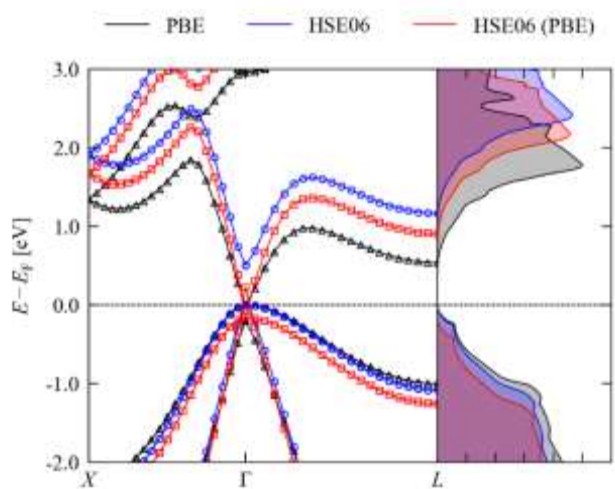

**Figure S4.12** Electronic band dispersion and density of states $g(E)$ for InSb calculated using PBE (black), HSE06 (blue) and non-self-consistent HSE06 calculations using the PBE orbitals (red). The PBE functional predicts a metallic electronic structure for this compound, resulting in notably larger errors in the non-self-consistent calculations (see text).

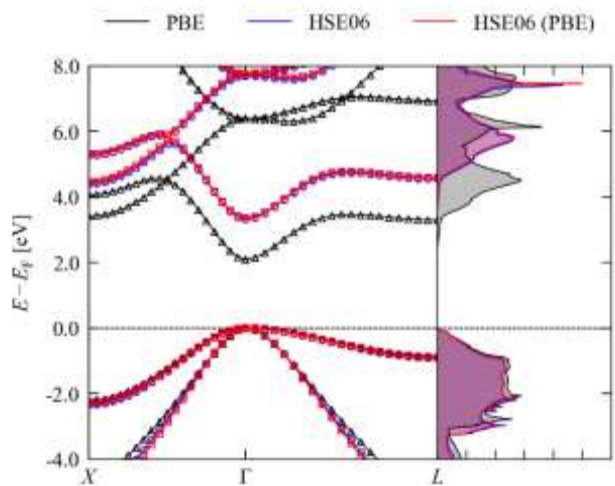

**Figure S4.13** Comparison of the electronic band dispersion and density of states $g(E)$ for ZnS calculated using PBE (black), HSE06 (blue) and non-self-consistent HSE06 calculations using the PBE orbitals (red).

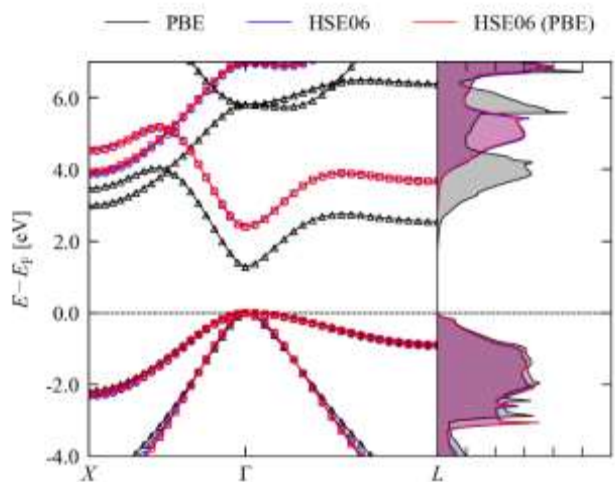

**Figure S4.14** Comparison of the electronic band dispersion and density of states $g(E)$ for ZnSe calculated using PBE (black), HSE06 (blue) and non-self-consistent HSE06 calculations using the PBE orbitals (red).

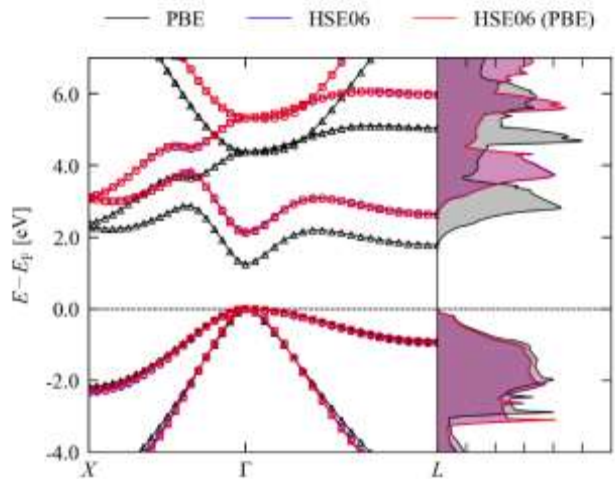

**Figure S4.15** Comparison of the electronic band dispersion and density of states $g(E)$ for ZnTe calculated using PBE (black), HSE06 (blue) and non-self-consistent HSE06 calculations using the PBE orbitals (red).

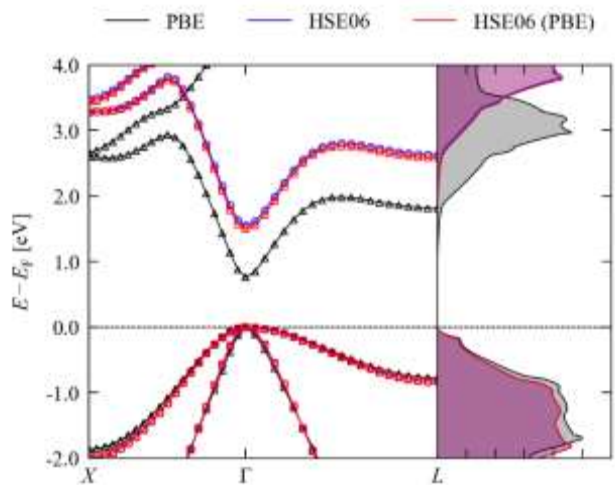

**Figure S4.16** Comparison of the electronic band dispersion and density of states $g(E)$ for CdTe calculated using PBE (black), HSE06 (blue) and non-self-consistent HSE06 calculations using the PBE orbitals (red).

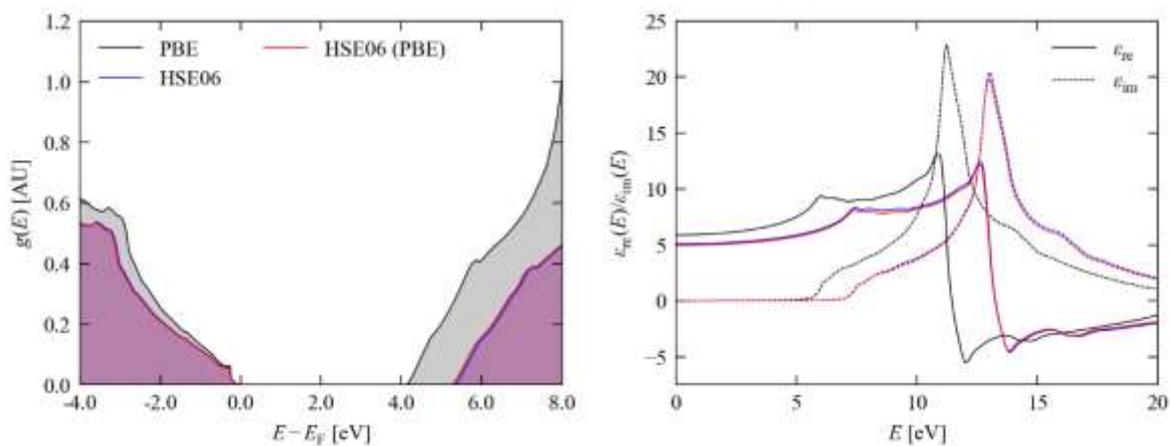

**Figure S4.17** Comparison of the density of states $g(E)$ (a) and real and imaginary energy-dependent dielectric functions $\varepsilon_{\mathrm{re}}(E)/\varepsilon_{\mathrm{im}}(E)$ (b) of C calculated using PBE (black), HSE06 (blue) and non-self-consistent HSE06 calculations using the PBE orbitals (red).

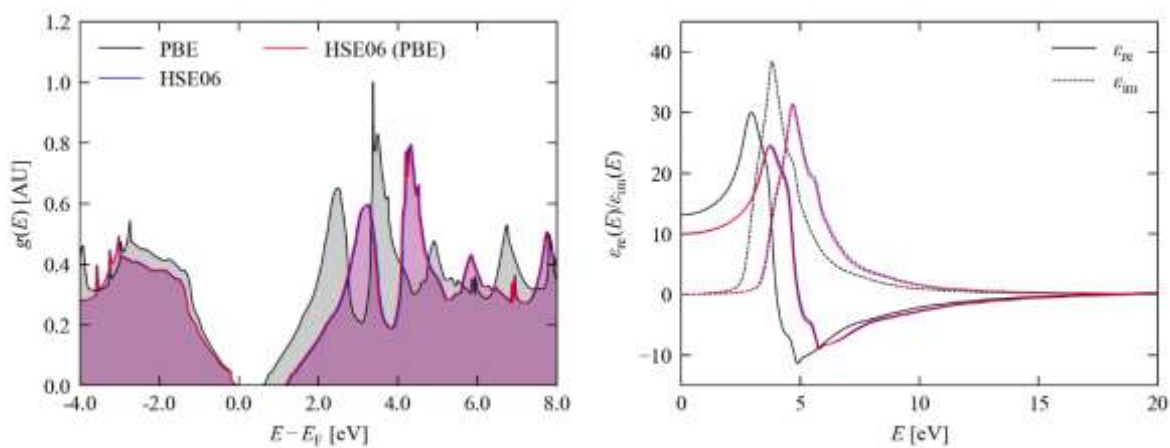

**Figure S4.18** Comparison of the density of states $g(E)$ (a) and real and imaginary energy-dependent dielectric functions $\varepsilon_{\mathrm{re}}(E)/\varepsilon_{\mathrm{im}}(E)$ (b) of Si calculated using PBE (black), HSE06 (blue) and non-self-consistent HSE06 calculations using the PBE orbitals (red).

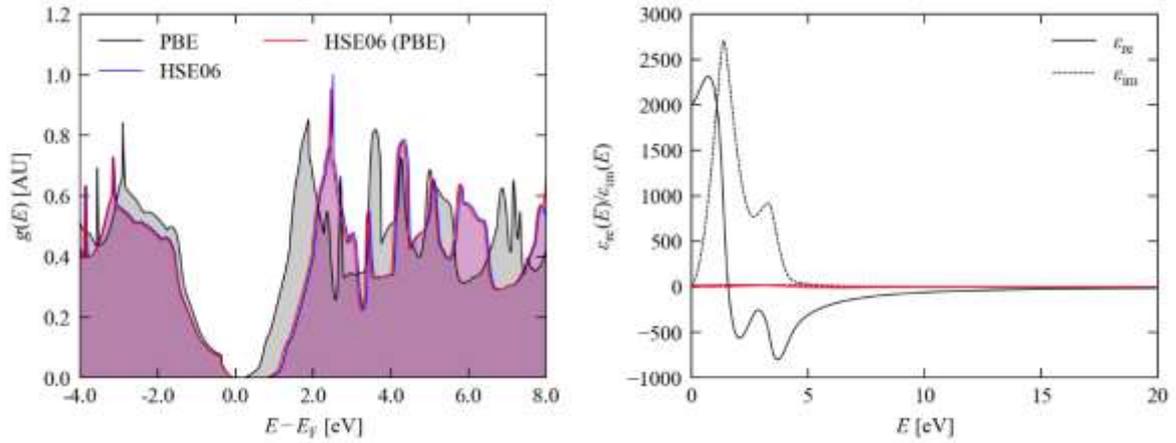

**Figure S4.19** Comparison of the density of states $g(E)$ (a) and real and imaginary energy-dependent dielectric functions $\varepsilon_{\text{re}}(E)/\varepsilon_{\text{im}}(E)$ (b) of Ge calculated using PBE (black), HSE06 (blue) and non-self-consistent HSE06 calculations using the PBE orbitals (red).

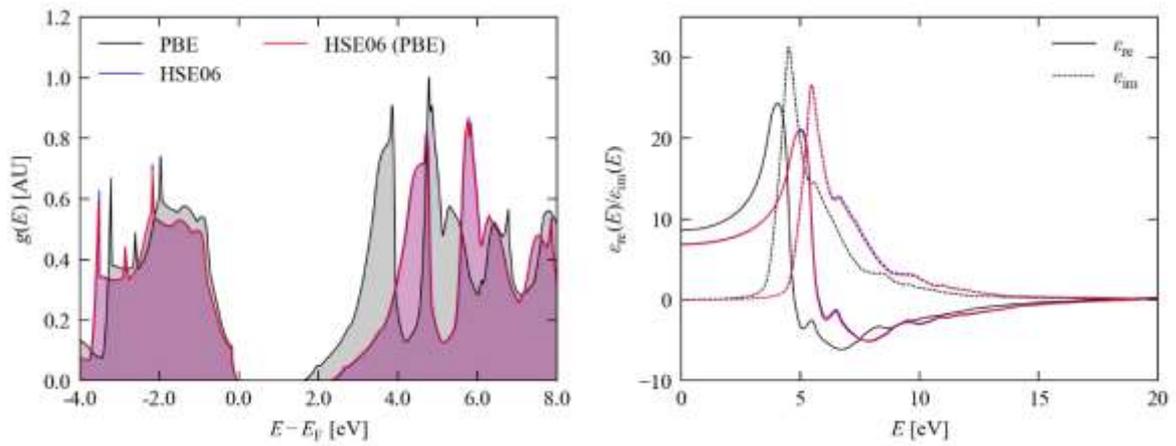

**Figure S4.20** Comparison of the density of states $g(E)$ (a) and real and imaginary energy-dependent dielectric functions $\varepsilon_{\text{re}}(E)/\varepsilon_{\text{im}}(E)$ (b) of AlP calculated using PBE (black), HSE06 (blue) and non-self-consistent HSE06 calculations using the PBE orbitals (red)

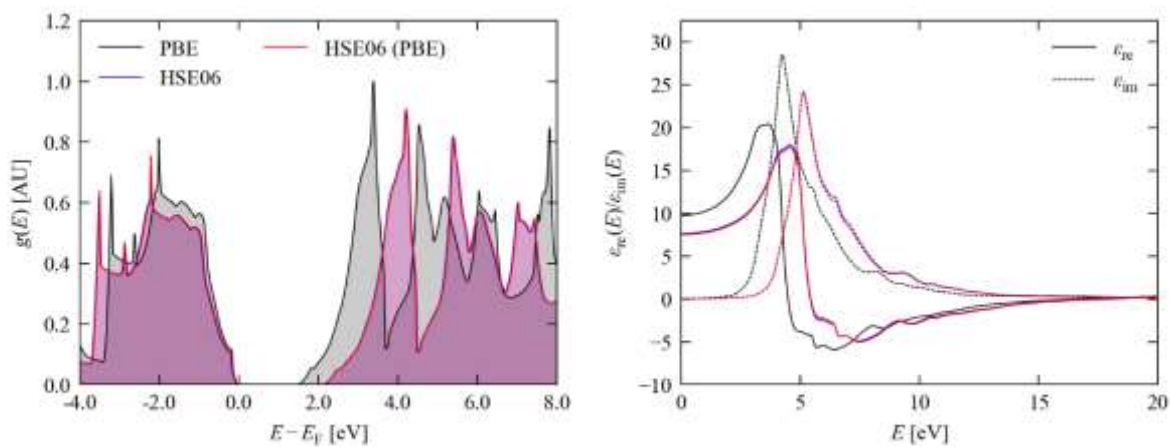

**Figure S4.21** Comparison of the density of states $g(E)$ (a) and real and imaginary energy-dependent dielectric functions $\varepsilon_{re}(E)/\varepsilon_{im}(E)$ (b) of AlAs calculated using PBE (black), HSE06 (blue) and non-self-consistent HSE06 calculations using the PBE orbitals (red).

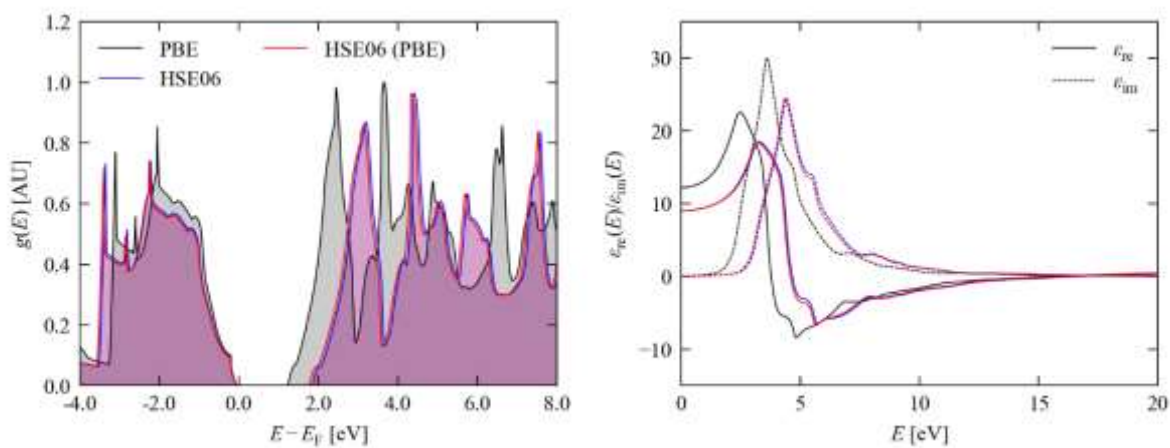

**Figure S4.22** Comparison of the density of states $g(E)$ (a) and real and imaginary energy-dependent dielectric functions $\varepsilon_{re}(E)/\varepsilon_{im}(E)$ (b) of AlSb calculated using PBE (black), HSE06 (blue) and non-self-consistent HSE06 calculations using the PBE orbitals (red).

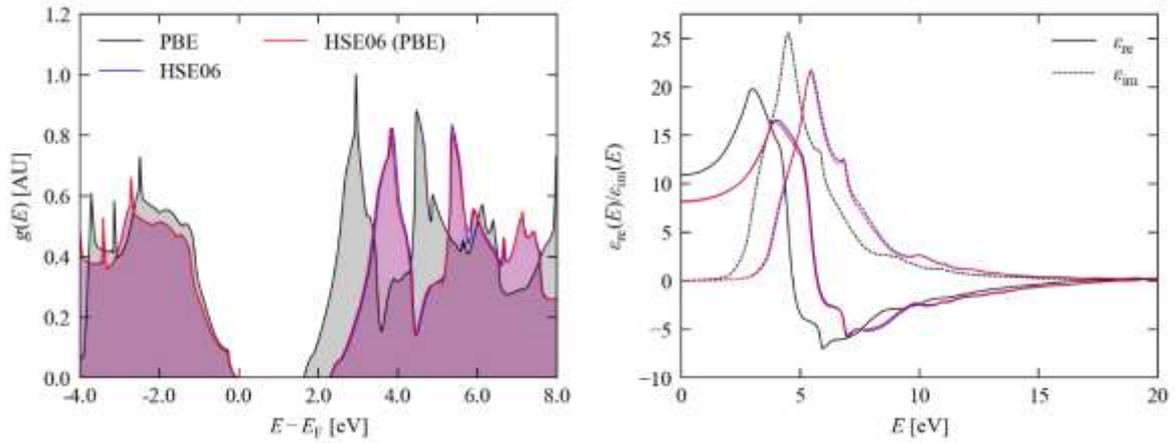

**Figure S4.23** Comparison of the density of states $g(E)$ (a) and real and imaginary energy-dependent dielectric functions $\varepsilon_{\mathrm{re}}(E)/\varepsilon_{\mathrm{im}}(E)$ (b) of GaP calculated using PBE (black), HSE06 (blue) and non-self-consistent HSE06 calculations using the PBE orbitals (red).

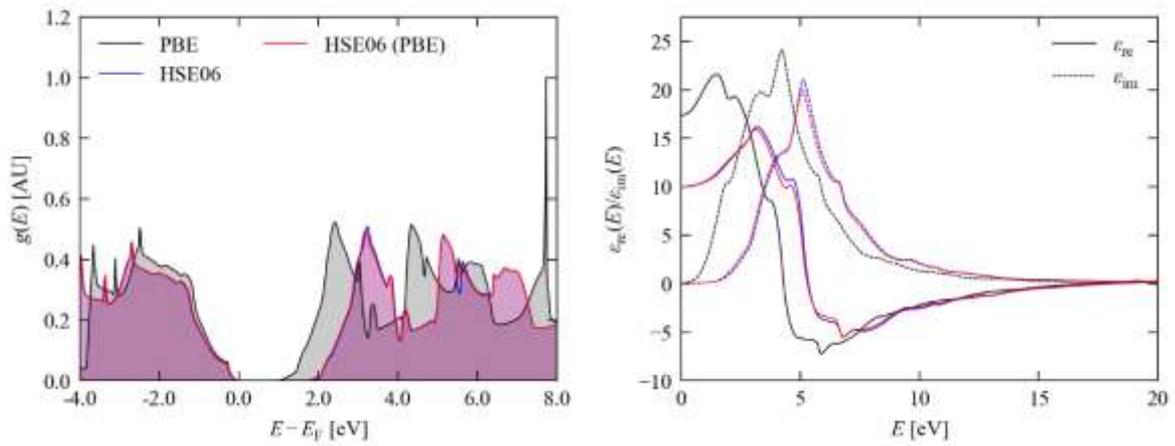

**Figure S4.24** Comparison of the density of states $g(E)$ (a) and real and imaginary energy-dependent dielectric functions $\varepsilon_{\mathrm{re}}(E)/\varepsilon_{\mathrm{im}}(E)$ (b) of GaAs calculated using PBE (black), HSE06 (blue) and non-self-consistent HSE06 calculations using the PBE orbitals (red).

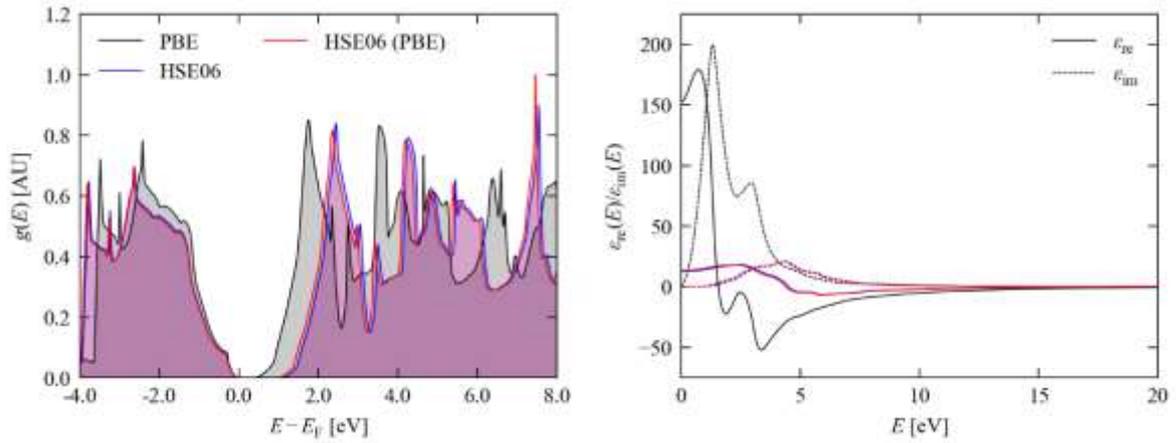

**Figure S4.25** Comparison of the density of states $g(E)$ (a) and real and imaginary energy-dependent dielectric functions $\varepsilon_{\mathrm{re}}(E)/\varepsilon_{\mathrm{im}}(E)$ (b) of GaSb calculated using PBE (black), HSE06 (blue) and non-self-consistent HSE06 calculations using the PBE orbitals (red).

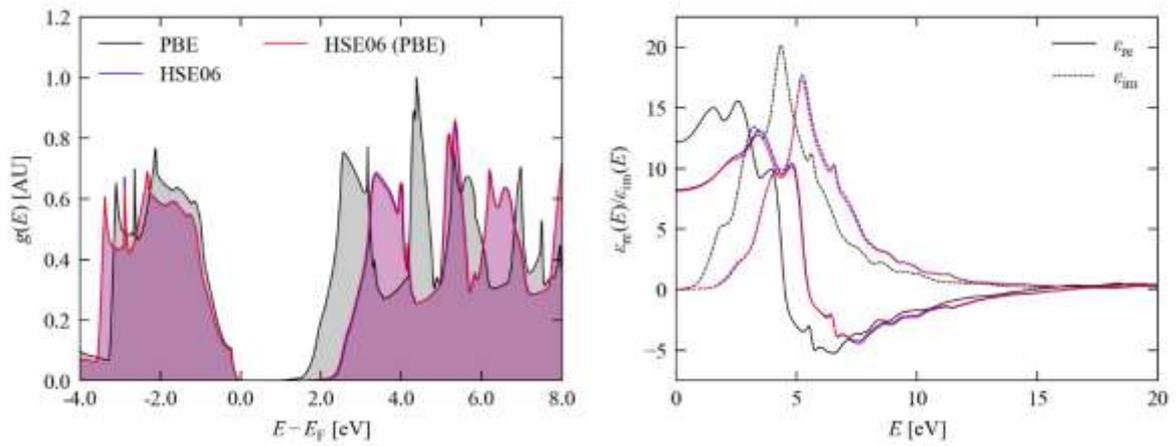

**Figure S4.26** Comparison of the density of states $g(E)$ (a) and real and imaginary energy-dependent dielectric functions $\varepsilon_{\mathrm{re}}(E)/\varepsilon_{\mathrm{im}}(E)$ (b) of InP calculated using PBE (black), HSE06 (blue) and non-self-consistent HSE06 calculations using the PBE orbitals (red).

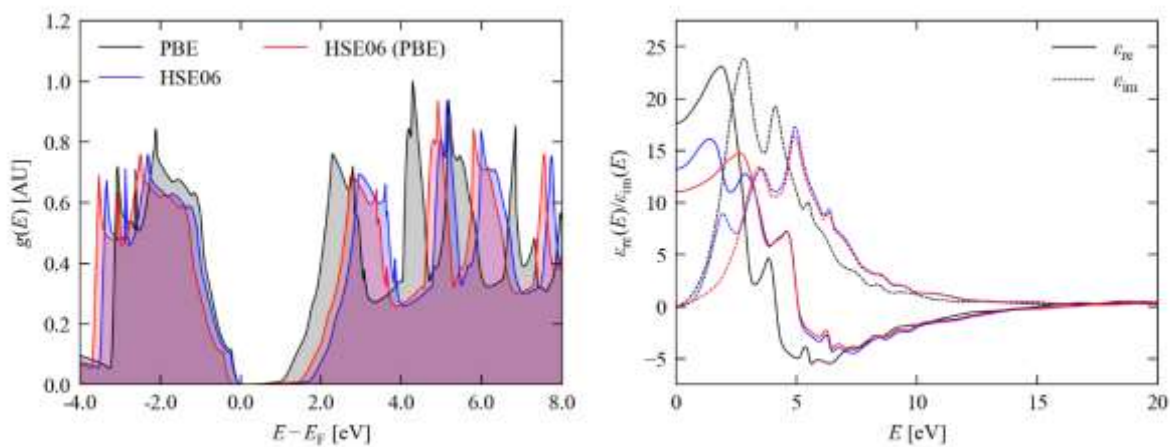

**Figure S4.27** Comparison of the density of states $g(E)$ (a) and real and imaginary energy-dependent dielectric functions $\varepsilon_{\mathrm{re}}(E)/\varepsilon_{\mathrm{im}}(E)$ (b) of InAs calculated using PBE (black), HSE06 (blue) and non-self-consistent HSE06 calculations using the PBE orbitals (red). The PBE functional predicts a metallic electronic structure for this compound, resulting in notably larger errors in the non-self-consistent calculations (see text).

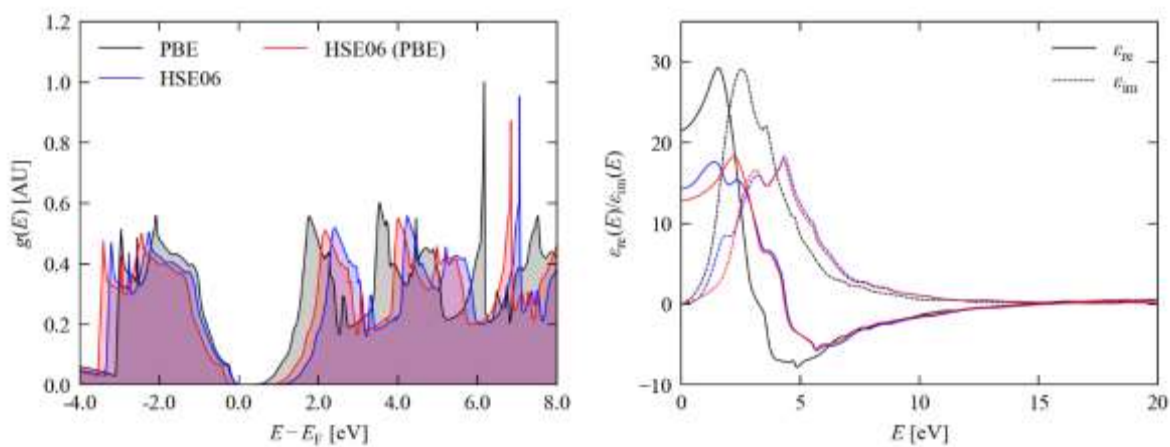

**Figure S4.28** Comparison of the density of states $g(E)$ (a) and real and imaginary energy-dependent dielectric functions $\varepsilon_{\mathrm{re}}(E)/\varepsilon_{\mathrm{im}}(E)$ (b) of InSb calculated using PBE (black), HSE06 (blue) and non-self-consistent HSE06 calculations using the PBE orbitals (red). The PBE functional predicts a metallic electronic structure for this compound, resulting in notably larger errors in the non-self-consistent calculations (see text).

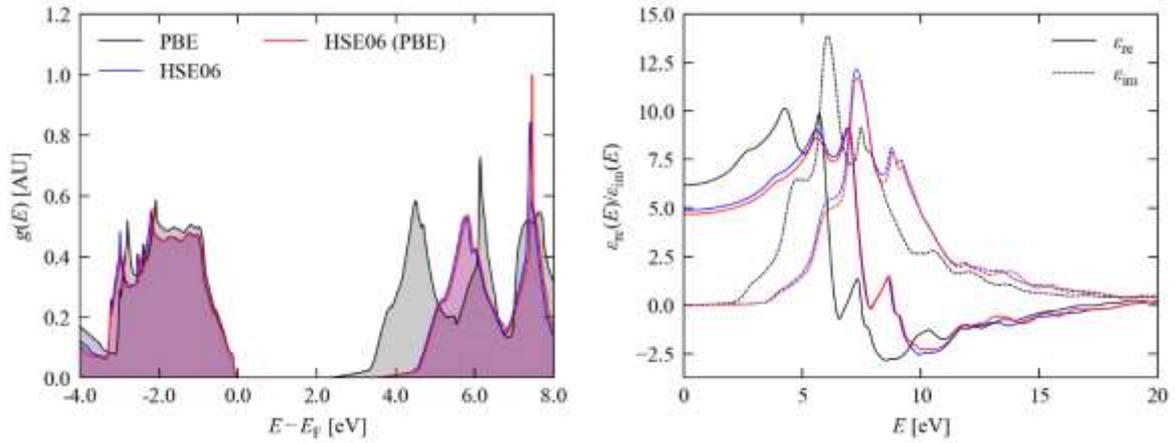

**Figure S4.29** Comparison of the density of states $g(E)$ (a) and real and imaginary energy-dependent dielectric functions $\varepsilon_{re}(E)/\varepsilon_{im}(E)$ (b) of ZnS calculated using PBE (black), HSE06 (blue) and non-self-consistent HSE06 calculations using the PBE orbitals (red).

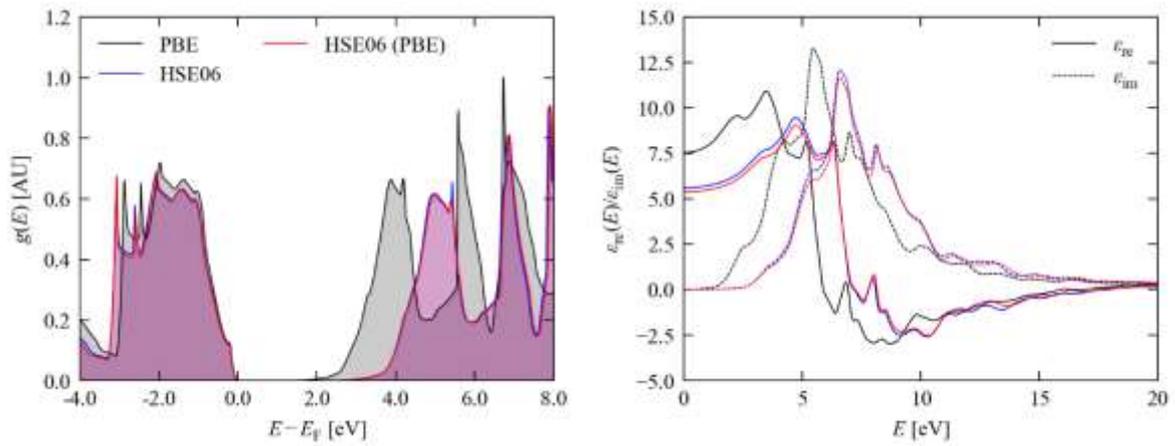

**Figure S4.30** Comparison of the density of states $g(E)$ (a) and real and imaginary energy-dependent dielectric functions $\varepsilon_{re}(E)/\varepsilon_{im}(E)$ (b) of ZnSe calculated using PBE (black), HSE06 (blue) and non-self-consistent HSE06 calculations using the PBE orbitals (red).

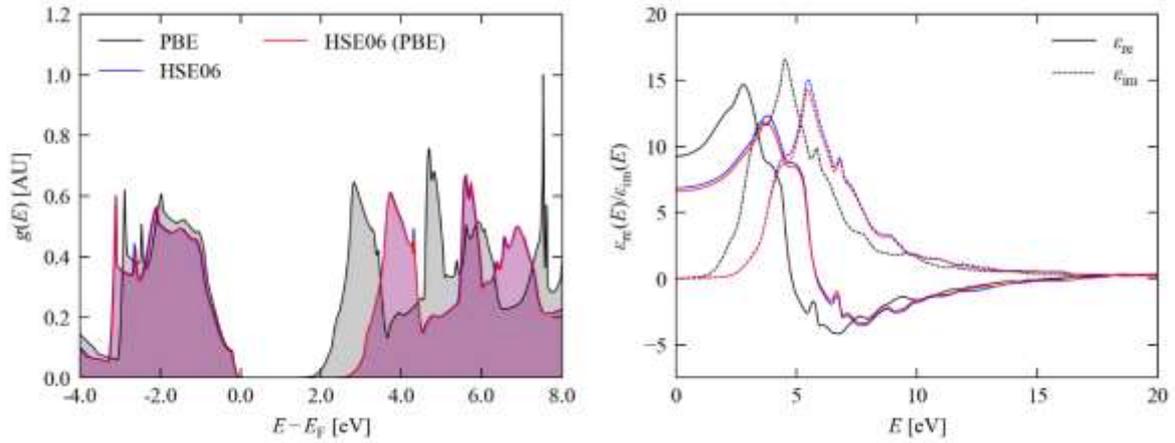

**Figure S4.31** Comparison of the density of states $g(E)$ (a) and real and imaginary energy-dependent dielectric functions $\varepsilon_{\text{re}}(E)/\varepsilon_{\text{im}}(E)$ (b) of ZnTe calculated using PBE (black), HSE06 (blue) and non-self-consistent HSE06 calculations using the PBE orbitals (red).

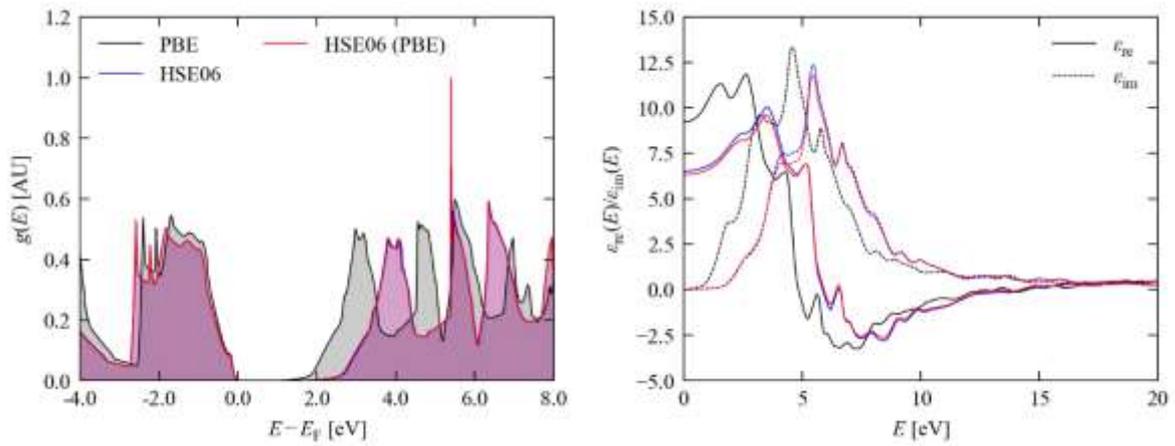

**Figure S4.32** Comparison of the density of states $g(E)$ (a) and real and imaginary energy-dependent dielectric functions $\varepsilon_{\text{re}}(E)/\varepsilon_{\text{im}}(E)$ (b) of CdTe calculated using PBE (black), HSE06 (blue) and non-self-consistent HSE06 calculations using the PBE orbitals (red).

**S5. Total Energies and Equations of State**

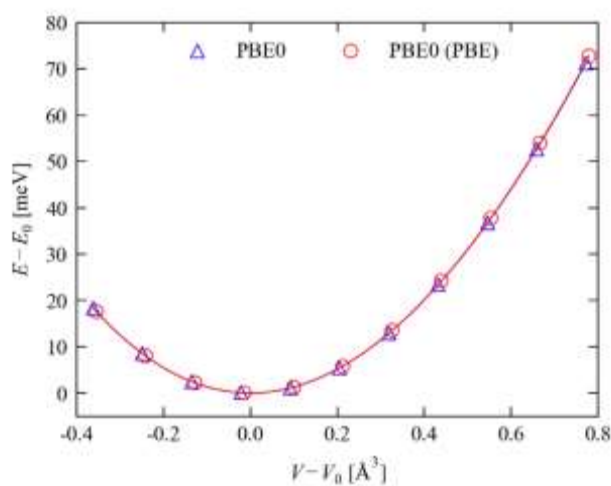

**Figure S5.1** Energy-volume curve for C obtained with PBE0[8] using self-consistent total energies (blue) and non-self-consistent energies calculated using the PBE[1] orbitals (red). The markers show the calculated energies and the solid lines are fits to the Birch-Murnaghan equation of state[2] (Eq. 6 in the text).

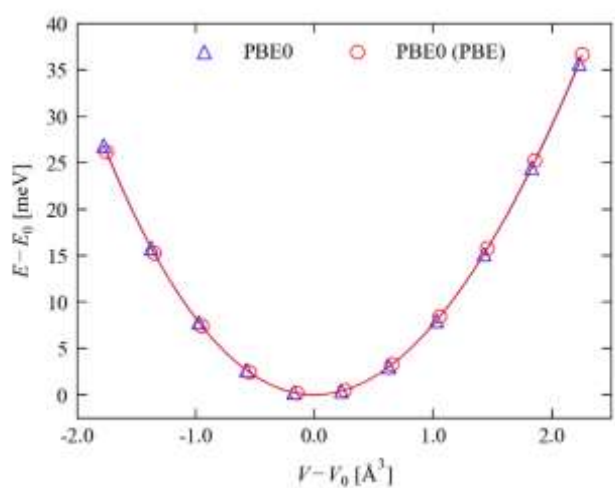

**Figure S5.2** Energy-volume curve for Si obtained with PBE0 using self-consistent total energies (blue) and non-self-consistent energies calculated using the PBE orbitals (red). The markers show the calculated energies and the solid lines are fits to the Birch-Murnaghan equation of state (Eq. 6 in the text).

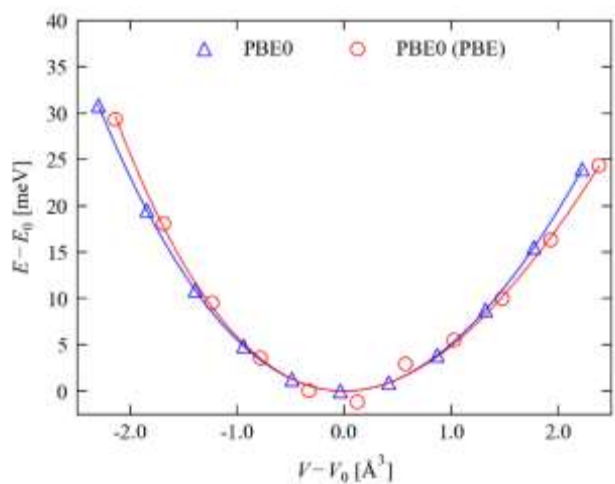

**Figure S5.3** Energy-volume curve for Ge obtained with PBE0 using self-consistent total energies (blue) and non-self-consistent energies calculated using the PBE orbitals (red). The markers show the calculated energies and the solid lines are fits to the Birch-Murnaghan equation of state (Eq. 6 in the text). PBE predicts a metallic electronic structure at some of the volume points, resulting in larger differences between the self-consistent and non-self-consistent results than in other compounds (see text).

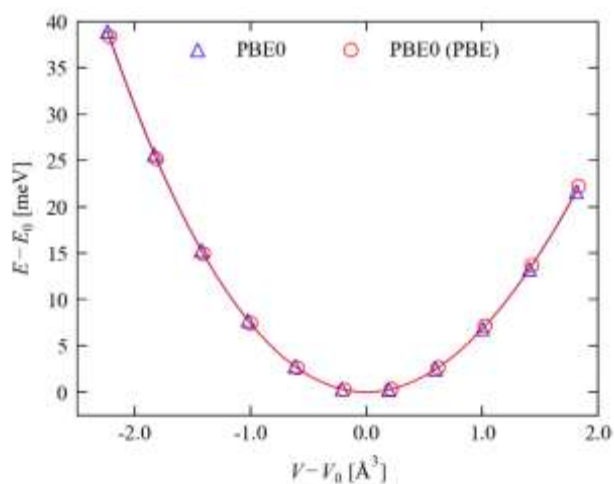

**Figure S5.4** Energy-volume curve for AlP obtained with PBE0 using self-consistent total energies (blue) and non-self-consistent energies calculated using the PBE orbitals (red). The markers show the calculated energies and the solid lines are fits to the Birch-Murnaghan equation of state (Eq. 6 in the text).

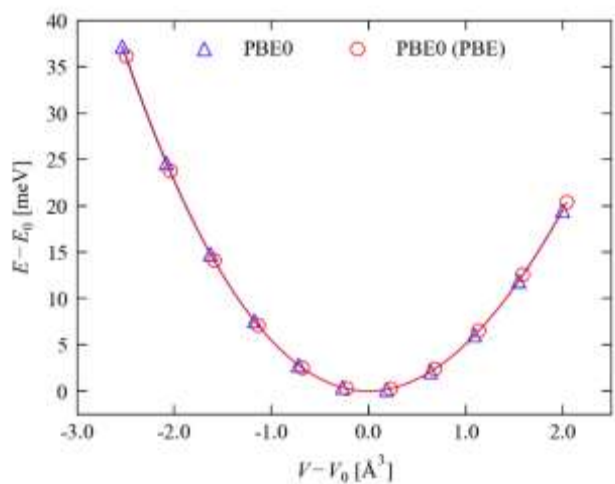

**Figure S5.5** Energy-volume curve for AlAs obtained with PBE0 using self-consistent total energies (blue) and non-self-consistent energies calculated using the PBE orbitals (red). The markers show the calculated energies and the solid lines are fits to the Birch-Murnaghan equation of state (Eq. 6 in the text).

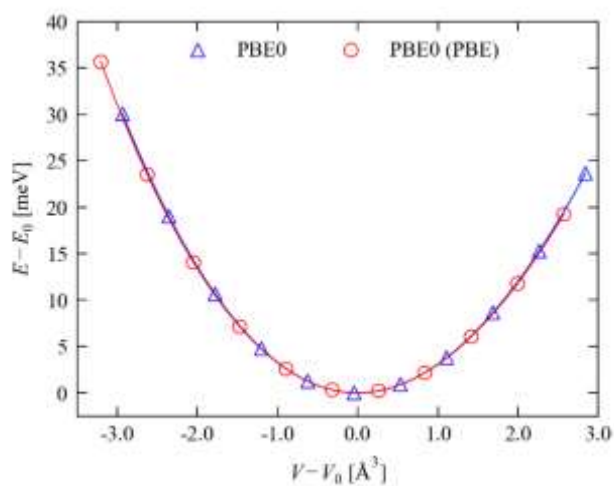

**Figure S5.6** Energy-volume curve for AlSb obtained with PBE0 using self-consistent total energies (blue) and non-self-consistent energies calculated using the PBE orbitals (red). The markers show the calculated energies and the solid lines are fits to the Birch-Murnaghan equation of state (Eq. 6 in the text).

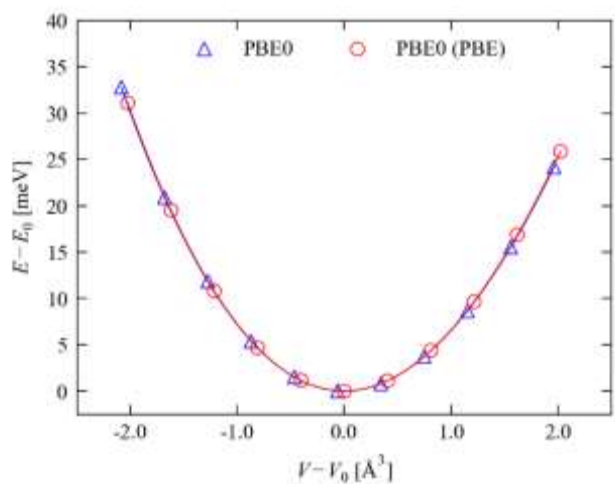

**Figure S5.7** Energy-volume curve for GaP obtained with PBE0 using self-consistent total energies (blue) and non-self-consistent energies calculated using the PBE orbitals (red). The markers show the calculated energies and the solid lines are fits to the Birch-Murnaghan equation of state (Eq. 6 in the text).

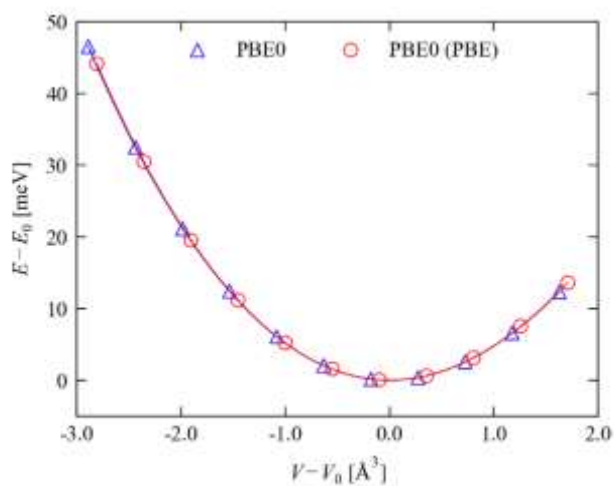

**Figure S5.8** Energy-volume curve for GaAs obtained with PBE0 using self-consistent total energies (blue) and non-self-consistent energies calculated using the PBE orbitals (red). The markers show the calculated energies and the solid lines are fits to the Birch-Murnaghan equation of state (Eq. 6 in the text).

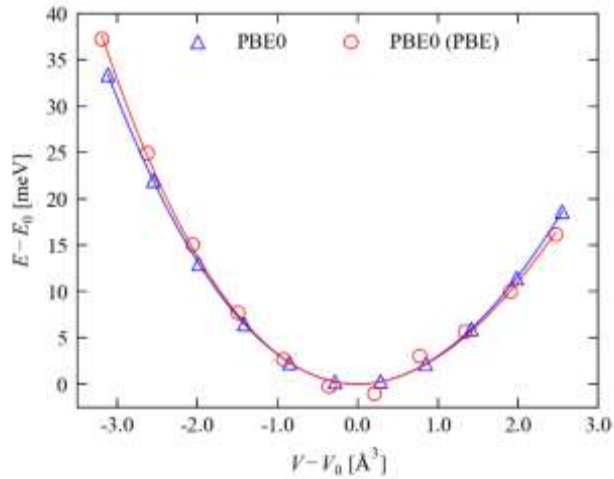

**Figure S5.9** Energy-volume curve for GaSb obtained with PBE0 using self-consistent total energies (blue) and non-self-consistent energies calculated using the PBE orbitals (red). The markers show the calculated energies and the solid lines are fits to the Birch-Murnaghan equation of state (Eq. 6 in the text). PBE predicts a metallic electronic structure at some of the volume points, resulting in larger differences between the self-consistent and non-self-consistent results than in other compounds (see text).

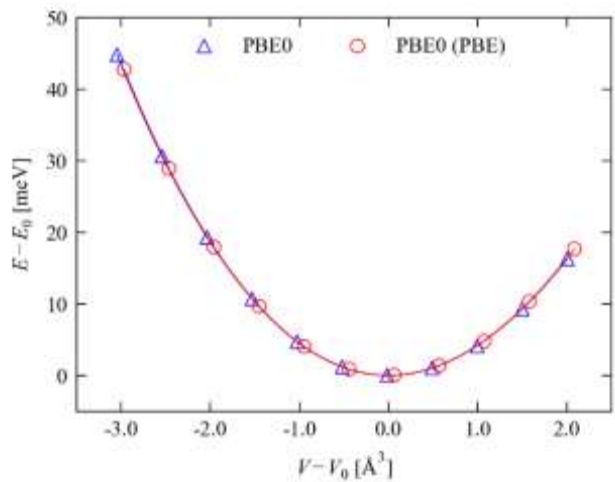

**Figure S5.10** Energy-volume curve for InP obtained with PBE0 using self-consistent total energies (blue) and non-self-consistent energies calculated using the PBE orbitals (red). The markers show the calculated energies and the solid lines are fits to the Birch-Murnaghan equation of state (Eq. 6 in the text).

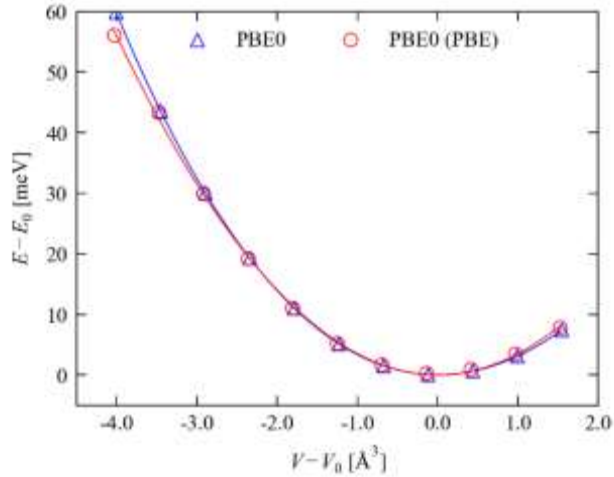

**Figure S5.11** Energy-volume curve for InAs obtained with PBE0 using self-consistent total energies (blue) and non-self-consistent energies calculated using the PBE orbitals (red). The markers show the calculated energies and the solid lines are fits to the Birch-Murnaghan equation of state (Eq. 6 in the text). PBE predicts a metallic electronic structure at some of the volume points, resulting in larger differences between the self-consistent and non-self-consistent results than in other compounds (see text).

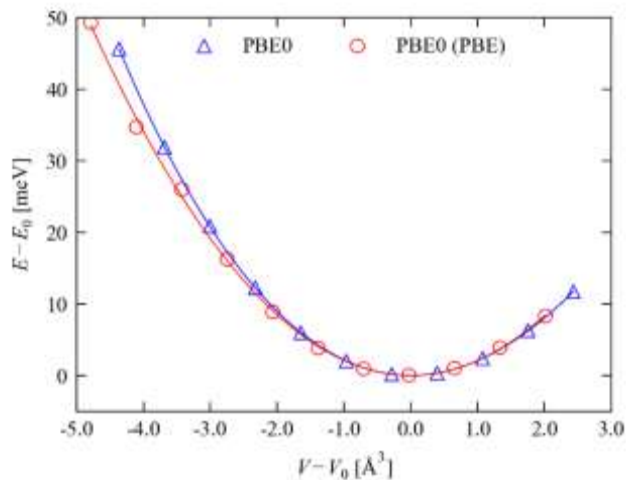

**Figure S5.12** Energy-volume curve for InSb obtained with PBE0 using self-consistent total energies (blue) and non-self-consistent energies calculated using the PBE orbitals (red). The markers show the calculated energies and the solid lines are fits to the Birch-Murnaghan equation of state (Eq. 6 in the text). PBE predicts a metallic electronic structure at some of the volume points, resulting in larger differences between the self-consistent and non-self-consistent results than in other compounds (see text).

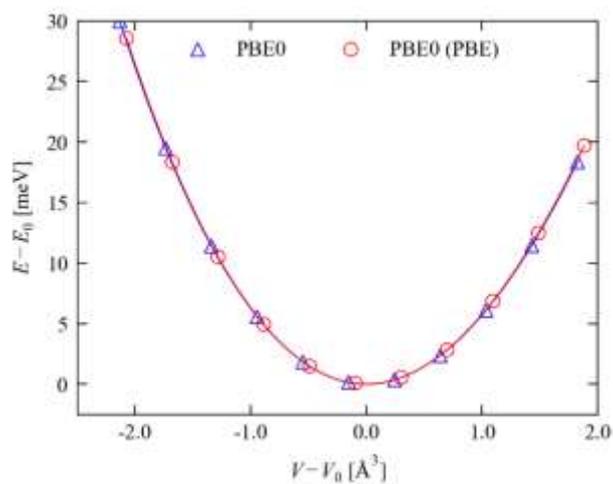

**Figure S5.13** Energy-volume curve for ZnS obtained with PBE0 using self-consistent total energies (blue) and non-self-consistent energies calculated using the PBE orbitals (red). The markers show the calculated energies and the solid lines are fits to the Birch-Murnaghan equation of state (Eq. 6 in the text).

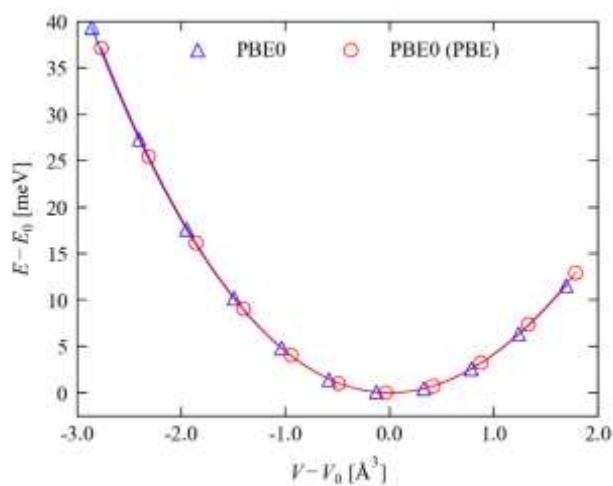

**Figure S5.14** Energy-volume curve for ZnSe obtained with PBE0 using self-consistent total energies (blue) and non-self-consistent energies calculated using the PBE orbitals (red). The markers show the calculated energies and the solid lines are fits to the Birch-Murnaghan equation of state (Eq. 6 in the text).

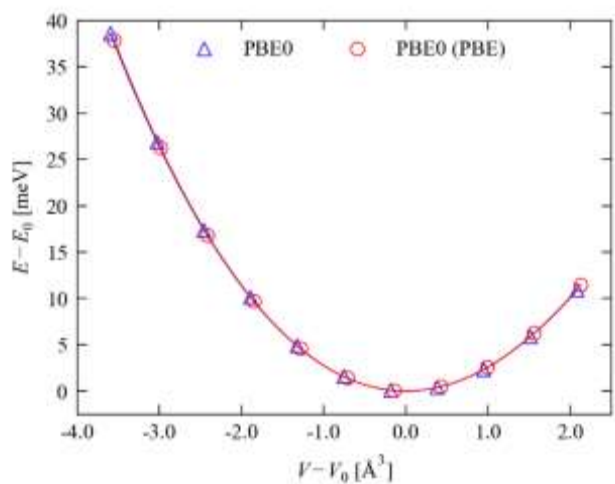

**Figure S5.15** Energy-volume curve for ZnTe obtained with PBE0 using self-consistent total energies (blue) and non-self-consistent energies calculated using the PBE orbitals (red). The markers show the calculated energies and the solid lines are fits to the Birch-Murnaghan equation of state (Eq. 6 in the text).

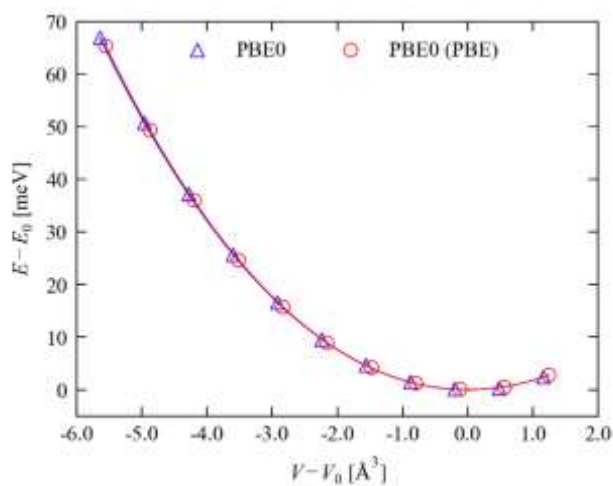

**Figure S5.16** Energy-volume curve for CdTe obtained with PBE0 using self-consistent total energies (blue) and non-self-consistent energies calculated using the PBE orbitals (red). The markers show the calculated energies and the solid lines are fits to the Birch-Murnaghan equation of state (Eq. 6 in the text).

**S6. Dielectric-Dependent Hybrid Functionals: scPBE0**

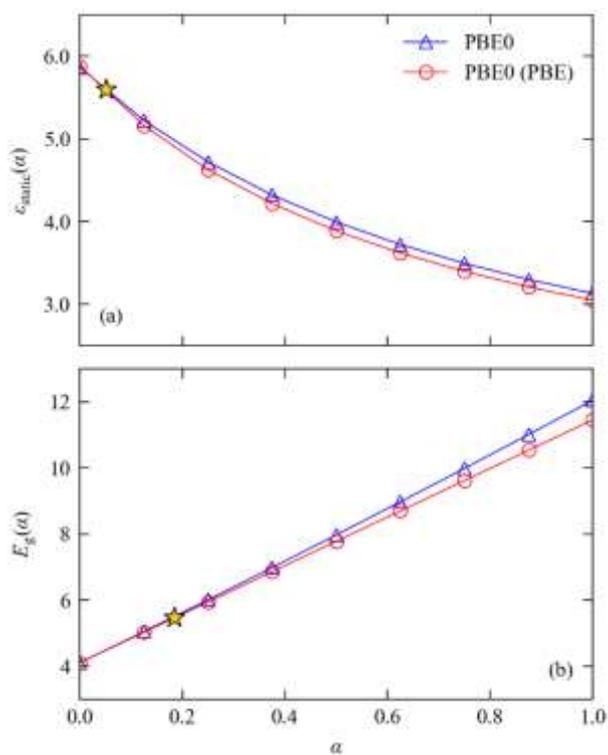

**Figure S6.1** Dependence of the calculated high-frequency dielectric constant $\varepsilon_\infty$ and bandgap $E_g$ of C on the fraction of exact exchange $\alpha$ used in the PBE0 hybrid functional (c.f. Eq. 7 in the text).[8] The self-consistent values and the non-self-consistent values obtained using the PBE[1] orbitals are shown as blue triangles and red circles, respectively, and the experimental values from Table 1 in the text are overlaid as gold stars.

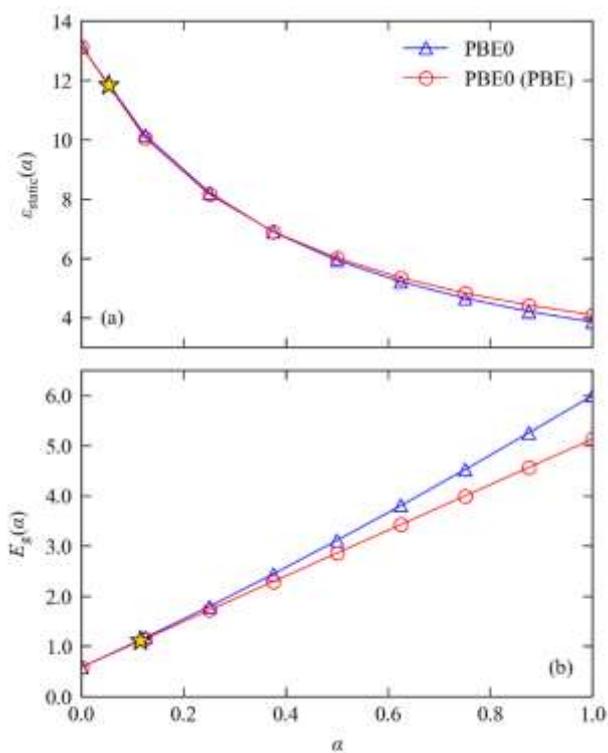

**Figure S6.2** Dependence of the calculated high-frequency dielectric constant $\varepsilon_\infty$ and bandgap $E_g$ of Si on the fraction of exact exchange $\alpha$ used in the PBE0 hybrid functional (c.f. Eq. 7 in the text). The self-consistent values and the non-self-consistent values obtained using the PBE orbitals are shown as blue triangles and red circles, respectively, and the experimental values from Table 1 in the text are overlaid as gold stars.

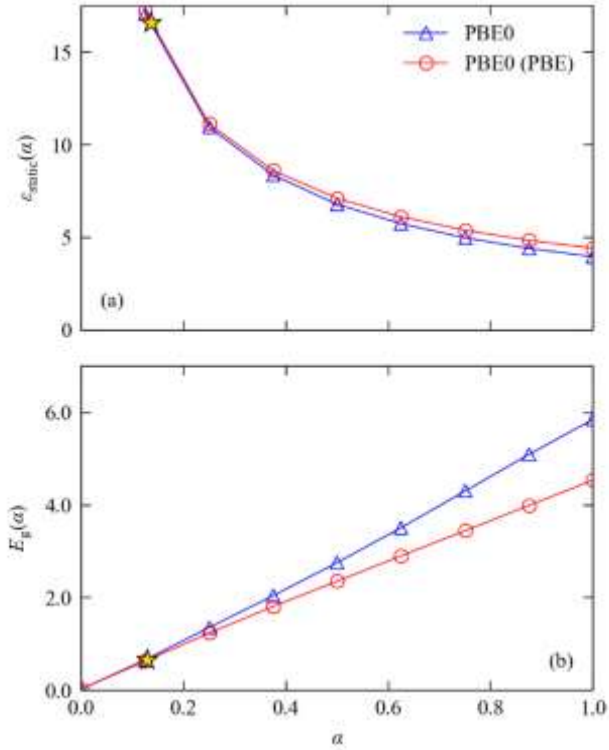

**Figure S6.3** Dependence of the calculated high-frequency dielectric constant $\varepsilon_\infty$ and bandgap $E_g$ of Ge on the fraction of exact exchange $\alpha$ used in the PBE0 hybrid functional (c.f. Eq. 7 in the text). The self-consistent values and the non-self-consistent values obtained using the PBE orbitals are shown as blue triangles and red circles, respectively, and the experimental values from Table 1 in the text are overlaid as gold stars.

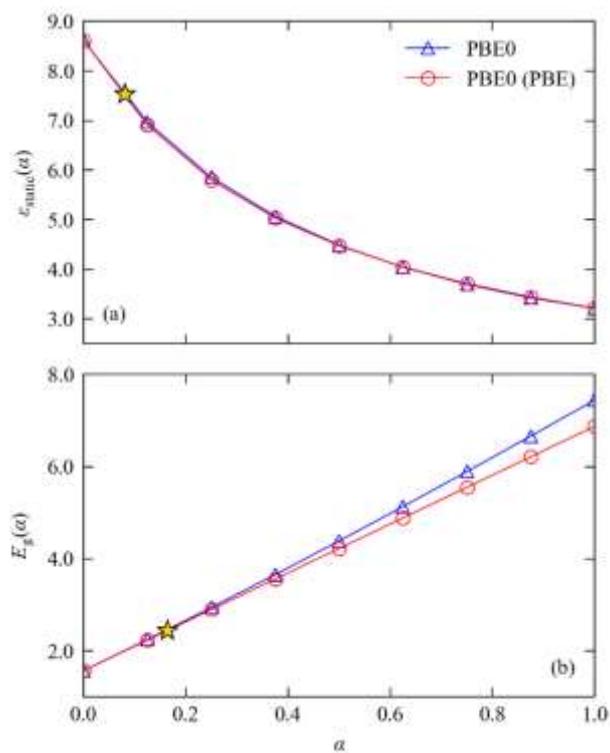

**Figure S6.4** Dependence of the calculated high-frequency dielectric constant $\varepsilon_\infty$ and bandgap $E_g$ of AlP on the fraction of exact exchange $\alpha$ used in the PBE0 hybrid functional (c.f. Eq. 7 in the text). The self-consistent values and the non-self-consistent values obtained using the PBE orbitals are shown as blue triangles and red circles, respectively, and the experimental values from Table 1 in the text are overlaid as gold stars.

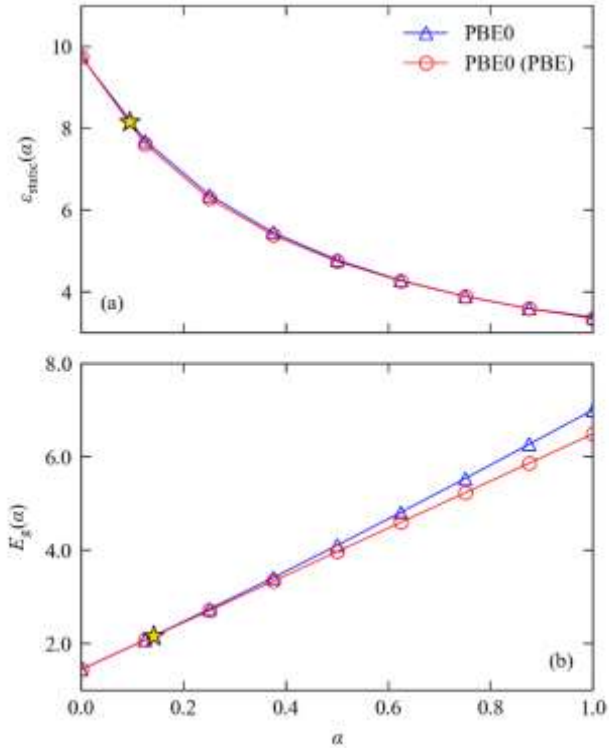

**Figure S6.5** Dependence of the calculated high-frequency dielectric constant $\varepsilon_\infty$ and bandgap $E_g$ of AlAs on the fraction of exact exchange $\alpha$ used in the PBE0 hybrid functional (c.f. Eq. 7 in the text). The self-consistent values and the non-self-consistent values obtained using the PBE orbitals are shown as blue triangles and red circles, respectively, and the experimental values from Table 1 in the text are overlaid as gold stars.

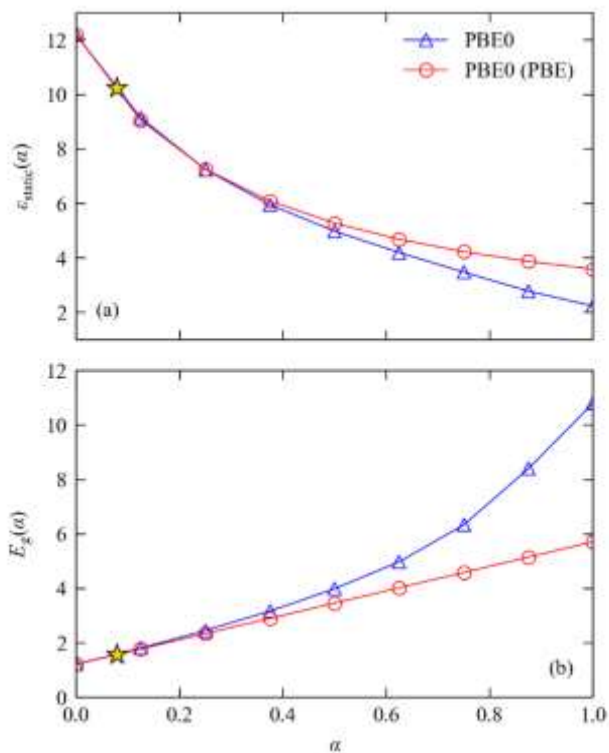

**Figure S6.6** Dependence of the calculated high-frequency dielectric constant $\varepsilon_\infty$ and bandgap $E_g$ of AlSb on the fraction of exact exchange $\alpha$ used in the PBE0 hybrid functional (c.f. Eq. 7 in the text). The self-consistent values and the non-self-consistent values obtained using the PBE orbitals are shown as blue triangles and red circles, respectively, and the experimental values from Table 1 in the text are overlaid as gold stars.

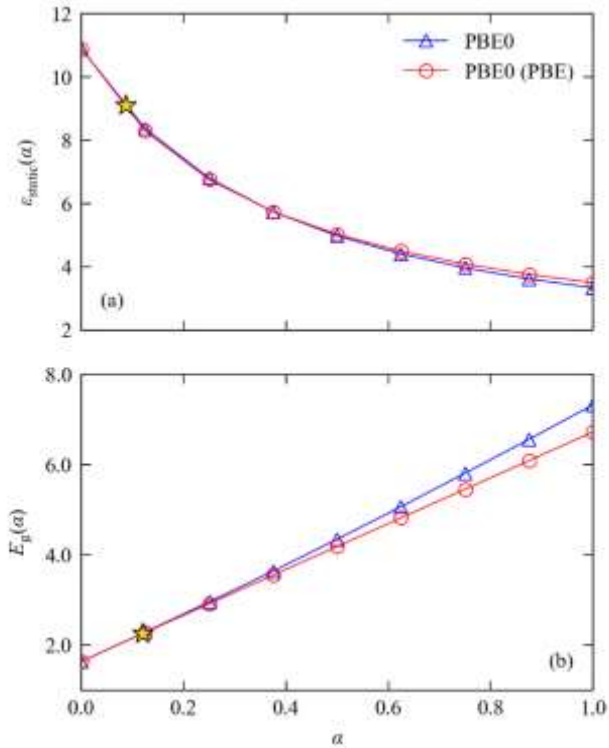

**Figure S6.7** Dependence of the calculated high-frequency dielectric constant $\varepsilon_\infty$ and bandgap $E_\mathrm{g}$ of GaP on the fraction of exact exchange $\alpha$ used in the PBE0 hybrid functional (c.f. Eq. 7 in the text). The self-consistent values and the non-self-consistent values obtained using the PBE orbitals are shown as blue triangles and red circles, respectively, and the experimental values from Table 1 in the text are overlaid as gold stars.

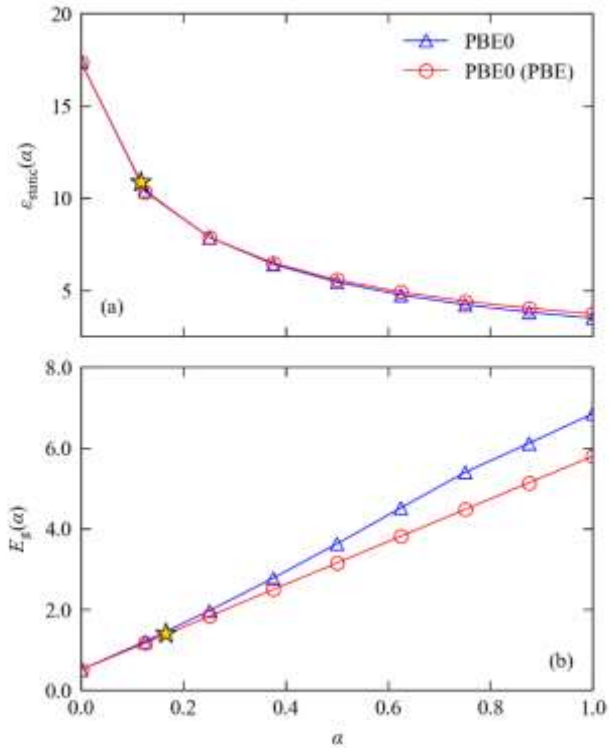

**Figure S6.8** Dependence of the calculated high-frequency dielectric constant $\varepsilon_\infty$ and bandgap $E_g$ of GaAs on the fraction of exact exchange $\alpha$ used in the PBE0 hybrid functional (c.f. Eq. 7 in the text). The self-consistent values and the non-self-consistent values obtained using the PBE orbitals are shown as blue triangles and red circles, respectively, and the experimental values from Table 1 in the text are overlaid as gold stars.

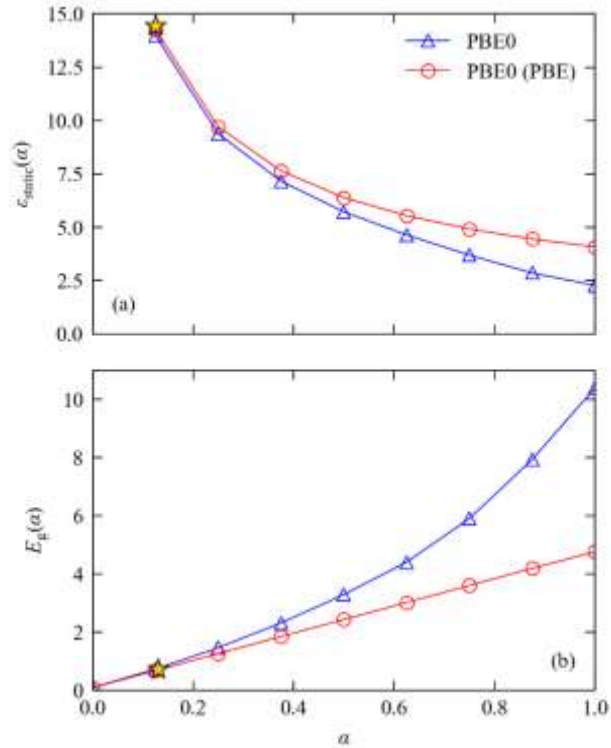

**Figure S6.9** Dependence of the calculated high-frequency dielectric constant $\varepsilon_\infty$ and bandgap $E_g$ of GaSb on the fraction of exact exchange $\alpha$ used in the PBE0 hybrid functional (c.f. Eq. 7 in the text). The self-consistent values and the non-self-consistent values obtained using the PBE orbitals are shown as blue triangles and red circles, respectively, and the experimental values from Table 1 in the text are overlaid as gold stars.

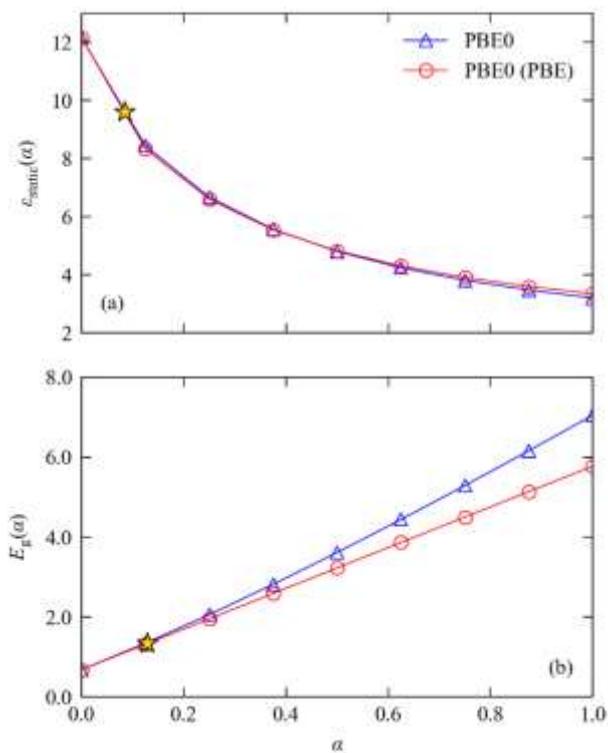

**Figure S6.10** Dependence of the calculated high-frequency dielectric constant $\varepsilon_\infty$ and bandgap $E_g$ of InP on the fraction of exact exchange $\alpha$ used in the PBE0 hybrid functional (c.f. Eq. 7 in the text). The self-consistent values and the non-self-consistent values obtained using the PBE orbitals are shown as blue triangles and red circles, respectively, and the experimental values from Table 1 in the text are overlaid as gold stars.

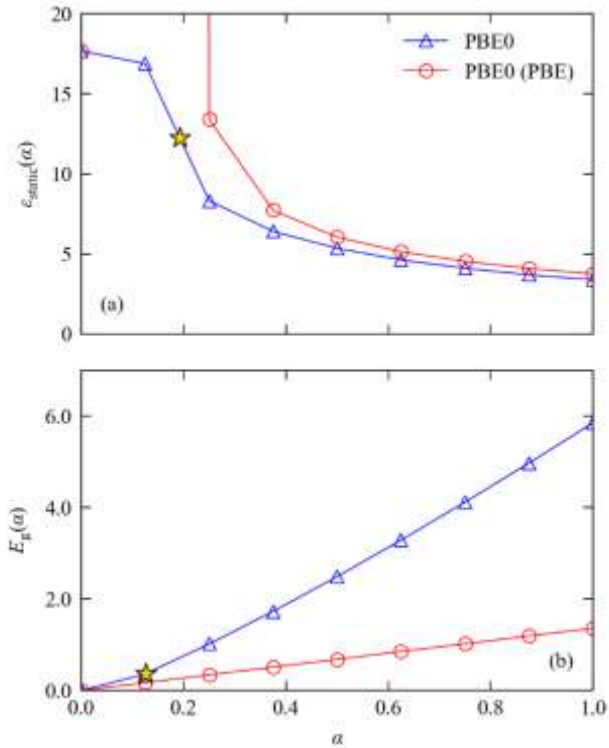

**Figure S6.11** Dependence of the calculated high-frequency dielectric constant $\varepsilon_\infty$ and bandgap $E_g$ of InAs on the fraction of exact exchange $\alpha$ used in the PBE0 hybrid functional (c.f. Eq. 7 in the text). The self-consistent values and the non-self-consistent values obtained using the PBE orbitals are shown as blue triangles and red circles, respectively, and the experimental values from Table 1 in the text are overlaid as gold stars. Note that PBE predicts a metallic electronic structure for this system, resulting in anomalously large errors in the non-self-consistent calculations (see text).

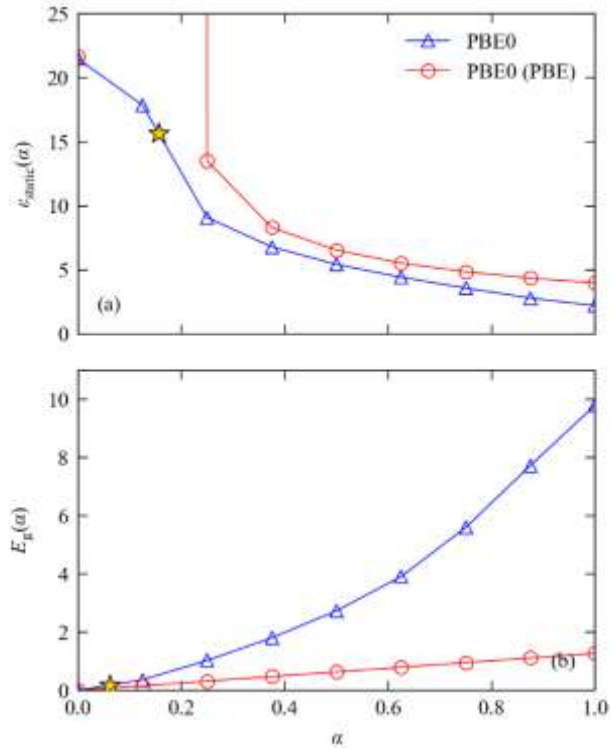

**Figure S6.12** Dependence of the calculated high-frequency dielectric constant $\varepsilon_\infty$ and bandgap $E_g$ of InSb on the fraction of exact exchange $\alpha$ used in the PBE0 hybrid functional (c.f. Eq. 7 in the text). The self-consistent values and the non-self-consistent values obtained using the PBE orbitals are shown as blue triangles and red circles, respectively, and the experimental values from Table 1 in the text are overlaid as gold stars. Note that PBE predicts a metallic electronic structure for this system, resulting in anomalously large errors in the non-self-consistent calculations (see text).

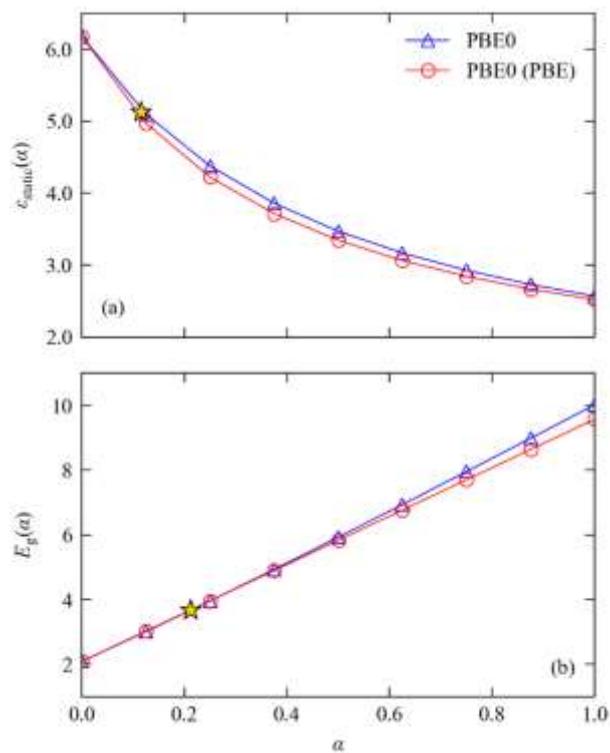

**Figure S6.13** Dependence of the calculated high-frequency dielectric constant $\varepsilon_\infty$ and bandgap $E_g$ of ZnS on the fraction of exact exchange $\alpha$ used in the PBE0 hybrid functional (c.f. Eq. 7 in the text). The self-consistent values and the non-self-consistent values obtained using the PBE orbitals are shown as blue triangles and red circles, respectively, and the experimental values from Table 1 in the text are overlaid as gold stars.

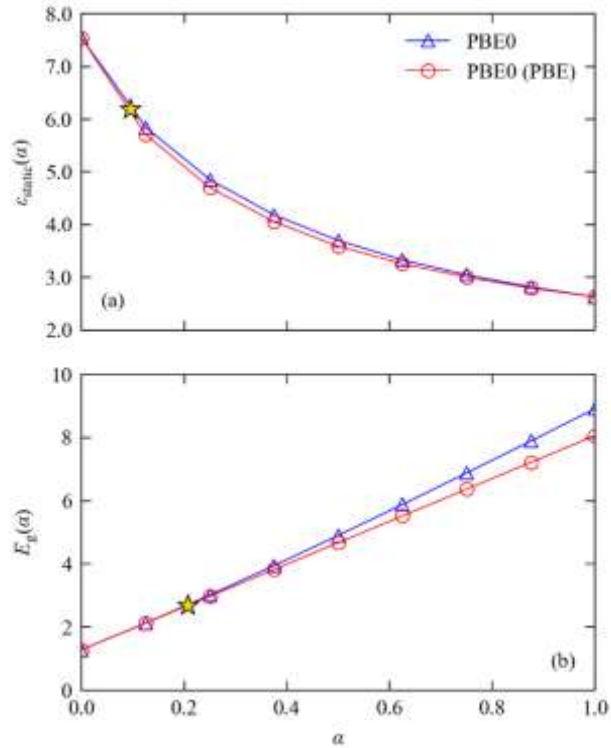

**Figure S6.14** Dependence of the calculated high-frequency dielectric constant $\varepsilon_\infty$ and bandgap $E_g$ of ZnSe on the fraction of exact exchange $\alpha$ used in the PBE0 hybrid functional (c.f. Eq. 7 in the text). The self-consistent values and the non-self-consistent values obtained using the PBE orbitals are shown as blue triangles and red circles, respectively, and the experimental values from Table 1 in the text are overlaid as gold stars.

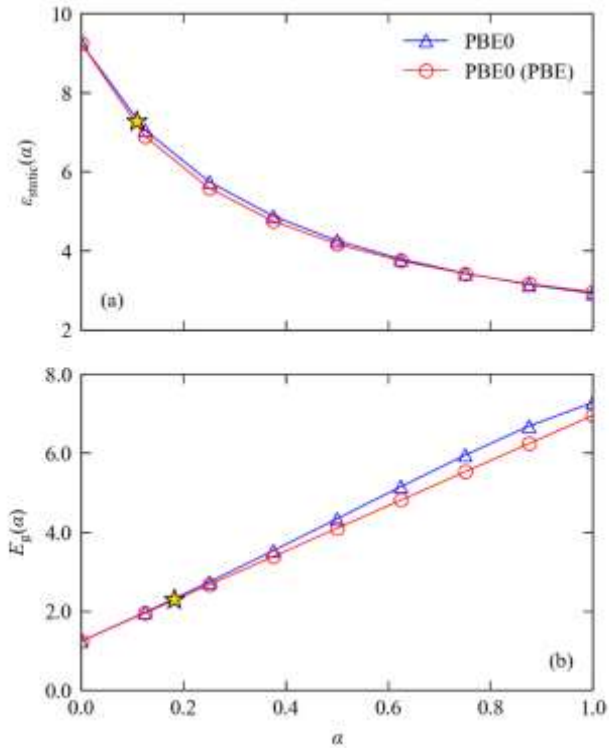

**Figure S6.15** Dependence of the calculated high-frequency dielectric constant $\varepsilon_\infty$ and bandgap $E_\mathrm{g}$ of ZnTe on the fraction of exact exchange $\alpha$ used in the PBE0 hybrid functional (c.f. Eq. 7 in the text). The self-consistent values and the non-self-consistent values obtained using the PBE orbitals are shown as blue triangles and red circles, respectively, and the experimental values from Table 1 in the text are overlaid as gold stars.

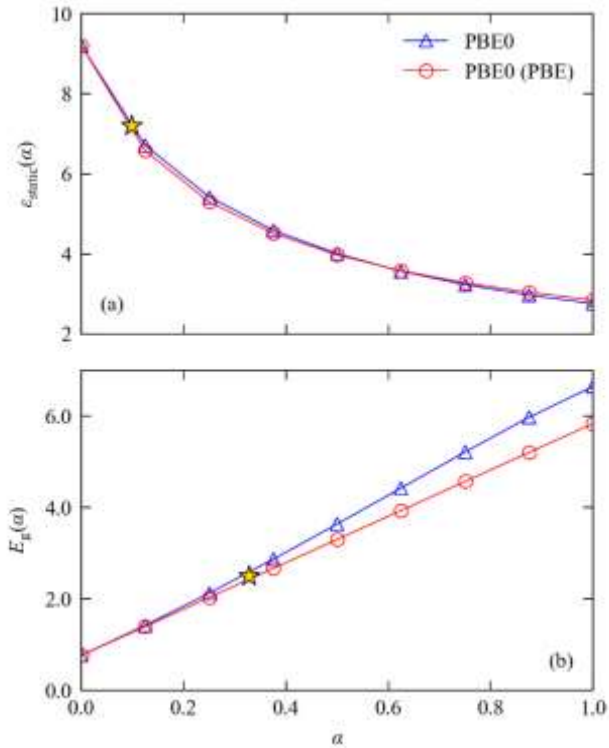

**Figure S6.16** Dependence of the calculated high-frequency dielectric constant $\varepsilon_\infty$ and bandgap $E_\mathrm{g}$ of CdTe on the fraction of exact exchange $\alpha$ used in the PBE0 hybrid functional (c.f. Eq. 7 in the text). The self-consistent values and the non-self-consistent values obtained using the PBE orbitals are shown as blue triangles and red circles, respectively, and the experimental values from Table 1 in the text are overlaid as gold stars.

**S7. Partial Self-Consistency**

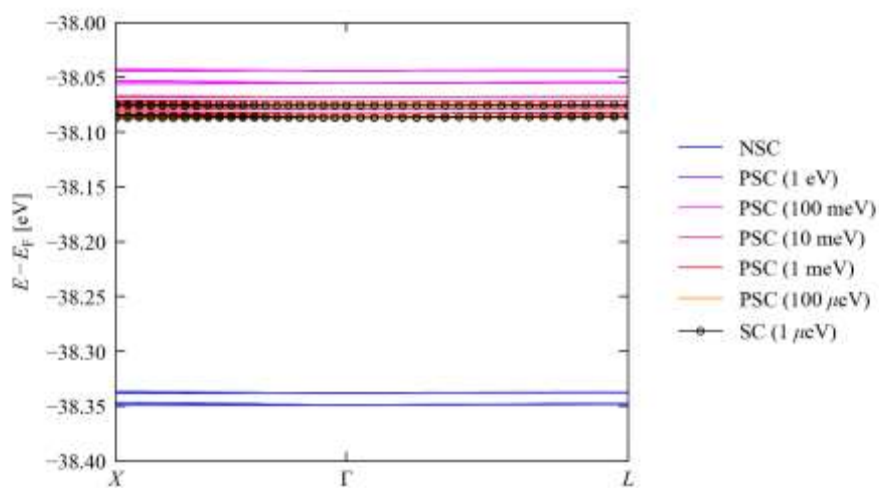

**Figure S7.1** Change in the dispersion of the semi-core As d bands in AlAs calculated using non-self-consistent and partially-self-consistent HSE06 starting from the PBE orbitals with SCF convergence thresholds of 1 eV to 100 $\mu$eV ($10^0$ - $10^{-4}$ eV). The black lines with markers show the reference dispersion obtained from fully self-consistent calculations performed to a tolerance of 1 $\mu$eV ($10^{-6}$ eV).

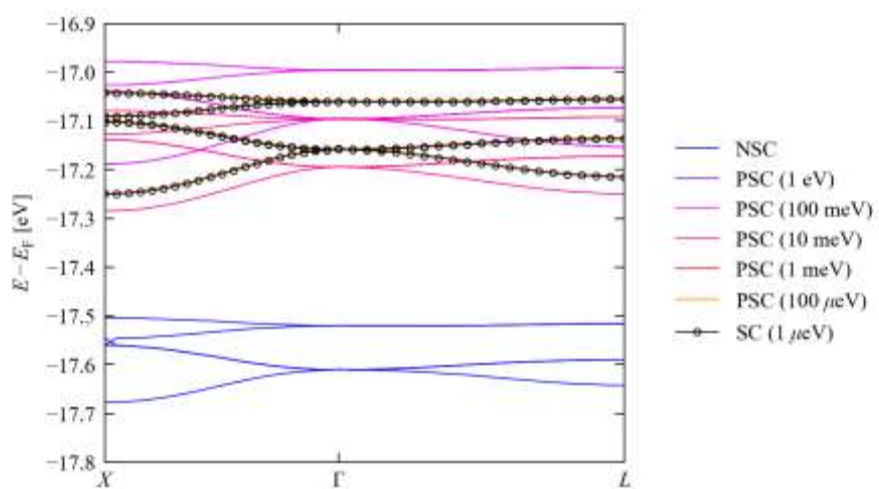

**Figure S7.2** Change in the dispersion of the semi-core Ga d bands in GaP calculated using non-self-consistent and partially-self-consistent HSE06 starting from the PBE orbitals with SCF convergence thresholds of 1 eV to 100 $\mu$eV ($10^0$ - $10^{-4}$ eV). The black lines with markers show the reference dispersion obtained from fully self-consistent calculations performed to a tolerance of 1 $\mu$eV ($10^{-6}$ eV).

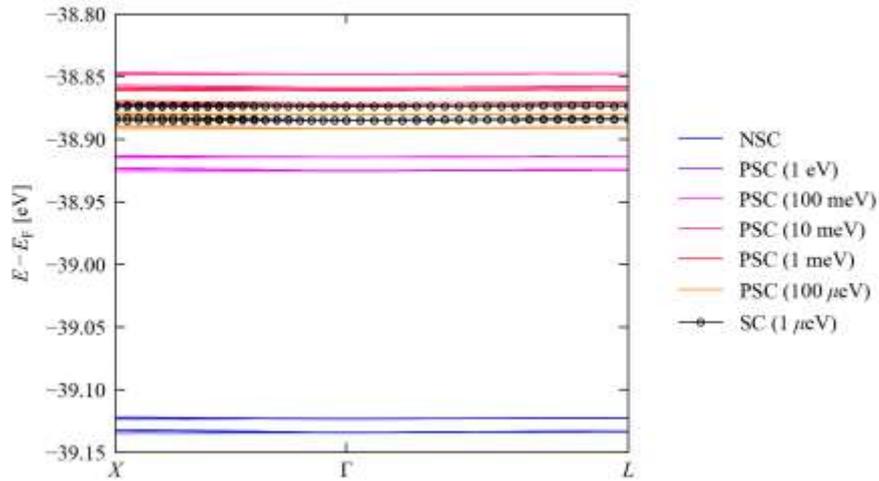

**Figure S7.3** Change in the dispersion of the semi-core As d bands in GaAs calculated using non-self-consistent and partially-self-consistent HSE06 starting from the PBE orbitals with SCF convergence thresholds of 1 eV to 100 $\mu$eV ($10^0$ - $10^{-4}$ eV). The black lines with markers show the reference dispersion obtained from fully self-consistent calculations performed to a tolerance of 1 $\mu$eV ($10^{-6}$ eV).

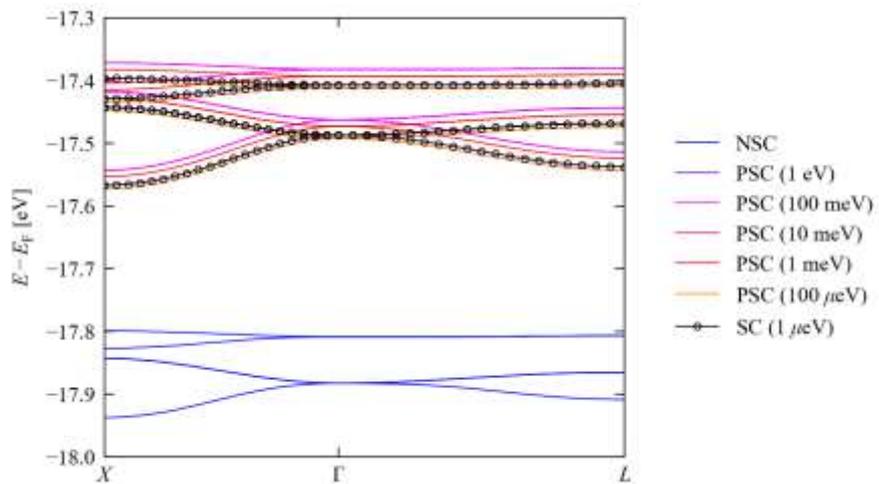

**Figure S7.4** Change in the dispersion of the semi-core Ga d bands in GaAs calculated using non-self-consistent and partially-self-consistent HSE06 starting from the PBE orbitals with SCF convergence thresholds of 1 eV to 100 $\mu$eV ($10^0$ - $10^{-4}$ eV). The black lines with markers show the reference dispersion obtained from fully self-consistent calculations performed to a tolerance of 1 $\mu$eV ($10^{-6}$ eV).

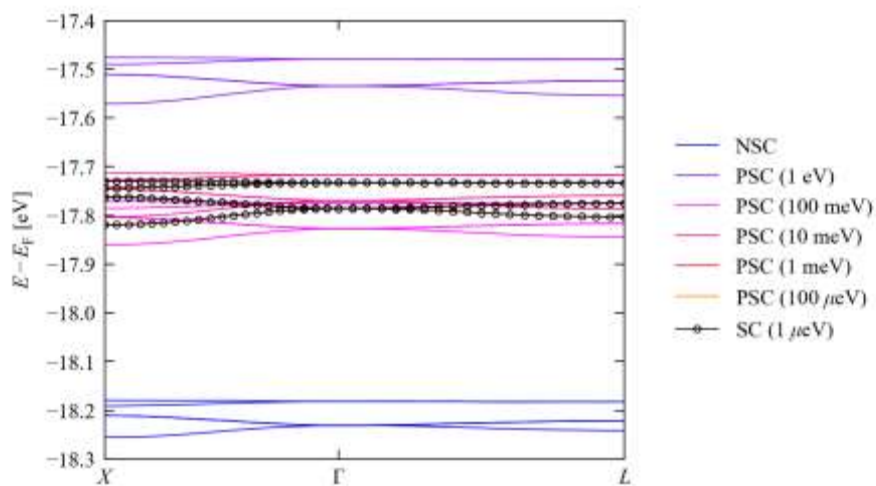

**Figure S7.5** Change in the dispersion of the semi-core Ga d bands in GaSb calculated using non-self-consistent and partially-self-consistent HSE06 starting from the PBE orbitals with SCF convergence thresholds of 1 eV to 100 $\mu$eV ($10^0$ - $10^{-4}$ eV). The black lines with markers show the reference dispersion obtained from fully self-consistent calculations performed to a tolerance of 1 $\mu$eV ($10^{-6}$ eV).

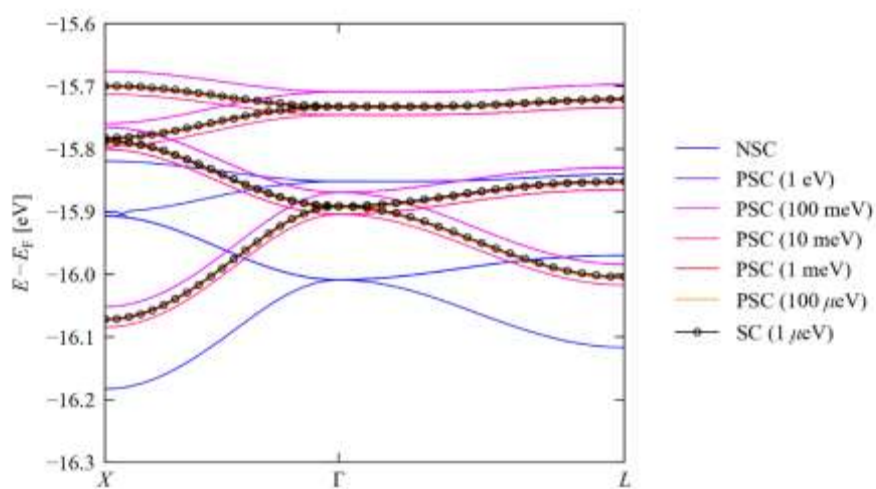

**Figure S7.6** Change in the dispersion of the semi-core In d bands in InP calculated using non-self-consistent and partially-self-consistent HSE06 starting from the PBE orbitals with SCF convergence thresholds of 1 eV to 100 $\mu$eV ($10^0$ - $10^{-4}$ eV). The black lines with markers show the reference dispersion obtained from fully self-consistent calculations performed to a tolerance of 1 $\mu$eV ($10^{-6}$ eV).

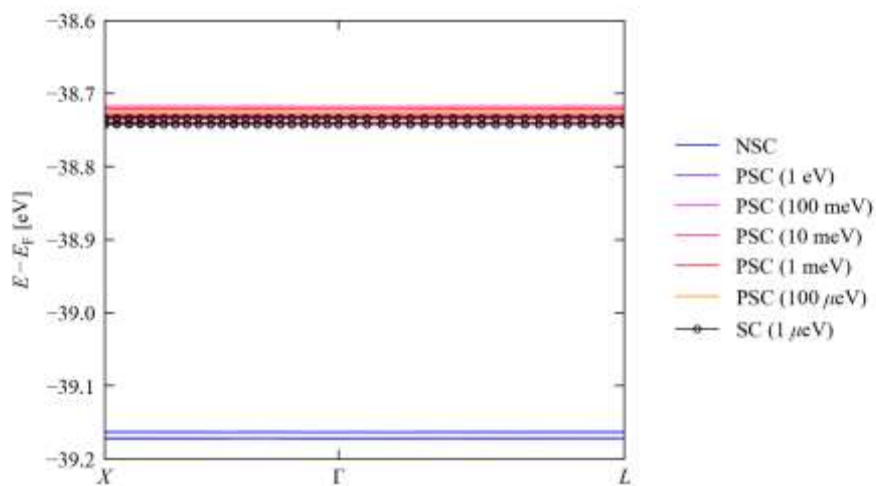

**Figure S7.7** Change in the dispersion of the semi-core As d bands in InAs calculated using non-self-consistent and partially-self-consistent HSE06 starting from the PBE orbitals with SCF convergence thresholds of 1 eV to 100 $\mu$eV ($10^0$ - $10^{-4}$ eV). The black lines with markers show the reference dispersion obtained from fully self-consistent calculations performed to a tolerance of 1 $\mu$eV ($10^{-6}$ eV).

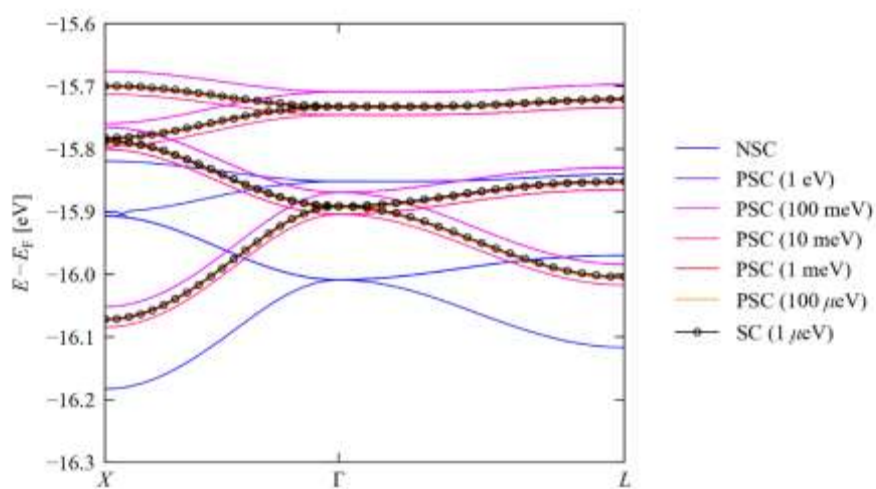

**Figure S7.8** Change in the dispersion of the semi-core In d bands in InAs calculated using non-self-consistent and partially-self-consistent HSE06 starting from the PBE orbitals with SCF convergence thresholds of 1 eV to 100 $\mu$eV ($10^0$ - $10^{-4}$ eV). The black lines with markers show the reference dispersion obtained from fully self-consistent calculations performed to a tolerance of 1 $\mu$eV ($10^{-6}$ eV).

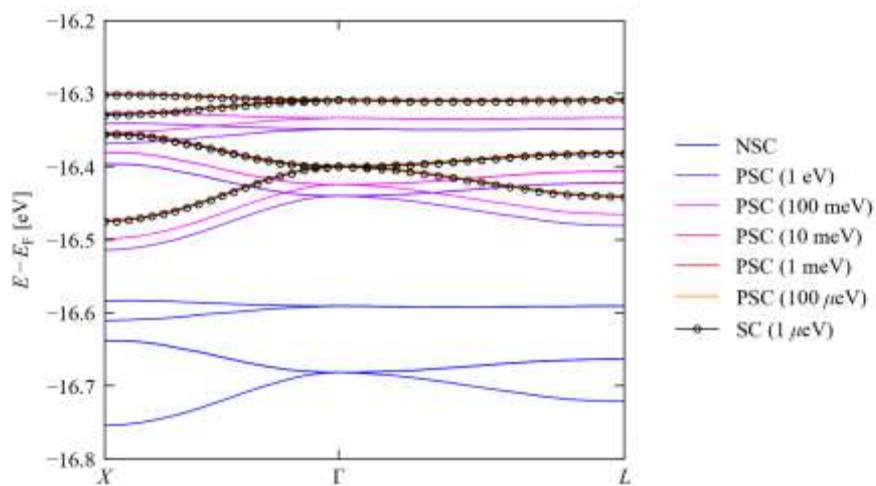

**Figure S7.9** Change in the dispersion of the semi-core In d bands in InSb calculated using non-self-consistent and partially-self-consistent HSE06 starting from the PBE orbitals with SCF convergence thresholds of 1 eV to 100 $\mu$eV ($10^0$ - $10^{-4}$ eV). The black lines with markers show the reference dispersion obtained from fully self-consistent calculations performed to a tolerance of 1 $\mu$eV ($10^{-6}$ eV).

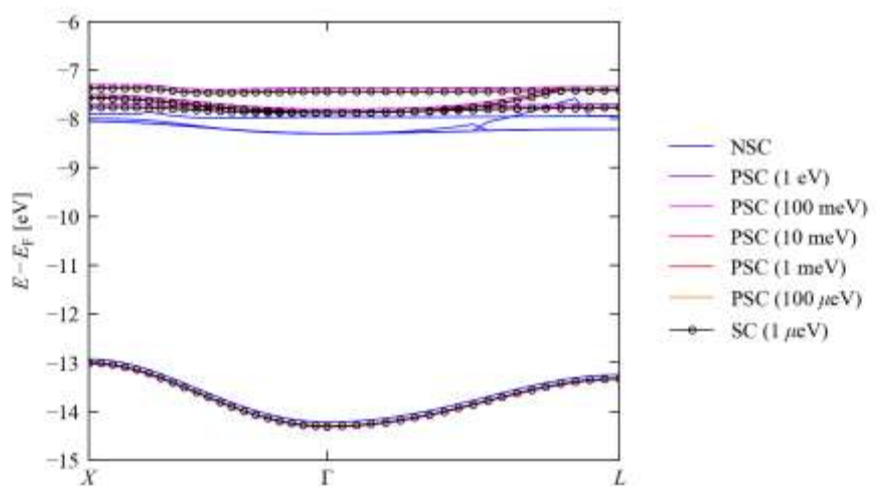

**Figure S7.10** Change in the dispersion of the Zn d bands in ZnS calculated using non-self-consistent and partially-self-consistent HSE06 starting from the PBE orbitals with SCF convergence thresholds of 1 eV to 100 $\mu$eV ($10^0$ - $10^{-4}$ eV). The black lines with markers show the reference dispersion obtained from fully self-consistent calculations performed to a tolerance of 1 $\mu$eV ($10^{-6}$ eV).

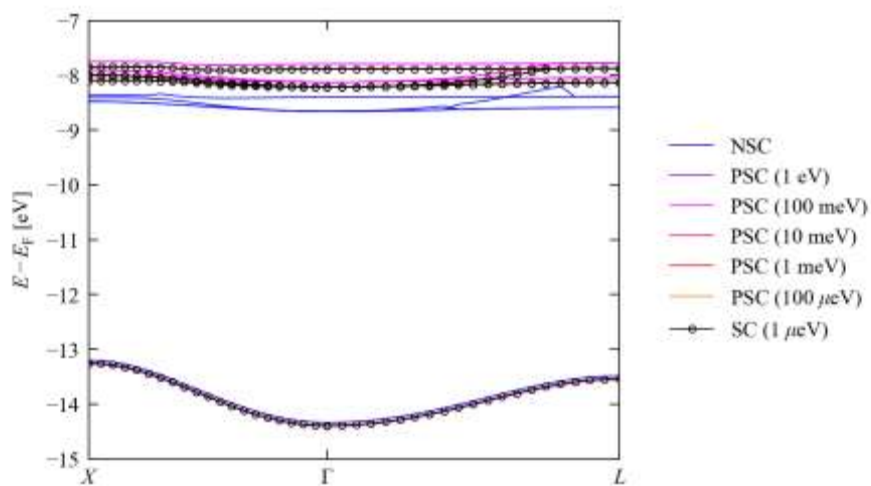

**Figure S7.11** Change in the dispersion of the Zn d bands in ZnSe calculated using non-self-consistent and partially-self-consistent HSE06 starting from the PBE orbitals with SCF convergence thresholds of 1 eV to 100 $\mu$eV ($10^0$ - $10^{-4}$ eV). The black lines with markers show the reference dispersion obtained from fully self-consistent calculations performed to a tolerance of 1 $\mu$eV ($10^{-6}$ eV).

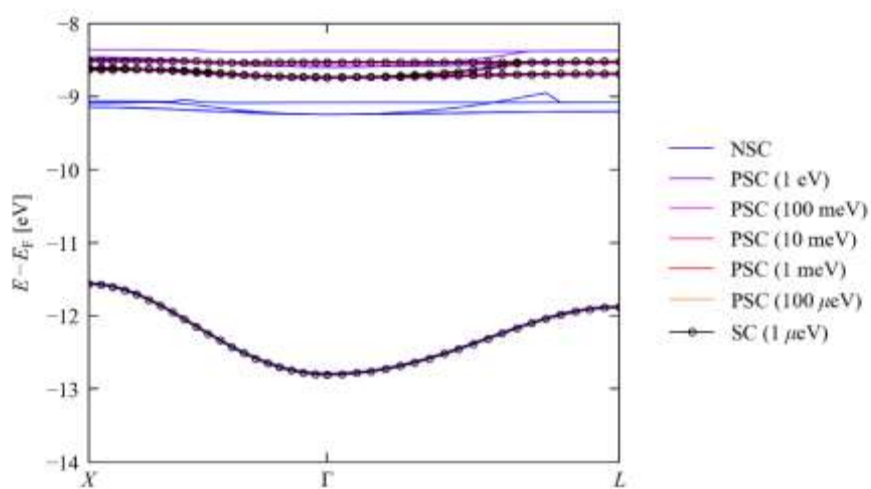

**Figure S7.12** Change in the dispersion of the Zn d bands in ZnTe calculated using non-self-consistent and partially-self-consistent HSE06 starting from the PBE orbitals with SCF convergence thresholds of 1 eV to 100 $\mu$eV ($10^0$ - $10^{-4}$ eV). The black lines with markers show the reference dispersion obtained from fully self-consistent calculations performed to a tolerance of 1 $\mu$eV ($10^{-6}$ eV).

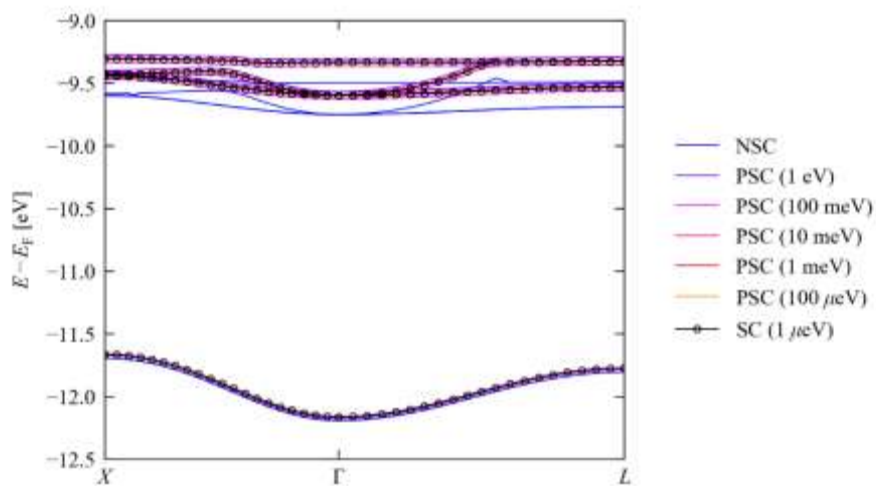

**Figure S7.13** Change in the dispersion of the Cd d bands in CdTe calculated using non-self-consistent and partially-self-consistent HSE06 starting from the PBE orbitals with SCF convergence thresholds of 1 eV to 100 $\mu$eV ($10^{0}$ - $10^{-4}$ eV). The black lines with markers show the reference dispersion obtained from fully self-consistent calculations performed to a tolerance of 1 $\mu$eV ($10^{-6}$ eV).

**S8. Transition-Metal Oxides: CoO and NiO**

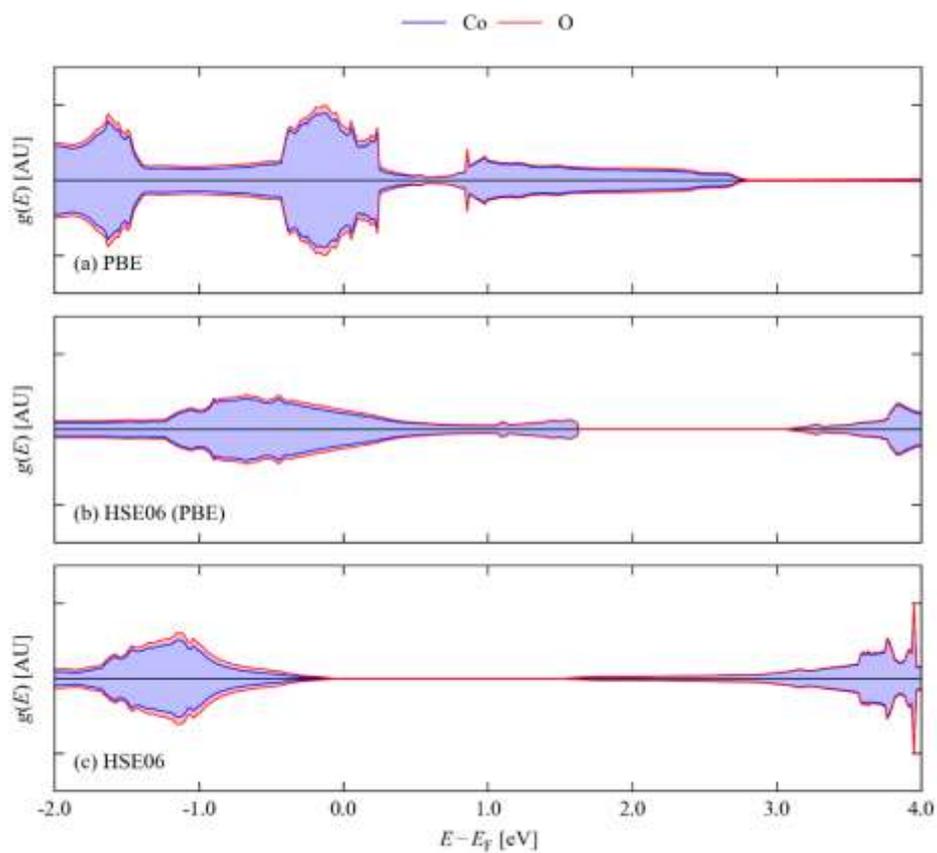

**Figure S8.1** Electronic density of states $g(E)$ (DoS) of CoO calculated using PBE[1] (a), non-self-consistent HSE06[9] using the PBE orbitals (b) and self-consistent HSE06 (c). The DoS is drawn as a stacked-area plot showing the projection onto Co (blue) and O (red) atomic states.

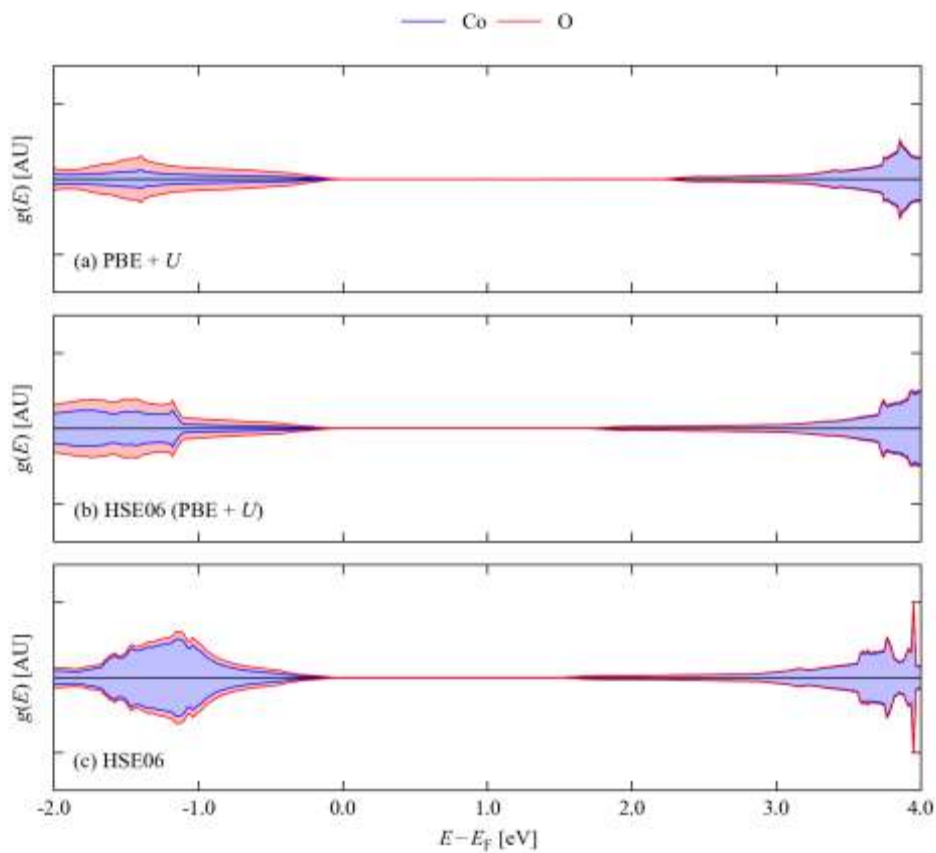

**Figure S8.2** Electronic density of states $g(E)$ (DoS) of CoO calculated using PBE + $U$ (a), non-self-consistent HSE06 using the PBE + $U$ orbitals (b) and self-consistent HSE06 (c). Details of the PBE + $U$ calculations are given in the text. The DoS is drawn as a stacked-area plot showing the projection onto Co (blue) and O (red) atomic states.

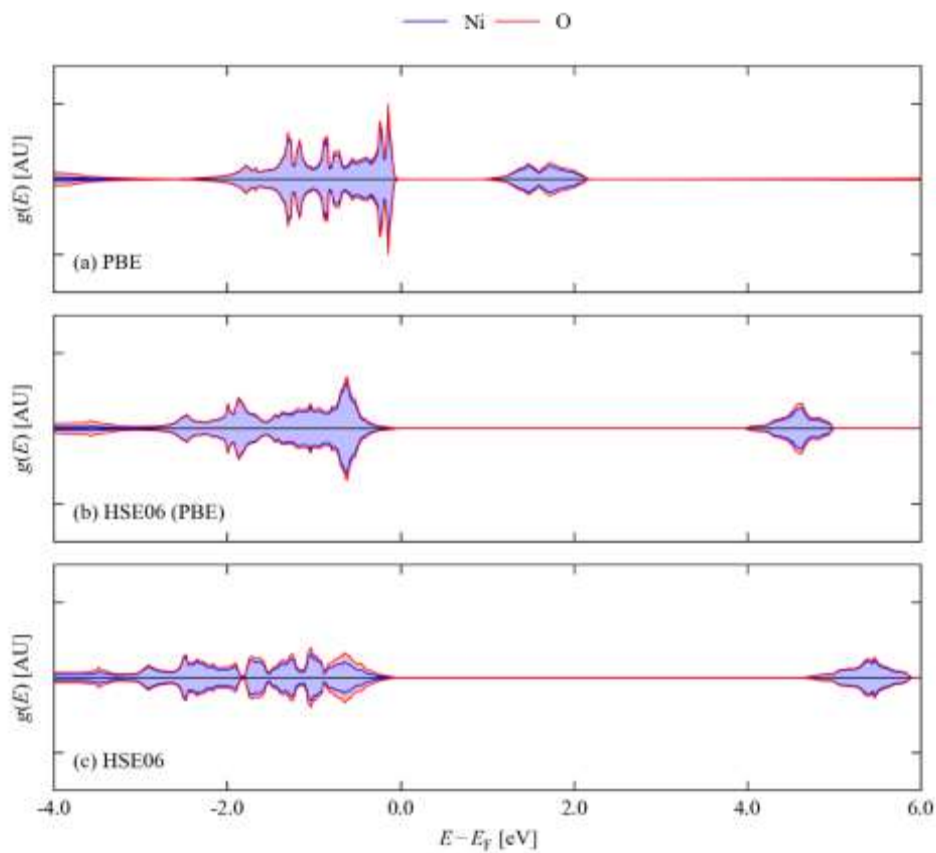

**Figure S8.3** Electronic density of states $g(E)$ (DoS) of NiO calculated using PBE (a), non-self-consistent HSE06 using the PBE orbitals (b) and self-consistent HSE06 (c). The DoS is drawn as a stacked-area plot showing the projection onto Ni (blue) and O (red) atomic states.

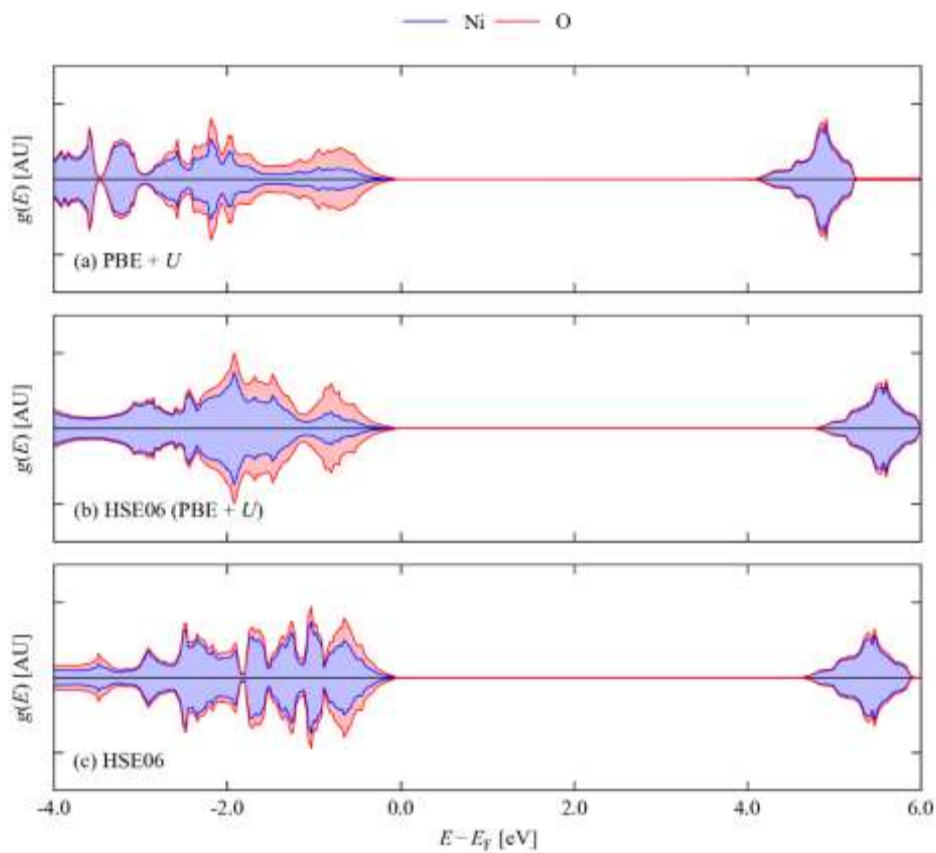

**Figure S8.4** Electronic density of states $g(E)$ (DoS) of NiO calculated using PBE + $U$ (a), non-self-consistent HSE06 using the PBE + $U$ orbitals (b) and self-consistent HSE06 (c). Details of the PBE + $U$ calculations are given in the text. The DoS is drawn as a stacked-area plot showing the projection onto Ni (blue) and O (red) atomic states.

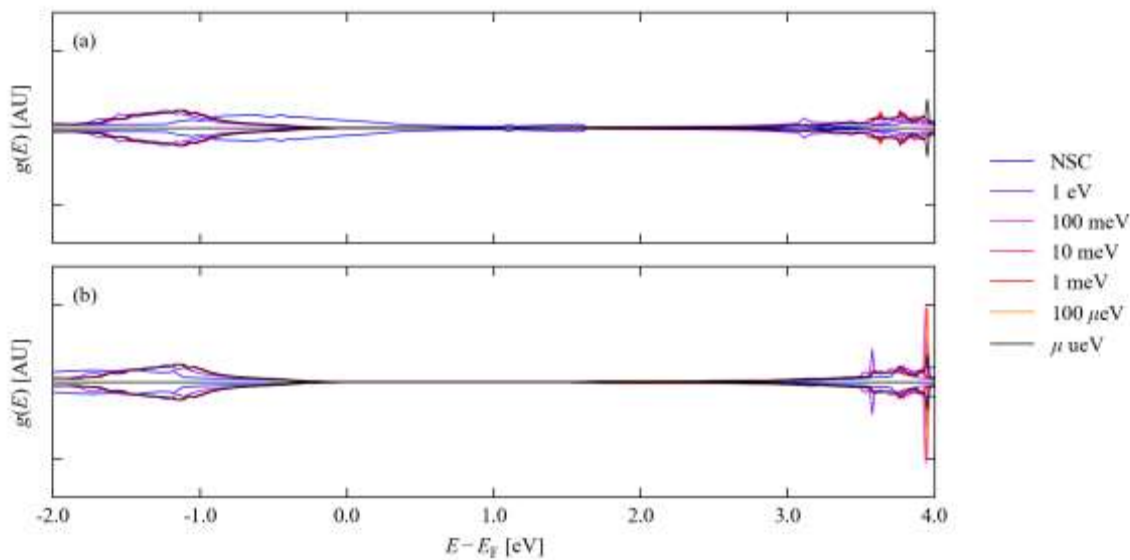

**Figure S8.5** Change in the valence band density of states $g(E)$ (DoS) of CoO calculated using HSE06 starting from the PBE (a) and PBE + $U$ orbitals (b) with SCF tolerances of 1 eV - 100 $\mu$eV ($10^0$-$10^{-4}$ eV) on the total energy. Details of the PBE + $U$ calculations are given in the text. The lines are colour coded from blue to orange to indicate successively tighter convergence thresholds. The blue line shows the DoS obtained from non-self-consistent calculations using the initial orbitals, and the black line shows the reference DoS obtained from the fully self-consistent calculation performed to a tolerance of 1 $\mu$eV ($10^{-6}$ eV).

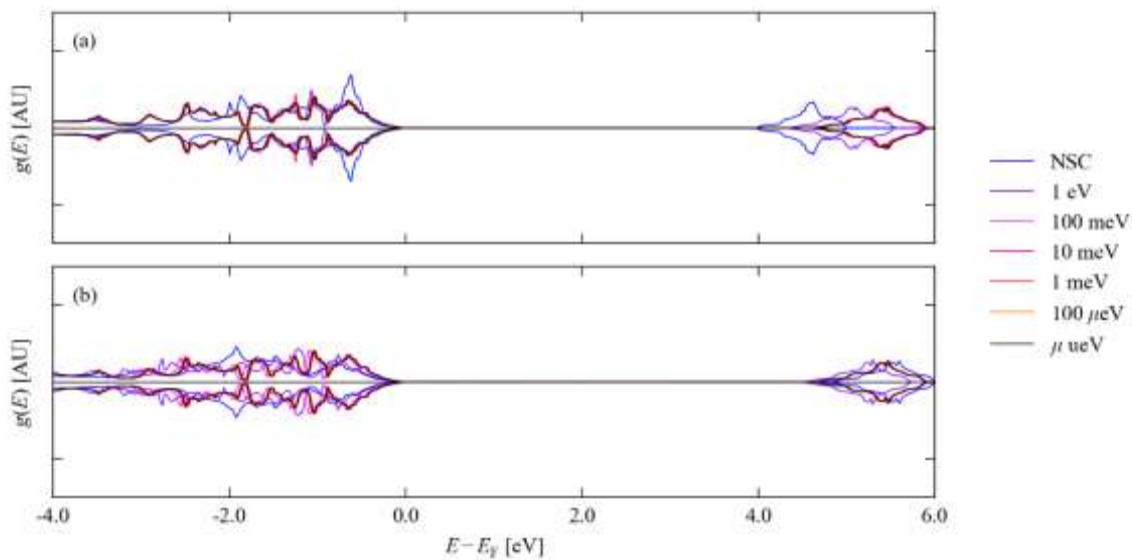

**Figure S8.6** Change in the valence band density of states $g(E)$ (DoS) of NiO calculated using HSE06 starting from the PBE (a) and PBE + $U$ orbitals (b) with SCF tolerances of 1 eV - 100 $\mu$eV ($10^{0}$-$10^{-4}$ eV) on the total energy. Details of the PBE + $U$ calculations are given in the text. The lines are colour coded from blue to orange to indicate successively tighter convergence thresholds. The blue line shows the DoS obtained from non-self-consistent calculations using the initial orbitals, and the black line shows the reference DoS obtained from the fully self-consistent calculation performed to a tolerance of 1 $\mu$eV ($10^{-6}$ eV).